\documentstyle[12pt]{article}
\begin{document}

\newcommand{\ket}[1] {\mbox{$ \vert #1 > $}}
\newcommand{\bra}[1] {\mbox{$ <
#1 \vert $}}
\newcommand{\bk}[1] {\mbox{$ \langle #1 \rangle $}}  \newcommand{\scal}[2]
{\mbox{$ < #1 \vert #2 > $}}
\newcommand{\expect}[3] {\mbox{$ \bra{#1} #2
\ket{#3} $}}  \newcommand{\ki}{\mbox{$ \ket{\psi_i} $}}
\newcommand{\bi}{\mbox{$ \bra{\psi_i} $}}  \newcommand{\p} \prime
\newcommand{\e}{\mbox{$ \epsilon $}}  \newcommand{\la} \lambda

\newcommand{\om}{\mbox{$ \omega $}}  \newcommand{\cc}{\mbox{$\cal C $}}
\newcommand{\w} {\hbox{ weak }} \newcommand{\al}{\mbox{$ \alpha $}}

\newcommand{\be}{\mbox{$ \beta $}}

\overfullrule=0pt \def\sqr#1#2{{\vcenter{\vbox{\hrule height.#2pt
          \hbox{\vrule width.#2pt height#1pt \kern#1pt
           \vrule width.#2pt}
           \hrule height.#2pt}}}}
\def\square{\mathchoice\sqr68\sqr68\sqr{4.2}6\sqr{3}6} \def\lrpartial{\mathrel
{\partial\kern-.75em\raise1.75ex\hbox{$\leftrightarrow$}}} \begin{flushright}
ULB-TH 94/02\\ LPTENS 95/07\\February 1995\\
\end{flushright} \vskip 1. truecm
\centerline{\LARGE\bf{From Vacuum Fluctuations to Radiation:}}
\vskip .2 truecm
\centerline{\LARGE\bf{Accelerated Detectors and Black Holes.}}
\vskip 1. truecm
 \centerline{  S. Massar\footnote{e-mail:
smassar @ ulb.ac.be}$\/^{,}$ \footnote{Boursier IISN}} \centerline{Service de
Physique Th\'eorique, Universit\'e Libre de Bruxelles,} \centerline{Campus
Plaine, C.P. 225, Bd du Triomphe, B-1050 Brussels, Belgium} \vskip 5 truemm
\centerline{R. Parentani\footnote{e-mail: parenta@physique.ens.fr}}
\centerline{Laboratoire de Physique Th\'eorique de l'Ecole
Normale Sup\'erieure\footnote{Unit\'e propre de recherche du C.N.R.S.
associ\'ee \`a l'
ENS et \`a l'Universit\'e de Paris Sud.}}
\centerline{24 rue Lhomond
75.231 Paris CEDEX 05, France}
\vskip 1.5 truecm
\vskip 1.5 truecm
{\bf Abstract }
The energy and particle fluxes emitted
by an accelerated two level atom are analysed in detail.
It is
shown both perturbatively and non perturbatively that the total number of
emitted photons is equal to the number of transitions characterizing thermal
equilibrium thereby confirming that each internal transition is accompanied by
the emission of a Minkowski quantum.
The mean fluxes are then
decomposed according to the final state of the atom and the notion of
conditional flux is introduced.
This notion is
generalized so as to study
the energy content of the
vacuum fluctuations that induce the transitions of the accelerated
atom. The physical relevance of these conditional fluxes is displayed
and contact is made with the formalism of Aharonov et al.
The same
decomposition is then applied to isolate,
in the context of black hole radiation,
the energy content of the particular vacuum
fluctuations which are converted
into on mass shell quanta.
It is shown that initially these fluctuations
are located around the light like geodesic
that shall generate the horizon and
have exponentially
large energy densities. Upon
exiting from the star they break up into two pieces. The external one is red
shifted and becomes an
on mass shell
quantum, the other, its ''partner", ends up in
the singularity.  We avail ourselves of this analysis to study
back reaction effects
to the production of a
single quantum.
\vfill \newpage

\section{Introduction}\label{Intro}

The history of Hawking radiation and of uniformly accelerated detectors in
Minkowski vacuum have to some extent evolved in parallel. Indeed
Hawking's discovery\cite{hawk2} that black holes emit particles at
temperature $1/8\pi M$ ($M$ being the mass of the black hole)
 closely preceded and inspired Unruh's
discovery\cite{Unruh} that a uniformly accelerated detector will thermalize at
a
temperature $a/2\pi $ ($a$ being the acceleration of the detector).
The global structure of the geometry (the
presence of horizons), the singular behavior of the modes and the mechanism of
particle emission (excitation of the detector) are in close analogy in both
problems.

In this article we exploit this isomorphism to obtain a quantum description
of the fluctuations of the energy density in both problems. To this end,
we first describe the energy content of the
field configurations correlated to excitations of
an accelerated system.
We then apply the same analysis to obtain
the energy distribution of field
configurations (the vacuum fluctuations) which get converted into Hawking
quanta. The main point of this analysis is that it
goes beyond the description of the mean (averaged) energy of the field (see
ref. \cite{birreld} for a review): we compute the energy of the
fluctuations around the mean
and evaluate specific gravitational back reaction effects
induced by these vacuum fluctuations.

In Part 2,
following the work of references
\cite{Unruh}\cite{UnWa}-\cite{AM},
we analyze the mean energy density emitted by a uniformly
accelerated detector.
We
emphasize
the difference between global quantities (the total energy, the number of
quanta) and local quantities (the energy density) and we exhibit the essential
role of transients in ensuring that global properties are respected
(such as the positivity of the total energy).
Our results generalize previous perturbative\cite{AM} and
non perturbative\cite{Unru2} analyses. In particular, we  show that the
total mean number of Minkowski photons emitted is equal to the mean number of
internal transitions of the two level atom which have occurred during
the interacting period.

In Part 3, we decompose the mean flux analyzed in Part 2.
 It is written as a sum of two terms\cite{UnWa}\cite{Grove}\cite{AM}.
The first term is the energy density  if the atom has made a transition
and the second term is the energy density has not
made a transition. From these energy densities we introduce the central notion
of  conditional
energy emitted. This notion is then generalized so as to investigate others
correlations.
We
decompose the mean vacuum energy (which is zero) into the energy densities
correlated to a future transition (or to the absence of a future transition)
of the two level atom.
These conditional
energy densities describe the vacuum fluctuations
which shall give rise to transitions of the atom.

This description follows the treatment
of ref. \cite{bmpps}
 wherein the mean current density carried by created pairs in an
external electric field was decomposed into conditional current densities
so as to describe the current carried by a specific pair of quanta.

As this approach to isolate certain
vacuum fluctuations based on
conditional values of operators
is
rather new
and leads to
peculiar
results such as
complex energy densities
we shall dwell on it somewhat. In
Section \ref{deco} we show how the non diagonal matrix elements which
describe these vacuum fluctuations can be obtained by decomposing the mean
value of the energy density.
Their
physical relevance are then revealed by introducing an additional quantum
system coupled to $T_{\mu\nu}$. Indeed, these complex densities
have a natural interpretation in Quantum Mechanics only.
In perturbation theory, one finds that the modification of the
wave function of the additional system is controlled by these matrix elements.
(In particular, in
the black hole situation, the first order response of the metric to the
creation of a particular Hawking photon is controlled by these
matrix elements.)
This was discussed
by Aharonov and collaborators\cite{aharo} who considered the additional
system to be a measuring device as in the von Newmann approach to measurement
theory\cite{Vonn}. For the sake of completeness, at
the end of this article, in Part 5, we show how
the approach of Aharonov et al. can be adapted and
generalized to the present cases where we are more concerned
by backreaction effects than measurability problems.

The
analysis of vacuum fluctuations is applied to  black
hole radiation in Part 4.
We obtain
the energy density of the
vacuum fluctuations which are converted by the time dependent geometry of a
collapsing star into Hawking photons.
The properties of these energy densities give rise to specific
quantum gravitational back reaction effects which are not present in the mean
theory.
At the present time, the only mathematically consistent
treatment  of the backreaction is the semiclassical theory (see refs.
\cite{Bardeen}\cite{PT}\cite{Massar2}).
In this approximation,
the external field remains purely classical and only
the mean value of the matter current operator acts on it as a source.
The quantum
properties of the matter, i.e. its fluctuations and correlations, are
completely ignored.
But it is the fluctuations which are problematic. Indeed it has been stressed
by t'Hooft\cite{THooft} and Jacobson \cite{Jacobson} that Hawking's
derivation of black hole
radiation is no longer valid
as soon as gravitational interactions are taken into
account because it makes appeal to the structure of the vacuum on exponentially
small scales.
In the present work this is seen in
particularly vivid fashion. Indeed we shall show that if a particle is emitted
by the black hole a time $u$ after collapse
with
asymptotic
frequency $\la$, then there
was a vacuum fluctuation inside the star of energy density $O(\om^2)$
(where $\om=\la e^{u/4M}$) located on a distance scale of order
$\om^{-1}$.
After a time $u=O(4M \ln M)$ for a typical
$\la=O(M^{-1})$, $\om$ is greater than a Planck frequency and the free field
theory of Hawking is unjustified. This aspect of the fluctuations
has been
presented
together with F.
Englert in ref. \cite{EMP}.

There have been two attitudes in the literature to confront this situation.
The first is to try to guess what is the physics at the Planck
scale near the horizon and how the Hawking radiation emerges therefrom
\cite{sus2}-\cite{Engl}.
The second has been to use Einstein equations to
investigate how back reaction effects modify the production of Hawking
photons \cite{Verlinde}\cite{Wilceck}.
The present
article places itself in this latter vein. We  show at the end of
Section \ref{bbb} how our results can be applied to study some simple back
reaction effects.  We first evaluate the change in
probability of finding a specific Hawking photon due to a modification of the
background metric.
We then show how the creation of a particular photon modifies
the probability of finding subsequent photons.
In both cases,
one sees explicitly that the change in probability is entirely
controlled
by
the conditional value
of the energy
momentum tensor.
The more difficult problem of
the self interaction of a Hawking photons with itself as it is created
necessitates the analysis of loop corrections and we hope to report on it in a
subsequent paper.

\section{The Energy Emitted by an Accelerated Atom}\label{EnAt}

\subsection{Introduction}\label{Intro2}

It is now well known that a uniformly accelerated two level atom
thermalizes in Minkowski vacuum at temperature $a/2 \pi$\cite{Unruh}.
But it is much more complicate to obtain a complete description
of the fluxes emitted by this thermalized atom.

As first pointed out by Grove\cite{Grove}, when thermal equilibrium is
reached there is no net  emission of energy.
His argument is the following. The accelerator feels the
effect
of a thermal bath.
So first consider the inertial two level atom in thermal equilibrium. The
time independence of the Hamiltonian and the
stationarity of the
state of the atom (in the thermodynamic sense) guarantee no mean flux,
since each
absorbed photon is re-emitted  with the
same energy.
This argument is immediately applicable to the accelerator since the
fact that $a={\rm constant}$ implies that his physics is translationally
invariant in his proper time $\tau $
(i.e. invariance under boosts). Since Minkowski
vacuum is
also an eigenstate of the boost operator, the total
Rindler energy is conserved.

A detailed picture of the steady state emerges from the following
consideration. Focus on the ground state of the accelerator which excites
by absorbing a Rindler quanta (a rindleron) coming in from its left. Then the
field
configurations  to its right is depleted of this rindleron. Since this
rindleron carried positive energy, its removal  can be described as the
emission of  negative energy to the right. In equilibrium there is also to be
considered the process of disexcitation corresponding to the emission of
positive energy   to the right. The Einstein relation
guarantees that the two cancel.
This implies no net energy flux.

However the accelerated atom is
in Minkowski vacuum, hence all perturbations of the radiation state
lead
to
the
production of Minkowski quanta.
Then, how do we reconcile the (certainly) positive energy of
these
produced quanta with the absence of radiated energy in thermal equilibrium?
The
answer
lies in a
global treatment of the radiation field which also takes into account the
transients due to switching on and off the detector (or equivalently
the transients which occur when the detector passes from an
inertial trajectory to the accelerated one).

That transients may have a global content
which
depends on the whole history also occurs in the problem of the classical
electromagnetic field emitted by a uniformly accelerated charge in 3
dimensional Minkowski space (see ref. \cite{Boul2}). We point out that
this situation is
particular to uniformly accelerated systems and makes explicit appeal to the
exponential Doppler shift between accelerated and inertial reference
frames. Indeed (anticipating some notations introduced in the next sections)
one can express the total Minkowski energy emitted (in
V-modes) as \begin{eqnarray}
E_M = \int_0^{+\infty}
dV \langle T_{VV}\rangle
=
\int_{ - \infty}^{+ \infty} dv
\langle T_{vv}\rangle {dv \over dV}
= C_i e^{-a \tau_i}-
C_f e^{-a \tau_f} \label{fournul} \end{eqnarray}
where the first integral
is
limited to the domain $V=t+z>0$ because the accelerator remains in the quadrant
$z>|t|$.
In the second equation we have used the Rindler coordinate $v$
related to $V$ by $dv / dV = e^{-av}$; and in the
last equation we have made appeal to Grove's theorem which states that the
integrand vanishes everywhere except at the endpoints $v_i$ ($= \tau_i $) and
$v_f$ ($= \tau_f $); $C_i$ and $C_f$ depend on the exact form of the
transients at $\tau_{i,f}$.

 Equation (\ref{fournul}) can also be written as an
integral over the rates of absorption and of emission of Rindler
photons taking into account that a transition at time $\tau$ is accompanied by
the emission of a Doppler shifted Minkowski photon of frequency $\om(\tau) = m
dv/dV=me^{-a\tau}$ where $m$ is the resonant frequency of the atom. Hence the
energy emitted is \begin{eqnarray}
E_M &=& \int_{\tau_i}^{\tau_f}\! d\tau \ R\/
\om(\tau)
\nonumber\\
&=&{R\/m \over a}( e^{-a \tau_i} - e^{-a\tau_f } ) +( C^\prime_i e^{-a \tau_i}
- C^\prime_f e^{-a \tau_f}
)
\label{fourtwo}
\end{eqnarray}
Here $R$ is the number of transitions per unit proper time.  The constants
$C^\prime_i$ and $C^\prime_f$
 depend on the detailed way the interaction is turned on and off at
$\tau_{i,f}$. So the integral eq. (\ref{fourtwo}) is
of the same form as eq. (\ref{fournul}).  These two   expressions are
compatible because all the photons  in  eq. (\ref{fourtwo}) interfere in such
a way that  their energy is only found at the edges  of
the
interaction period as in eq. (\ref{fournul}).

This part is organized as follow. Section 2.2 is devoted to recalling
the main properties of Rindler quantization. In Section 2.3 we present the
model of the accelerated two level atom. In Section 2.4  we obtain formal
expressions for the energy emitted.
We decompose the flux according to the
final state of the atom thereby introducing the central notion of conditional
energy emitted.  The
properties of these conditional fluxes
as well as their physical meaning shall be displayed in Part 3.
As a warm up,
we present in Sections 2.5 - 2.7
the various properties of the mean flux
 insisting on the role of the transients  in guaranteeing that
global properties are respected.
The mean fluxes as the atom thermalizes are discussed in Section 2.5 and then
in thermal equilibrium in Section 2.6. Finally in Section 2.7 we use the exact
solvable model of ref. \cite{RSG} to extend the previous perturbative
results to all order
in $g$.

\subsection{The Rindler Quantization in 1+1 Dimensions}

In this section, we review the relevant properties of the Rindler
quantization of the scalar field in Minkowski space time in 1+1 dimensions.
The conformal invariance of the massless scalar field is best
exploited by using the light like coordinates $U,V$ defined by
\begin{eqnarray}
U=t-z
\nonumber\\
V=t+z
\label{UV}
\end{eqnarray}
Whereupon the Klein-Gordon equation takes the form $\partial_U\partial_V \phi
= 0$ and any solution can be written as
\begin{equation} \phi(U,V) = \phi(U) + \phi(V) \label{threethreei}
\end{equation} From now on we shall drop the right moving piece and consider
the ''$V$" term only. It is
obvious that all conclusions shall be equally valid
for
the
right
movers.

The second quantized field can be decomposed into the orthonormal
complete basis of
Minkowski modes
\begin{equation} \phi(V) = \int_0^{\infty} d\omega  \left( a_{\om}
\varphi_{\om} (V) + a_{\om}^\dagger \varphi_{\om}^* (V)\right)
\label{threethreeii} \end{equation} \begin{equation} \varphi_{\om} (V) = {
e^{-i\om V} \over \sqrt{4 \pi \om}} \label{threethreeiii}
\end{equation}
Minkowski vacuum $\ket{0_M}$ is the state
 annihilated by all the $a_{\om}$'s. The
propagator (Wightman function)
in Minkowski vacuum is \begin{eqnarray}
G_+(V,V^\prime)
&=&\expect{0_M}{ \phi(V)\phi(V^\prime)}{0_M} \nonumber\\
&=&  \int_0^{\infty} \! d\om \varphi_{\om} (V) \varphi_{\om}^*(V^\prime) =
-{1 \over 4 \pi} \log (V-V^\prime
- i \e) \label{threethreeiv} \end{eqnarray} The (normal ordered)
hamiltonian of the field is
(for left movers) \begin{equation} H_M= \int_{-\infty}^{+\infty} \! dV T_{VV}
=  \int_0^{\infty} \! d\om {\om }  (
a_{\om}^\dagger  a_{\om} ) \label{threethreev} \end{equation}
where
\begin{equation}
 T_{VV} = \partial_V \phi  \partial_V \phi
\label{threetvv}
\end{equation}
 Therefore Minkowski vacuum is the ground state of the hamiltonian $H_M$:
$H_M \ket{0_M} =0$.

The uniformly accelerated observer
will be
taken to be in the right
(R) Rindler
quadrant $U<0,V>0$. In this quadrant one defines Rindler coordinates $\rho ,
\tau$ by
\begin{equation} \left \{ \begin{array}{ll} t=\rho\ \mbox{sinh} a\tau
\\ x=\rho\ \mbox{cosh} a\tau \end{array} \right. \label{threethreevbis}
\end{equation}
The accelerated observer follows the trajectory
$\rho = 1/a$ (where $a$ is its acceleration)
and its proper time is $\tau$
(see figure 1).
In this quadrant one introduces also the light like
Rindler coordinates $u,v$ defined by \begin{eqnarray}
u = \tau - a^{-1} \ln a\rho
\nonumber\\
 v= \tau + a^{-1} \ln a\rho
\end{eqnarray}
These are related to the light like Minkowski coordinates $U, V$ by
\begin{eqnarray}
\left \{ \begin{array}{ll}
U=-a^{-1}\ e^{-au} \\ V=a^{-1}\ e^{av} \end{array} \right.
\label{threethreevi} \end{eqnarray}
 The
coordinates eq. (\ref{threethreevbis})  may be extended to the left (L) Rindler
quadrant by the analytic continuation $\tau \rightarrow \tau \pm i \pi/a$. One
introduces in L the light like coordinates $u_L, v_L$ given by
\begin{eqnarray}
\left \{ \begin{array}{ll}
U=a^{-1}\ e^{au_L} \\ V=- a^{-1}\ e^{- a v_L}\end{array} \right.
\label{threethreeviL}
\end{eqnarray}

The natural basis of quantization the uniformly accelerated observer would
choose  is the Rindler basis which consists of plane waves in the variables
$u,v$ (Rindler modes). This is because $u, v$ are related to its proper time
$\tau$ as $U, V$
are
related to the Minkowski time $t$. The Rindler $v$ modes are thus
given, in strict analogy with eq. (\ref{threethreeiii}), by
\begin{equation} \varphi_{\la, R} (v) = {e^{-i\la v} \over \sqrt{4 \pi \la}}
\label{threethreeiiiv}
\end{equation}
Since the $\varphi_{\la, R}$ constitute a complete set in R ($V>0$)
only, they cannot be related to the Minkowski basis by a unitary
transformation.
One must also introduce Rindler modes living in the left quadrant.
But one finds that the Bogoljubov transformation relating the
  Minkowski
modes to the Rindler modes is singular  at $V=0$ \cite{tmunu}
 and care must be
taken to define it as a limit if the Minkowski properties of the theory are
to be satisfied.
To this end it is useful to first introduce an alternative basis
of positive frequency Minkowski modes,
eigenmodes of $iaV\partial_V$ ($=i\partial_v$ for $V>0$),
and defined for all $V$
\cite{Unruh}:
\begin{eqnarray}  \varphi_{\la,M}(V) &=& \int_0^{\infty}\! d\omega\
\gamma_{\la,\om} \varphi_{\om}(V) \nonumber\\ &=& {\left[ a(\e +
iV)\right]^{-i \la /a} \over  \sqrt{ ( e^{\pi \la /a} - e^{-\pi \la /a})4 \pi
\la}}\nonumber\\  &=_{\e\rightarrow 0}&
{e^{\pi \la/2a}\over  \sqrt{\vert e^{\pi \la /a} -
e^{-\pi \la /a}\vert}} { \theta(V)(aV)^{-i \la /a}
\over \sqrt{4 \pi \vert \la\vert}} +
{e^{-\pi \la/2a}\over \sqrt{\vert e^{\pi \la /a} - e^{-\pi \la /a}\vert}}
{\theta(-V) (-aV )^{-i \la /a} \over \sqrt{ 4 \pi \vert \la\vert}}
\nonumber\\
\label{threethreevii} \end{eqnarray}
where \begin{equation}  \gamma_{\la,\om}=
\left({1\over \Gamma({i\la/ a})}  \sqrt{a \pi \over {\la }
\mbox{sinh}{\pi \la /a} }\right) {1 \over \sqrt{2 \pi a \om}} \left(
{\om \over
a } \right)^{i\la/ a} e^{-\om \e} \label{threethreeviii} \end{equation}

The first factor in $\gamma_{\la,\om}$ is a pure phase introduced for
convenience. The factor $ e^{-\om \e}$ is the crux of the construction.
It
defines the integral eq. (\ref{threethreevii}), regularizes the modes
$\varphi_{\la,M}(V)$ at $V=0$  and ensures the correct
Minkowski
properties of the theory. For instance it gives the correct pole
prescription
at $V=V^\prime$ of the propagator eq. (\ref{threethreeiv}) when expressed in
terms
of the modes $\varphi_{\la,M}$ as $G_+(V,V^\prime) = \int_{-\infty}^{+
\infty}
d \la   \varphi_{\la,M}(V)\varphi_{\la,M}^*(V^\prime)$. The limit $\e
\rightarrow 0$ is to be taken at the end of all calculations.

The  annihilation and creation operators corresponding to
the modes $ \varphi_{\la,M}$ are $a_{\la,M}$ and $a_{\la,M}^\dagger $:
\begin{equation} \phi(V) =\int_{\infty}^{+\infty} d\la \left( a_{\la,M}
\varphi_{\la,M}(V) + a_{\la,M}^{\dagger} \varphi^*_{\la,M}
(V)\right)
 \end{equation}

The right and left Rindler modes  $\varphi_{\la,R} (V)$ and $\varphi_{\la,L}
(V)$ can now be defined by the linear unitary
transformation
\begin{equation}
\left\{
\begin{array}{ll} \varphi_{\la , M} = \alpha_{\la} \varphi_{\la,R} +
\beta_{\la} \varphi_{\la,L}^* \\ \varphi_{- \la , M} = \beta_{\la}
\varphi_{\la,R}^* + \alpha_{\la} \varphi_{\la,L} \label{threethreeix}
\end{array} \right. \quad \la>0 \label{threethreex} \end{equation}
with
\begin{equation}  \alpha_{\la} = { e^{\pi \la / 2 a} \over \sqrt { e^{\pi \la
/ a} - e^{- \pi \la / a}}} \quad , \quad
 \beta_{\la} =  { e^{-\pi \la / 2 a} \over
\sqrt { e^{\pi \la / a} - e^{- \pi \la / a}}}
\quad \mbox{and} \quad \alpha_{\la}^2 -
\beta_{\la}^2 = 1 \label{threethreexi}
\end{equation}
In the limit $\e \rightarrow 0$ the
Rindler modes take the familiar form (see eq. (\ref{threethreeiiiv}))
\begin{equation} \begin{array}{ll} \varphi_{\la,R} (V) =
\theta(V) (aV)^{-i\lambda/a} /{ \sqrt{4 \pi \la}}
 =
 e^{-i\la v}/\sqrt{4 \pi \la} \\ \varphi_{\la,L} (V) =
\theta(-V) (-aV)^{i\lambda/a}/{ \sqrt{4 \pi \la}} =
 e^{-i\la v_L}/\sqrt{4 \pi \la}
\end{array}
\label{threethreexii} \end{equation}
For finite $\e$ they differ from these
limiting forms only
when $V \leq \e$.
The Rindler destruction and creation operators $a_{\la,R},
a_{\la,R}^\dagger $ and $a_{\la,L}, a_{\la,L}^\dagger $
associated to these modes are related to the Minkowski operators $a_{\la,M}$
by the following Bogoljubov transformation
\begin{equation}
\left\{
\begin{array}{ll}
a_{\la,R}  = \alpha_{\la} a_{\la,M} + \beta_{\la} a_{-\la,M}^{\dagger}\\
a_{\la,L} = \alpha_{\la} a_{-\la,M} + \beta_{\la}
a_{\la,M}^{\dagger}\label{threethreeixbog} \end{array} \right.\end{equation}
(by virtue of the orthonormal character of the two sets of modes
$\varphi_{\la,R} ,\varphi_{\la,L}$ and $\varphi_{\la , M}$ as well as
$\alpha_{\la}^2 -\beta_{\la}^2 = 1 $).
One immediately deduces that the mean number of Rindler quanta present in
Minkowski vacuum is given by the Bose Einstein distribution
\begin{equation}
\bra{0_M} a_{\la,R}^{\dagger} a_{\la^{\prime},R} \ket{0_M} = {1 \over
e^{2 \pi \la /a} - 1} \delta (\la - \la^{\prime})
\label{meannum}
\end{equation}
It is useful to introduce the generator of boosts $ H_R $ since
it generates translations in $\tau$ and is therefore the Hamiltonian for
the accelerated observer.
\begin{eqnarray} H_R &=& \int_{-\infty}^{+\infty}\!\! dV
 aV T_{VV} \nonumber\\
&=& \int_{-\infty}^{+\infty}\!\! dv T_{vv} \ -
\   \int_{-\infty}^{+\infty}\!\! dv_L T_{v_Lv_L} \nonumber\\
&=& \int_0^{+\infty}\!
d\la \la (a_{\la,R}^\dagger a_{\la,R} - a_{\la,L}^\dagger a_{\la,L})
\nonumber\\
&=& \int_{0}^{+\infty}\!
d\la \la (a_{\la,M}^\dagger a_{\la,M} - a_{-\la,M}^\dagger a_{-\la,M})
\label{threefouriv} \end{eqnarray}
One sees that quanta in L carry
negative ``Rindler energy''. This is because $\tau$ (defined by arctanh$(t/z)$)
 goes backwards in time in L ($d
\tau / dt <0$ in $L$).
Furthermore since Minkowski vacuum
is invariant under boosts it is annihilated by the Rindler energy operator:
\begin{equation}
H_R\ket{0_M}=0
\label{threefouriv2}
\end{equation}
Using the above Bogoljubov transformation it is easy to show that
\begin{equation} \ket{0_M} = \prod_\la {1 \over \alpha_\la} e^{-{ \beta_\la
\over \alpha_\la} a^\dagger_{\la,L} a^\dagger_{\la,R}} \ket{0_{RL}}
\label{threethreexiii} \end{equation}
where $\ket{0_{RL}}= \ket{0_{R}} \otimes\ket{0_{L}}$ is Rindler vacuum in
both the right(R) and left(L) quadrants.
Thus, from eq. (\ref{threefouriv}), one sees that the
 pairs of Rindler quanta (whose mean number is given by eq. (\ref{meannum}))
present in Minkowski vacuum carry zero Rindler energy.
One sees also that upon tracing over the left quanta the
reduced density matrix in the right Rindler quadrant is an exact thermal
distribution of right rindlerons \cite{birreld}.

Eq. (\ref{threethreexiii}) will be  useful in the next chapters since
it provides an easy way of calculating the probability ($\vert \bk{\psi_R
\vert 0_M} \vert^2$) of
finding in Minkowski vacuum a state ( $\ket{\psi_R}$) containing a given number
of
Rindler quanta.

\subsection{The Uniformly Accelerated Two Level Atom}

The situation we consider is a uniformly accelerated two level atom coupled
to the massless field $\phi$ introduced in the previous section.
The trajectory of the two level atom is
given by eq. (\ref{threethreevbis}) with $\rho =a^{-1}$:
\begin{eqnarray} t_a (\tau) =a^{-1}
\mbox{sinh} a \tau & ,&  x_a (\tau) =a^{-1} \mbox{cosh} a \tau
\nonumber\\
V_a (\tau) = a^{-1} e^{a\tau} & ,& U_a (\tau) = -a^{-1} e^{-a\tau}
\label{threefivexiibis}
\end{eqnarray}
The time integral of the interaction hamiltonian is
\begin{eqnarray} \int\! dt dx\ H_{\rm int} (t,x) &=& g m \int\!
d\tau \left[ \left( f(\tau) e^{-im\tau} A  +
f^*(\tau) e^{im\tau} A^{\dagger}
\right) \phi(t_a(\tau),x_a (\tau)) \right]
\label{threefiveiiter} \end{eqnarray}
where $g$ is a dimensionless coupling constant that shall be
taken for simplicity  small enough that second order perturbation theory be
valid, $m$ is the difference of energy between the ground and the
excited state
of the atom,
$A$ is the lowering operator that induces a transition
 from the excited state to the
ground state of the atom
and $f(\tau)$ is a dimensionless function that governs when and how the
interaction
is turned on and off. The factor $m$ on the r.h.s. of eq.
(\ref{threefiveiiter})  is
introduced for dimensional reasons.
In addition, we shall assume that the $V$-part of the $\phi$ field only
is coupled to the atom. This is a legitimate truncation owing to the
chiral character of the field in $1+1$ dimensions.
For simplicity of notation it is convenient to rewrite
eq. (\ref{threefiveiiter})
 as
\begin{eqnarray} \int\! dt dx\ H_{\rm int} (t,x) =
 g m  \left[ A \phi_m^\dagger  +  A^\dagger \phi_m
\right] \label{threefiveii} \end{eqnarray} where
\begin{equation} \phi_m = \int_{-\infty}^{+\infty}\!
d\tau e^{+im\tau} f^*(\tau) \phi(\tau) \label{threefiveiii} \end{equation}

We shall be most interested in the situation where $f(\tau)=1$ inside a long
interval $\tau_i < \tau < \tau_f$ and $f(\tau)$ tends to
zero outside this interval.
Then in the limit $\tau_f-\tau_i =T \to \infty$
with $g^2 T$ finite (i.e. in the Golden Rule limit)
the concept of a transition rate
emerges
and is due to the resonance of the Minkowski vacuum fluctuations with the
fixed Rindler frequency $m$ \cite{pbt}. The
fundamental reason
why we need to work with an explicit switch function is that we want  finite
energy
momentum densities everywhere including the horizon.
We shall show that this is possible only for
sufficiently rapidly decreasing $f(\tau)$ (which
amounts to deal only with wave packets
of Rindler modes which are well defined in the ultraviolet --see eqs.
(\ref{threethreevii},
 \ref{threethreeviii})).

When
$e^{-im\tau} f(\tau)$ contains
no negative frequency in its Fourier transform with respect to $\tau$,
eq. (\ref{threefiveii}) defines
 a Lee model:
were it inertial it would only respond to the presence
of Minkowski
particles.
For a Lee model $f(\tau)$ must necessarily decrease
less
rapidly then an exponential when $\tau \rightarrow \pm \infty$. Alas
this condition is too
strong and will lead to singularities on the horizon in the
uniformly accelerating situation. Hence
we shall be obliged to consider
non Lee models which can spontaneously excite. However by
choosing $f(\tau)$ such that the negative frequency part of $e^{-im\tau}
f(\tau)$ is exponentially small the spontaneous  excitations are
exponentially . Moreover the spontaneous excitations occurs
only at
switch
on and switch off transitory periods
but do not occur during the steady regime when $f(\tau)=1$, i.e. these
excitations
do not contribute to rates.

Let us consider first the situation in which both
 the atom and the field are initially in their ground state. The state
$\ket{\psi_-}$
at $t=-\infty$ is
thus
\begin{eqnarray}
\ket{\psi_-(t=-\infty)} = \ket{0_M} \ket{-}
\end{eqnarray}
where $\ket{-}$ ($\ket{+}$) designates the ground (excited)
state of the atom.
At $t=+\infty$, when the interaction has been switched off, the state
can again be expressed in terms of the uninteracting states and it is
given, to order $g^2$,
by
\begin{eqnarray}
\ket{\psi_-(t=+\infty)} &=&  e^{-i \int dt dx H_{int}}\ket{0_M} \ket{-}
\nonumber \\
&=& \ket{0_M} \ket{-}\nonumber\\
& & -ig m \int_{-\infty}^{+\infty}\!
d\tau   f^*(\tau) e^{+im\tau}  \phi(\tau)\ket{0_M} \ket{+}\nonumber \\
& & - g^2 m^2
\int_{-\infty}^{+\infty}\!
d\tau
\int_{-\infty}^{\tau}\!
d\tau^\p f(\tau) e^{-im\tau}  \phi(\tau)
  f^*(\tau^\p) e^{+im\tau^\p}  \phi(\tau^\p)\ket{0_M}
\ket{-}\nonumber \\
&=& \ket{0_M} \ket{-}  -ig m \phi_m \ket{0_M}\ket{+}- {1 \over 2} g^2 m^2
\phi_m^\dagger \phi_m \ket{0_M}\ket{-}
\nonumber \\& &-  g^2 m^2
{\cal D}  \ket{0_M}\ket{-}
\label{state}
\end{eqnarray}
where
\begin{equation}
{\cal D} = {1 \over 2}
\int \! d\tau_2
\int \! d \tau_1 \epsilon(\tau_2 - \tau_1) e^{-i m \tau_2} f(\tau_2)
\phi(\tau_2) e^{+i m \tau_1} f^*(\tau_1) \phi(\tau_1)
\label{calD}
\end{equation}
and where $\epsilon(\tau_2 - \tau_1) = \theta(\tau_2 - \tau_1) -
\theta(\tau_1 - \tau_2 )$.
We have split the $g^2$ term in two pieces
in order to isolate the steady regime
part of the interaction.
The latter is given by the term proportional
to $\phi_m^\dagger \phi_m$
whereas the ${\cal {D}}$ term is concerned with the transitory periods
associated with the switch on and off. Indeed to order $g^2$ the  ${\cal D}$
term
does not contribute to the energy density emitted in the
steady state regime (i.e. its energy density scales like $g^2/T$ rather than
like $mg^2$).
Furthermore it carries no Minkowski nor Rindler energy. This is
proven
in the Appendix. We shall therefore drop this term
in the sequel.
To get a flavor of this, the reader can
already verify that $\cal{D}$ does not contribute to the probability
to remain in the ground state. Only the term proportional to
$\phi_m^\dagger \phi_m$ does so.

The probability $P_e$ for the two level atom to get excited is, in second
order perturbation theory
  \begin{eqnarray}
P_e =
g^2 m^2 \expect{0_M}{\phi_m^\dagger  \phi_m}{0_M} \label{threefiveiv}
\end{eqnarray}
Had we coupled the atom to both the $U$ and $V$ parts of $\phi$, the
probability would have been twice $P_e$.
When $f(\tau)$ is equal to $1$ between $\tau_i$ and $\tau_f$
and $\tau_f - \tau_i = T \rightarrow +\infty$ while $g^2 m T$ remains finite,
the operator $\phi_m^\dagger  \phi_m
$ appearing in eq. (\ref{threefiveiv}) becomes the counting operator for
rindlerons of energy $m$ ($= a_{m,R}^\dagger  a_{m,R}$) multiplied by
$\pi/ m$. A direct golden rule calculation
then shows that the probability for the
uniformly accelerated atom to get excited is
\cite{Unruh} \begin{equation} P_{e} =
(1/2)
g^2 m
T N_m \label{threefivexiii}
\end{equation}
where $N_m = 1/ (e^{2 \pi m/a} -1)$ is
the mean number of Rindler quanta present in Minkowski vacuum, see
eq. (\ref{meannum}). This proves that the atom maintained on an
accelerated trajectory  reacts to
the mean number of Rindler quanta as
the same atom, put on an inertial
trajectory, would have reacted to the mean number of Minkowski quanta.

If the initial state is the product of Minkowski vacuum and the excited state
for the atom
\begin{equation}\ket{\psi_+(t=-\infty)}= \ket{0_M}\ket{+}
\end{equation}
then $\ket{\psi_+}$ at $t=+\infty$ is
\begin{eqnarray}
\ket{\psi_+(t=+\infty)} &=& \ket{0_M} \ket{+}
-ig m \phi_m^\dagger \ket{0_M}\ket{-}-  {1\over 2} g^2
m^2  \phi_m \phi_m^\dagger \ket{0_M}\ket{+}
\nonumber\\
& &+  g^2 m^2
{\cal D}  \ket{0_M}\ket{+}
\label{state+}
\end{eqnarray}
where the operator ${\cal D}$ which appears is the same as
in eq. (\ref{state}).
The probability to be found in the ground state at $t=+\infty$ (i.e.
the probability of disexcitation) in the golden
rule limit is
\begin{equation} P_{d} = g^2 m^2 \expect{0_M}{\phi_m \phi_m^\dagger}{0_M}
= (1/2) g^2 m
T ( N_m + 1) \label{threefivexiiid}
\end{equation}

Hence, at equilibrium, by Einstein's famous argument,
 the probabilities $P_+, P_-$ to be in the excited or
ground states are given by
\begin{equation}
{P_+ \over P_-}={P_e\over P_d}={N_m\over N_m + 1} = e^{-2 \pi m / a}
\label{P/P}
\end{equation}
that is a thermal distribution at temperature $T= a/ 2\pi$.

\subsection{The Mean Fluxes to order $g^2$}

We investigate now the properties of the flux emitted by the accelerated atom
and its relations with the transition probabilities $P_e$ and $P_d$.
Since the field is massless, the conservation law reads
\begin{equation}
\partial_U T_{VV} =0
\label{conserv}
\end{equation}
Thus since the interaction, eq. (\ref{threefiveiiter}),
 occurs only on the accelerated trajectory (eq. (\ref{threefivexiibis})),
 one has in the past of this line,
 or to the right of it by virtue of eq. (\ref{conserv}),
\begin{equation}
\langle T_{VV}(V, U<U_a(\tau)) \rangle =0
\label{meanTI-}
\end{equation}
The mean value means that the initial state can be any combination
of $\ket{\psi_-}$ and $\ket{\psi_+}$. (When the initial state does contain
Minkowski quanta, the mean value
eq. (\ref{meanTI-}) should by understood as the modification of the
mean induced by the coupling eq. (\ref{threefiveii}).)

 We consider first the situation where
the atom is in its ground state at $t=-\infty$ and the field
in Minkowski vacuum (i.e. the state of the
system is $\ket{\psi_-}$). After the interaction is switched off,
the mean flux emitted on ${\cal{I}}^+$
(i.e. on $U= + \infty$) or equivalently to
the
left of the trajectory by virtue of eq. (\ref{conserv})
 is, to order $g^2$,
\begin{eqnarray}
\langle T_{VV}(V) \rangle_{\psi_-} &=&
\bra{\psi_-(t=+\infty)} T_{VV}(V, U>U_a(\tau)) \ket{\psi_-(t=+\infty)}
\nonumber\\
&=&
g^2m^2 \bra{0_M} \phi_m^\dagger T_{VV} \phi_m \ket{0_M}
-  g^2m^2\mbox{ Re}\left[
 \bra{0_M}  T_{VV} \phi_m^\dagger \phi_m \ket{0_M} \right]
\label{tVV}
\end{eqnarray}
where we have used eq. (\ref{state}) and dropped the contribution of the
$\cal{D}
$ term.
The physical meaning of the two terms on the r.h.s. of eq. (\ref{tVV})
was first discussed in \cite{UnWa}.
Before discussing it we first
rewrite
eq. (\ref{tVV}) as
\begin{eqnarray}
\langle T_{VV}(V) \rangle_{\psi_-} &=&
P_e \langle T_{VV} \rangle _{e}
+  P_g \langle T_{VV} \rangle _{g}
\label{tVVB}
\end{eqnarray}
where we
have defined
\begin{eqnarray}
\langle T_{VV}\rangle _{e}
 &=& g^2m^2 \bra{0_M} \phi_m^\dagger T_{VV} \phi_m
\ket{0_M}/ P_e\nonumber\\
\langle T_{VV}\rangle _{g}
&=& - g^2m^2\mbox{ Re}\left[
 \bra{0_M}  T_{VV} \phi_m^\dagger \phi_m \ket{0_M} \right]/ P_g
\label{45b}
\end{eqnarray}
where $P_e$ and $P_g$ are the probabilities to find the atom
in the excited
or ground state at $t=+\infty$.  $P_e$ is given in eq. (\ref{threefivexiii})
and $P_g
=1-P_e$.

 The interpretation of the two quantities $\langle T_{VV}\rangle _{e}$ and
$\langle T_{VV} \rangle _{g}$ is clear when one
recalls their origin. $\langle T_{VV}\rangle _{e}$
comes from the square of the second term of eq.
(\ref{state}) (linear in $g$) whereas $\langle T_{VV} \rangle _{g}$ comes
from an interference between the first term of eq. (\ref{state})
(unperturbed) and the third term (in which the interaction has acted twice).
Hence $\langle T_{VV}\rangle _{e}$ is the energy emitted
if the atom is found excited at $t=+\infty$ and $\langle T_{VV}\rangle _{g}$
is the energy emitted if the atom
is found in the ground state at $t=+\infty$. These fluxes
have been normalized so as to express
the r.h.s. of eq. (\ref{tVVB})  as the probability of finding the atom in
a final state times the energy
emitted if that final state is realized.
Thus $\langle T_{VV} \rangle _{g}$  and $\langle T_{VV}\rangle _{e}$
are the conditional "mean" energy emitted. (The word "mean"  is understood
here in its quantum sense, i.e. as the average over repeated realizations
of the same situation: the same initial state $\ket{\psi_-}$ and the same
final state of the atom --see Part 5 for further comments on this point).

Similarly, when the initial state of the system is $\ket{\psi_+}$,
the mean energy emitted is
\begin{eqnarray}
\langle T_{VV}(V)\rangle _{\psi_+} &=&
\bra{\psi_+(t=+\infty)} T_{VV} (V, U>U_a(\tau))\ket{\psi_+
(t=+\infty)}
\nonumber\\
&=&
g^2m^2 \bra{0_M} \phi_m  T_{VV} \phi_m^\dagger\ket{0_M}
-  g^2m^2 \mbox{ Re}\left[
\bra{0_M}  T_{VV} \phi_m  \phi_m^\dagger\ket{0_M}
\right]
\label{tVV+}
\end{eqnarray}
where we have used eq. (\ref{state+}) and dropped the $\cal{D}$ term as well.
As in eq. (\ref{tVVB}), we rewrite this flux as
\begin{eqnarray}
\langle T_{VV}\rangle _{\psi_+} = P_d \langle T_{VV} \rangle _{d}
+  P_h \langle T_{VV} \rangle _{h}
\label{tVVB+}
\end{eqnarray}
where $P_d$ is the disexcitation probability given in eq.
(\ref{threefivexiiid})
and where
$P_h$ is
the probability to be found in the excited state at $t=+\infty$, hence
$
P_h =  1- P_d
$.
The conditional fluxes $\langle T_{VV}\rangle _{d}$ and
$\langle T_{VV}\rangle _{h}$ are given by
\begin{eqnarray}
\langle T_{VV}\rangle _{d}
 &=& g^2m^2 \bra{0_M} \phi_m T_{VV} \phi_m^\dagger
\ket{0_M}/ P_d\nonumber\\
\langle T_{VV}\rangle _{h}
&=& - g^2m^2\mbox{ Re}\left[
 \bra{0_M}  T_{VV} \phi_m \phi_m^\dagger \ket{0_M} \right]/ P_h
\label{50b}
\end{eqnarray}
These two quantities
are interpreted as the energy emitted when the atom is found
in the ground state (disexcitation $d$)
or in the excited state at $t=+\infty$ knowing that the atom was prepared
in the
excited state at $t=- \infty$.

When equilibrium is reached, the state of the atom is a thermal
superposition. The probabilities of finding it in the excited or
ground state are $P_+$ and $P_-$ given in eq. (\ref{P/P}) with
$P_++P_-=1$.
Since Minkowski vacuum is a thermal distribution of Rindler quanta, one
can approximate the state at equilibrium by
\begin{equation}
\ket{\psi_{therm.}} = P_- \ket{\psi_-} + P_+ \ket{\psi_+}
\label{eqstate}
\end{equation}
This neglects the dressing of the states due to high orders in $H_{int}$
but gives correctly the properties of the fluxes. The
conclusions obtained using this naive state will be proven to be true to all
orders in $g^2$ in Section \ref{allg}.

In the state eq. (\ref{eqstate}) the energy flux is given given by the weighted
sum of
$\langle T_{VV}\rangle _{\psi_-}$ and $\langle T_{VV}\rangle _{\psi_+}$:
\begin{equation}
\langle T_{VV}(V)\rangle _{therm.} =
P_-
\langle T_{VV}\rangle _{\psi_-}
+ P_+
\langle T_{VV}\rangle _{\psi_+}
\label{eqT}
\end{equation}
(This stems from the fact that
the energy momentum operator changes the photon number by an even number and
that the interaction hamiltonian changes the photon number by an odd number
while changing the state of the atom).
Hence all the matrix elements of $T_{VV}$ we shall want to calculate can be
expressed in terms of $\langle T_{VV}\rangle _i$ where $i$ stands
for
$e,
g,
d$ and $h$.

At this point, we remark that
each each of these matrix elements $\langle T_{VV}\rangle _i$
are acausal, for instance they are non vanishing in the left Rindler quadrant
$V<0$,
$U>0$. This will be discussed and interpreted in Section \ref{conded}.
 However the
mean  energies
$\langle T_{VV}\rangle _{\psi_j}$
(where $j= +,-,therm.$) are causal. This follows from the following very
general
argument\cite{UnWa}. If $V,U$ is separated from the trajectory of the
atom $U_a(\tau),
V_a(\tau)$
by a space like distance then $T_{VV}(V,U)$ commutes with $H_{int}$
and one can
rewrite
the mean value as
\begin{eqnarray}
\langle T_{VV}(V,U) \rangle_{\psi_j} &=&
< \psi_j \vert e^{+i\int dt H_{int}} T_{VV}(V,U) e^{-i\int dt
H_{int}}\vert\psi_j > \nonumber\\
&=& < \psi_j \vert  T_{VV}(V,U) e^{+i\int dt H_{int}}e^{-i\int dt
H_{int}}\vert\psi_j > \nonumber\\
&=& < \psi_j \vert  T_{VV}(V,U) \vert\psi_j > =0
\label{van}
\end{eqnarray}
Furthermore, by virtue of  the conservation eq. (\ref{conserv}), the
 zone where $\langle T_{VV}(V,U)
\rangle_{\psi_j}$ vanishes
can be extended to regions which are not space likely
separated
from
the trajectory but to which $V$ modes are unable to propagate.
We note that a  similar causality argument applies in regions where $\langle
T_{VV}(V,U)
\rangle_{\psi_j} \neq 0$ to guarantee that it only depends on $H_{int}(\tau)$
for $\tau$'s
such
that
$V(\tau) < V$, i.e. that it only depends on the form of $H_{int}(\tau)$ in the
past
light
cone of $(V,U)$.

We also point out here a global property of the matrix elements
$\langle T_{VV}\rangle _{e,  g}$ which will have important consequences
in the sequel. Namely that the Minkowski energy carried by $\langle
T_{VV}\rangle _e$ is strictly positive  \begin{eqnarray}
\int^{+\infty}_{-\infty}\! dV \langle T_{VV}\rangle _e
= g^2m^2 \bra{0_M} \phi_m^\dagger H_M \phi_m
\ket{0_M}/ P_e\ >\ 0
\label{pos1}
\end{eqnarray}
since it is the expectation value of $H_M$ (defined in eq.
(\ref{threethreev})) in a
state which is not Minkowski vacuum. On the other hand the Minkowski energy
carried by $\langle T_{VV}\rangle _g$ vanishes identically
\begin{eqnarray}
\int^{+\infty}_{-\infty}\! dV \langle T_{VV}\rangle _g
= -g^2m^2 \mbox{ Re} \left[ \bra{0_M}  H_M \phi_m^\dagger \phi_m
\ket{0_M}\right]/ P_g \ =\ 0
\label{pos2}
\end{eqnarray}
since $H_M \ket{0_M}=0$. The same results are also true for
$\langle T_{VV}\rangle _{d}$ and $\langle T_{VV}\rangle _{h}$.

In preparation for the next sections,
we remark that
all the matrix elements $\langle T_{VV}\rangle _i$  $(i=e, g, d, h)$
can all be expressed in terms of the following two functions (once the ${\cal
{D}}$
term
is dropped)
\begin{eqnarray}
\cc_+ (V) &=& \expect{0_M} {\phi(V) \phi_m^\dagger }{0_M}
=
\int \! d\tau G_+(V,V_a (\tau)) e^{-i m \tau} f(\tau) \nonumber\\ \cc_-(V) &=&
\expect{0_M} {\phi(V) \phi_m}{0_M}
=
\int \! d\tau
G_+(V,V_a (\tau)) e^{+i m \tau} f^*(\tau) \label{threefivexiv} \end{eqnarray}
where $G_+(V,V')$ is given in eq. (\ref{threethreev}).
Indeed using eqs. (\ref{45b}) and (\ref{50b}), one has
\begin{eqnarray}
\langle T_{VV}\rangle _{e} &=&2 ({ g^2m^2 \over P_e})
\left( \partial_V \cc_- \right) \left( \partial_V \cc_-^* \right)
\nonumber\\
\langle T_{VV}\rangle _{d} &=&2 ({ g^2m^2 \over P_d})
\left( \partial_V \cc_+\right) \left( \partial_V \cc_+^* \right)
\nonumber\\
\langle T_{VV} \rangle _{g}
&=& 2 ({g^2m^2 \over P_g}) \mbox{ Re} \left[\left(  \partial_V \cc_- \right)
\left( \partial_V \cc_+
\right)\right]
= ({P_h \over P_g} ) \langle T_{VV} \rangle _{h}
\label{TVVi}
\end{eqnarray}

For these matrix elements of $T_{VV}$ not to be singular the functions $
\partial_V
\cc_+(V) $ and $
\partial_V \cc_-(V) $ must be regular.
The function $ \partial_V
\cc_+(V) $ can be expressed as
\begin{equation} \partial_V \cc_+(V) = -{1 \over 4
\pi} \int \! d \tau {1 \over V - a^{-1}e^{a\tau} - i \e} f(\tau)e^{-im\tau}
\label{threefivexvi} \end{equation}
It can be singular only for $V=0$ where
it takes the form  \begin{equation} \partial_V \cc_+(V) = -{1 \over 4 \pi} \int
\! d
\tau {1 \over -
a^{-1}e^{a\tau} - i \e } f(\tau)e^{-im\tau} \simeq  {a \over 4 \pi} \int \! d
\tau e^{-a\tau} f(\tau)e^{-im\tau} \label{threefivexvii} \end{equation} The
last integral is finite if and only if $f(\tau)$ decreases for $\tau
\rightarrow -\infty$ quicker than $e^{a\tau}$. Similarly if we had considered
right movers, the condition for finiteness on the future horizon would have
been sufficient rapid decrease of $f$ for $\tau \rightarrow +\infty$. Putting
all together the condition to not have singularities on the horizons is that
$f(\tau)$ decreases faster than
$e^{-a \vert \tau \vert}$.
This can be rewritten as
\begin{equation} \int \! d \tau {d t \over d \tau } \ \vert f(\tau)\vert =
\int \! dt \ \vert f(\tau(t))\vert < \infty \label{threefivexviii}
\end{equation}
ie. the interaction of the atom with the field must last a finite
Minkowski time.

In order to obtain explicit expressions for the functions $\cc_\pm$ we
reexpress $f(\tau)e^{-im\tau}$ in terms of its Fourier transform
\begin{equation} f(\tau)e^{-im\tau} =
\int_{-\infty}^{+\infty}\!\! d\la { c_\la \over 2 \pi} e^{-i \la \tau}
\label{threefivexix}
\end{equation}
The normalization is
\begin{equation}
\int d \tau \vert f(\tau) \vert ^2 =  \int d\la { \vert c_\la \vert^2 \over 2
\pi} = T =\hbox{total time of interaction} \label{threefivexx}
\end{equation}

The regularity condition eq. (\ref{threefivexviii})
is equivalent to having $c_\la$ be an
analytic function of $\la$ in the strip $-a < {\rm Im} \la < a$. Hence we
will not consider Lee models since they
have $c_\la=0$ for $\la<0$ and thus singularities on the horizons.
But in order that the behaviour of the uniformly accelerated two level atom
be physically
unambiguous, it is necessary that $c_\la$ be peaked around
$+m$
(the contribution of the negative frequency components of $c_\la$ should be
negligible) and  the golden rule probability of transition
eq. (\ref{threefivexiii}) be recovered. For this to
be the case $T$ must satisfy
 $T >>  m^{-1} $ and $T>>a^{-1} $.

 The first of these conditions is  that $f(\tau)$ be spread
over a distance at least equal to the inverse frequency $m^{-1}$ (the
time--energy uncertainty condition). The second condition, which corresponds to
$T$ being greater than the euclidean tunneling time $2 \pi a^{-1}$ \cite{pbt},
is
required for the probability $P_{e}$ to be linear in time and proportional to
the Bose distribution $N_m$. Both these conditions arise in the
standard textbook calculation for the probability of transition in a thermal
bath \begin{equation}
P_e= 4
g^2
m^2
\int
d\om
\rho (\om) {\sin^2(m-\om)T/2 \over (m-\om)^2}
\label{golder}
\end{equation}
 where $\rho$ is the density of states
(available  photon states for disexcitation or the density of photons present
for excitation). It is legitimate to
replace  $[\sin[(m-\om)T/2] / (m-\om)]^2$ by $
\pi T \delta (m-\om)/2$ in the integrand provided $T^{-1}d \ln \rho / d \om
<<1$. For disexcitation in vacuum this yields $T>>m^{-1}$. In a thermal bath at
inverse temperature $\beta$, the Boltzmann distribution yields the additional
condition
$T>> \beta$, hence $T>>a^{-1}$ in the uniformly accelerated case.

We now express the operator $\phi_m$ and the
functions  $\cc_\pm$
 in terms of
$c_\la$:
\begin{eqnarray}
\phi_m &=& \int_0^\infty\!\! d \la  {a_{\la,R} \over \sqrt{4
\pi \vert \la \vert }}c_\la^*  \ + \  \int_{-\infty}^0\!\! d \la
{a_{\vert\la\vert,R}^\dagger  \over \sqrt{4 \pi \vert \la \vert }}c_\la^*
\nonumber\\ &=&\int_{-\infty}^{+\infty}\!\! d\la c_\la^* {1 \over \sqrt{4 \pi
\la (e^{\pi \la/a} - e^{-\pi \la /a})}}  (e^{\pi \la/2a} a_{\la,M} + e^{-\pi
\la/2a} a_{-\la,M}^\dagger ) \nonumber\\
\cc_+(V) &=&
\int_{-\infty}^{+\infty}\!\! d\la c_\la {1 \over \sqrt{4 \pi  \la (e^{\pi
\la/a} - e^{-\pi \la /a})}}
 e^{\pi \la/2a} \varphi_{\la,M}(V) \nonumber\\  &=&
\int_{-\infty}^{+\infty}\!\! d\la c_\la
{1 \over 4 \pi  \la}  \left[
( \tilde n_\la + 1) (aV)^{-i\la/a} \theta(V) +
\tilde n_\la e^{\pi \la/a}
\vert aV \vert^{- i\la/a} \theta(-V)
\right]\nonumber\\
 \cc_-(V)  &=&
\int_{-\infty}^{+\infty}\!\! d\la c^*_\la {1 \over \sqrt{4 \pi  \la (e^{\pi
\la/a} - e^{-\pi \la /a})}}
 e^{-\pi \la/2a} \varphi_{-\la,M}(V)\nonumber\\
 &=& \int_{-\infty}^{+\infty}\!\! d\la c_\la^*
{1 \over 4 \pi \la}
\left[ \tilde n_\la  (aV)^{i\la/a} \theta(V) +
\tilde n_\la e^{\pi \la/a} \vert aV \vert^{i\la/a} \theta(-V)
\right] \label{threefivexxiii} \end{eqnarray}
where we have used eq. (\ref{threethreevii})
for the expression of $\varphi_{\la,M}(V)$.
And
$\tilde n_\la = 1/ (e^{2 \pi \la/a}-1)$
is equal to (see eq. (\ref{threethreexi}))
\begin{eqnarray} \tilde n_\la &=&N_{\la}= \beta_{\la}^2
\quad \hbox{ for $\la > 0$}\nonumber\\
\tilde n_\la &=&-(N_{\vert\la\vert}+1)= -\alpha_{\vert\la\vert}^2
\quad \hbox{ for $\la
< 0$.} \end{eqnarray}
The probability $P_{e}$ to excite can also be written
in terms of $c_\la$
\begin{equation} P_{e} = g^2 m^2
\int_{-\infty}^{+\infty}\!\! d\la { \vert c_\la \vert^2 \over 4 \pi \la}
\tilde n_\la  \label{threefivexxiv} \end{equation}

As one picture is worth a thousand words we take a particular form for
$c_\la$ such that all the integrals above are gaussian and  can be
evaluated explicitly
\begin{equation}  c_\la = D { \la \over m} e^{-(\la - m )^2 T^2 /2}
 (1 -
e^{-2\pi \la /a})
 \label{threefivexxv} \end {equation} where $D$ is a
normalization constant taken such as to verify eq. (\ref{threefivexx}).

We shall give throughout the text the exact expressions followed
by the approximate  expressions which are valid when $T$
satisfies the condition discussed above $T>> m^{-1}$ and $T>> a^{-1}$
as these
last are physically relevant  and are particularly easy to read and understand.
The approximate expressions are preceded by the symbol $\simeq$.
We conclude this section by the value of the switch off function $f$
(see figure 2)
\begin{eqnarray}
f(\tau) &=&
{D \over \sqrt{2 \pi} T} e^{-\tau^2/2 T^2} \left[
(1 - i {\tau \over mT^2})
- e^{- 2 \pi m/ a} e^{i 2 \pi \tau / a T^2} e^{2 \pi^2 / a^2 T^2}
(1 - i {\tau \over mT^2} - {\pi \over a m T^2})\right]
\nonumber\\
&\simeq &\pi^{-1/4}  e^{-\tau^2/2 T^2}
( 1+ N_m(1 - e^{i 2 \pi \tau / a T^2}))
\label{fapprox}
\end{eqnarray}
where the constant $D$ takes the form $ D \simeq
 2^{1/2}\pi^{1/4} T (N_m + 1)$.
Eq. (\ref{fapprox}) shows the almost gaussian character of the switch off
function whose width is $T$.
The plateau of the gaussian gives a good approximation of the steady state
regime which we which to study.

\subsection{Fluxes and Particles to Order $g^2$ During Thermalisation}

We briefly sketch the main results of this section.
During thermalization a
steady flux of negative Rindler energy is emitted:
  $\bk{ T_{vv}}_{\psi_-} \simeq - g^2 m^2
N_m /2$. This is understood from the isomorphism \cite{Grove}
 with the thermal bath: as the
atom gets exited it absorbs energy from the thermal bath, thus the minus sign.
The transcription of this flux to Minkowski quanta is more subtle. Oscillatory
tails in the Rindler flux are enhanced by the jacobian that converts from
Rindler to Minkowski energy with the net result that positive Minkowski energy
is emitted.
In the Minkowski description the origin of
the steady negative flux is due to a ''repolarization"
of the atom corresponding to
the fact that the probability of finding the atom in its exited level
decreases with time. This repolarization is similar (CPT conjugate)
 with that which occurs
when negative energy is {\it absorbed} by an inertial detector \cite{Grove2}.

We shall discuss both the adiabatic switch on and off presented in
the previous section (to reveal the oscillatory tails)
 and  a sudden switch on and off (to display the properties in the
stationary
regime).

We start with the adiabatic switch on and off.
In terms of the function
 $c_\la$ introduced in eq. (\ref{threefivexxv}) the mean energy
radiated by the two
level atom
initially in its ground state is
\begin{eqnarray}
\bk{T_{vv}(v)}_{\psi_-}  &=& -
g^2m^2\int d\la \int d\la^\prime c_\la c^*_{\la^\prime}
{1 \over (4 \pi)^2} (\tilde n_\la + \tilde n_{\la^\prime})
e^{-i(\la-\la^\prime)v}
\nonumber\\
&\simeq&{-
{g^2 m^2} \over 2  }
N_m  { e^{-v^2/T^2} \over\pi^{1/2}}
[ (N_m + 1) \cos (2 \pi v/aT^2) - N_m]
\label{rayii}
\end{eqnarray}
As announced it carries negative
Rindler energy:
 \begin{eqnarray}
\int^{+\infty}_{-\infty} \!dv \bk{T_{vv}(I_+)}_{\psi_-} &=& -  {g^2
m^2 \over 4 \pi} \int^{+\infty}_{-\infty}
d\la
\vert c_\la \vert^2  \tilde n_\la
 \nonumber\\  &\simeq&
-{1 \over 2}g^2 m^2 N_mT= -m P_{e}
\label{rayiii}
\end{eqnarray}
which is equal to the probability to be found excited times the absorbed
Rindler energy
$-m$.

The total Minkowski energy radiated is
\begin{eqnarray} \bk{H_M}_e &=& \int^{+\infty}_0
dV \bk{T_{VV}(V)}_{\psi_-} =
 \int^{+\infty}_{-\infty}  dv  e^{av} \bk{T_{vv}(v)}_{\psi_-}\nonumber\\
&\simeq& +{1 \over
2} g^2 m^2   N_m T e^{a \tau_0}(1 + 2 N_m)= +m P_{e} e^{a \tau_0}
(1 + 2 N_m)
\label{rayiv}
\end{eqnarray}
where $ e^{a \tau_0}$ is the mean Doppler
effect associated with the window
function $f(\tau)$. We define it by
\begin{equation}
\int\! dv e^{-av} e^{-v^2/T^2} \cos(2 \pi v/aT^2) = - e^{a \tau_0} T/\pi^{1/2}
\label{rayv}\end{equation}
The Minkowski energy is positive (as it should be), whereas the Rindler
energy
is negative. The flip in sign is due to the effect of the transients
around $v = aT^2$.
Indeed whereas
the transients
 are negligible upon computing the Rindler energy eq. (\ref{rayiii}),
upon computing
the Minkowski energy they are enhanced by the jacobian $dv/dV$ and give
rise to the flip in sign.
(Note that this sign flip of the Minkowski energy versus the
Rindler energy can be conceived as arising from the imaginary part of
the saddle point
of eq. (\ref{rayv}): $v_{sp} = -a T^2/4 + i \pi /a$ and is therefore on the
same footing as  that the flip of frequency which leads
to a non vanishing $\beta$ coefficient, see \cite{pbt}). The additional
factor $1+ 2N_m $ in eq. (\ref{rayiv}) and the difference with
eq. (\ref{fourtwo})
by a factor $Ta$
 comes from
the inherent ambiguity in defining $e^{a \tau_0}$ as the mean
Doppler shift associated to the  switch  function $f(\tau)$.

We insist on the fact that
the total Minkowski energy radiated can also be expressed as
\begin{eqnarray}
\bk{H_M}_e &=& P_e \int^{+\infty}_{-\infty} \! dV\langle T_{VV} \rangle_e
\label{aa23}
\end{eqnarray}
because of eq. (\ref{pos2}).
So the Minkowski energy can be conceived as coming from
$\langle T_{VV} \rangle_e$ only. When expressed in this fashion the integrand
in eq. (\ref{aa23}) is strictly positive but located essentially in the region
$ V<0$ (this is shown in Part 3). (We remark that eq. (\ref{fourtwo}) can be
viewed as a rewriting of eq. (\ref{aa23}) but with the integration taken
in the quadrant $V<0$). The  $\langle T_{VV} \rangle_g$ term  restores
causality and localizes all the energy in the transients.

Another case of interest is the golden rule limit for which $c_\la = 2 \pi
\delta(\la-m)$ corresponding to $f(\tau) =1$ for all $\tau$. In this case
their is a constant negative flux for all $V>0$. The transients are located on
the past horizon $V=0$ where they consist of a singular positive flux
\cite{Unru2}. Rather
than this case
we
now analyse the case where the time dependent coupling is
$f(\tau) = \theta(\tau) \theta (T - \tau)$. With this time dependence the
transients are  singular and will not be studied
(this divergent behavior is already
present for an inertial detector with the same switch function and has
nothing to do with the presence of a horizon). On
the contrary, the steady part is easily computed and corresponds exactly to
the intermediate values ($- aT^2  << \tau << a T^2   $)
found in the adiabatic situation described
above in eq. (\ref{rayii}). The reason for which we shall now belabour this
case is that it shows
explicitly the  relations between the
flux emitted and the  transition rate (not only the relation
between the probability and the
total Rindler energy as in
eq. (\ref{rayiii})).

The probability of
spontaneous emission is given by
 \begin{eqnarray}
P_{e}(T) &=& g^2 m^2 \int_0^T d\tau_1 \int_0^T d\tau_2
e^{-im(\tau_2 - \tau_1)}
\bk{\phi(\tau_2)\phi(\tau_1)} \nonumber\\ &\simeq& {1 \over 2} g^2 m  N_m
T
\label{rayvi}
\end{eqnarray}
The second line contains the golden rule result valid when
$aT\to \infty$ with $g^2 T$ finite. It is useful to
introduce the rate of transition, the derivative of $P_{e}(T)$:
\begin{eqnarray}
\dot P_{e}(T) &=& {d P_{e}(T) \over dT}
= g^2 m^2 2 \mbox{ Re} \left[
\int_0^T d\tau e^{-im(T-\tau)} \bk{\phi(T)\phi(\tau)}\right] \nonumber\\
&\simeq& {1 \over 2} g^2 m N_m
\label{rayvii}
\end{eqnarray}
This rate is related to the (steady part of) the
stress energy tensor. Indeed one finds
\begin{eqnarray}
\langle T_{vv}(v
=T)\rangle_{\psi_-} &=&
g^2 m^2 2 \mbox{Re} \left[ \int_0^T d\tau_2
\int_0^{\tau_2} d\tau_1 e^{-im(\tau_2 - \tau_1)} \bk{ [\phi(\tau_2) ,
T_{vv}(T) ]_- \phi(\tau_1)}\right] \nonumber\\ &=& g^2 m^2 2 \mbox{Re}  \left[
\int_0^T d\tau e^{-im(T-\tau)} \bk{i \partial_v \phi(T) \phi(\tau)}\right]
\nonumber\\ &=& - m \dot P_{e}(T) + g^2m^2 2 \mbox{Re}  \left[ i e^{-imT} \bk{
\phi(T) \phi(0)}\right]
\label{rayviii}\end{eqnarray}
 The first equality follows
straightforwardly from the expansion of the evolution operator $e^{-i\!
\int\! H_{int} d \tau}$ in $g^2$. The second equality is obtained using
the commutator relation:
$[\phi(\tau_2) , T_{vv}(\tau_1) ]_- = i \partial_v \phi \delta (\tau_1 -
\tau_2)$. The third equality follows by
integration by parts. The final result
contains a steady part proportional to $- m \dot P_{e}(T)$ which
 tends  to $-{1
\over 2} g^2 m^2 N_m$ in the golden rule limit and an oscillatory term
(which is exponentially damped if a slight mass is given to $ \phi$). The
steady piece simply indicates that to an increase of the probability
 to make a
transition corresponds the absorption of the necessary Rindler energy to
provoke this increase.

We now turn to the Minkowski description of this steady piece.
We first rewrite these expressions in terms of the Minkowski
basis $e^{-i\omega V}/\sqrt{4 \pi \omega}$ (see eq. (\ref{threethreeiii})). The
probability of
transition eq. (\ref{rayvi})
reads
 \begin{eqnarray} P_{e}(T) &=&
g^2 m^2 \int_0^\infty\! d\omega \ \vert\! \int_0^T\! d\tau\ e^{-im\tau} {
e^{-i{\omega \over a} e^{a \tau}} \over \sqrt{4 \pi \omega}} \vert^2
\nonumber\\ &=& \int_0^\infty\! d \omega\ P_{e,\omega}(T)
\label{rayix}
\end{eqnarray}
Similarly the transition rate eq. (\ref{rayvii}) becomes
\begin{eqnarray} \dot P_{e}(T)
 &=& g^2
m^2 \int_0^\infty\! d\omega \ 2 \mbox{Re} \left[ \int_0^T d\tau e^{-im(T-\tau)}
{ e^{-i{\omega \over a} (e^{aT} - e^{a \tau})} \over 4 \pi \omega} \right]
\nonumber\\ &=& \int_0^\infty \! d\omega\  \dot P_{e,\omega}(T)
\label{rayx}
\end{eqnarray}
And the total Minkowski energy is given by
 \begin{eqnarray}
\bk{H_M(T)}_{e} =
\int_{-\infty}^{+\infty} dv e^{av} \bk{T_{vv}}_{\psi_-}
= \int_0^\infty\! d\omega\ \omega P_{e, \omega}(T)
\label{rayxi}
\end{eqnarray}
(Where in the first equality the integral is only over region of positive $V$
since by causality the mean energy is unaffected in the other quadrant,
see eq. (\ref{van})).
The
second equality follows from the diagonal character of the energy operator
$H_M$
see eqs. (\ref{pos1}), (\ref{pos2}).
The positivity of
$\bk{H_M(T)}_{e}$
is manifest since
all the $P_{e,\omega}(T)$
are
positive definite. Nevertheless the time derivative of $\bk{H(T)}_{e}$
is negative, within the steady regime,
\begin{eqnarray}
{d\bk{H_M(T)}_{e} \over dT} &=& \int_0^\infty\! d\omega\ \omega
\dot P_{e,\omega}(T)\nonumber\\ &=& e^{a v(T)} \bk{T_{vv}(v(T))}_{\psi_-}
\nonumber\\
&=& -m e^{av(T)}  \left[ \dot P_{e} (T) + \hbox{oscillatory "damped" term}
 \right]
\label{rayxii}
\end{eqnarray}
${d\bk{H_M} / dT}$
negative implies thus that,
for large $ \omega$ (since $ \dot P_e(T) >0$), some $\dot P_{e,\omega}$
are negative. This
corresponds to a ''repolarization" since all the $P_{e,\omega}$
are positive definite
and vanish for $\tau \leq 0$. This repolarization is
exactly the inverse process of the absorption of negative energy by an atom
described in \cite{Grove2}.

\subsection{Fluxes and Particles to Order $g^2$ at Equilibrium}

Before studying the equilibrium situation it behoves us first to consider the
flux emitted by an atom that makes a transition from excited to ground state.

The mean energy emitted when the initial state is $\ket{\psi_+}$ is
\begin{eqnarray}
\bk{T_{vv}}_{\psi_+} &=&
 g^2 m^2
\int\!
d\la \int \! d\la^\prime c_\la c_{\la^\prime}^*
{1 \over (4 \pi)^2}
\left (  \tilde n_\la + \tilde n_{\la^\prime}  + 2 \right)
e^{-i(\la-\la^\prime)v}\nonumber\\
&\simeq&{g^2 m^2 \over 2 \sqrt{\pi}}
(N_m + 1) e^{-v^2/T^2}
[ 1 - N_m \{ \cos (2 \pi v/a T^2) -1 \} ]
\label{rayxv}
\end{eqnarray}
and the total Rindler energy radiated is
\begin{eqnarray}
\int\! dv \bk{T_{vv}(v)}_{\psi_+} &=&   {g^2
m^2 \over 4 \pi} \int
d\la
\vert c_\la \vert^2  (\tilde n_\la + 1)
\nonumber\\  &\simeq&
{1 \over 2} g^2 m^2 (N_m+1)T   = m P_{d}
\label{rayxvi}
\end{eqnarray}
In the example for which the time dependent
coupling is $f(\tau) = \theta(\tau) \theta( T- \tau)$,
the relation between the derivative of the probability
$ \dot P_{d}(T)$ and the flux $ \bk{T_{vv}}_{\psi_+}$ is
 \begin{equation}
 \bk{T_{vv}(T)}_{\psi_+} =
+ m \dot P_{d}(T) + \hbox{oscillatory ''damped" term}
\label{rayxvii}
\end{equation}
The sign in front of $ \dot P_d(T)$ is now positive (contrary to the one
in eq. (\ref{rayviii})).
Disexcitation  consists in emitting the energy stored in the atom.

The total
Minkowski
energy emitted
is
\begin{eqnarray}
 \int^{+\infty}_{0}\! dV \ \bk{T_{V V}}_{\psi_+}
\simeq {g^2 m^2 \over 2}(N_m+1)T e^{a \tau_0}(2 N_m + 1) =
m P_{d} e^{a \tau_0}(2 N_m + 1)
\end{eqnarray}
For the disexcitation, the integrated Rindler and Minkowski energies
have the same sign and are related by the mean Doppler shift
$e^{a \tau_0}$ times $(2 N_m + 1)$.

We now turn to the thermal equilibrium situation.
We recall that the energy radiated is the sum of the fluxes
emitted when the atom is initially
in its ground state
and
when the atom is initially in its
exited state weighted by
their initial probabilities.
Hence one has
 \begin{eqnarray}
\bk{T_{vv}}_{therm.} &=&
P_-  \bk{T_{vv}}_{\psi_-} +  P_+  \bk{T_{vv}}_{\psi_+}
 \nonumber\\ &\simeq&
-m P_-  \dot P_{e} + m P_+  \dot P_{d} =  0
\label{rayxx}
\end{eqnarray}
The steady fluxes cancel exactly each other because at thermal
equilibrium $P_{\pm}$ satisfy eq. (\ref{P/P}).
This is Grove theorem in $g^2$ \cite{Grove}\cite{mpbrsg}.
 Only the oscillatory transients remain.
They read
\begin{eqnarray}
 \bk{T_{vv}}_{therm.} &=&  g^2 m^2
{1 \over 2 N_m +1}
  \int d\la \int d\la^\prime  {c_\la
c_{\la^\prime}^*
 \over (4 \pi)^2}
\nonumber\\
&\ &\quad\quad
\quad\quad
\left[
N_m ( \tilde n_\la +\tilde n_{\la^\prime} + 2 )
- ( N_m + 1)  (\tilde n_\la +\tilde n_{\la^\prime} )
\right ]
e^{-i(\la- \la^\prime) v}
\nonumber\\
 &\simeq&  {g^2 m^2 \over  \sqrt{4 \pi}}
N_m ( N_m + 1) e^{-v^2/T^2} \left[
1 - \cos (2 \pi v/ aT^2)\right ]
\label{rayxxi}
\end{eqnarray}
 To illustrate the positive transients, we have plotted
$\bk{T_{vv}}_{therm.}$ in figure 3.

The total Rindler
energy emitted is
\begin{eqnarray}
 \int\! dv\ \bk{T_{vv}}_{therm.}&=&
{ g^2 m^2
\over 4 \pi}{1 \over 2 N_m + 1}
 \int d\la \vert c_\la \vert^2 (N_m - \tilde n_\la )\nonumber\\
&\simeq& {g^2 m^2 \over  2} N_m ( N_m + 1) { \pi^2 \over a^2 T^2}
\label{rayxxii}
\end{eqnarray}
It tends to zero
 as the time of interaction $T$
tends to $\infty$ i.e. as $c_\la$ tends to a $\delta$ function.
(In this case, the two level atom tends a Lee model. This can be seen
in eq. (\ref{threefivexxv}) where the negative frequencies are exponentially
suppressed.)

However, the total Minkowski energy
{\it increases}
with the interaction time $T$ and is
given by
 \begin{eqnarray} \int^{+\infty}_0 dV \bk{T_{VV}}_{therm.} &=&
P_-  \int^{+\infty}_0 dV \bk{T_{VV}}_{\psi_-} +  P_+
\int^{+\infty}_0 dV \bk{T_{VV}}_{\psi_+}
\nonumber\\ &\simeq&
m( P_-  \dot P_{e} + P_+  \dot P_{d,v}) T e^{a \tau_0}(2 N_m + 1)
\label{rayxxiii}
\end{eqnarray}
The Minkowski energy of the two fluxes coincide, by virtue of
eq. (\ref{P/P}) and sum up.
This result is what one might have "naively" guessed :
The total energy is the integral over the interacting period
of the rate of transition times the varying Doppler shift
times the energy gap $m$.

We now go to all order in $g$ to prove that
this emission of Minkowski quanta is not an artefact of the second
order perturbation theory.
The forthcoming section can be skipped by the reader mainly interested by the
study of vacuum fluctuations and the black hole problem. He can
go to
Part 3 directly.

\subsection{Fluxes and Particles to All Order in $g$}\label{allg}

  We use the exactly solvable model (RSG), used by Raine, Sciama and Grove
\cite{RSG}--\cite{mpbrsg},
to prove that one does recover, to all
order in $g$, that every quantum jump of the
accelerated oscillator, in thermal equilibrium in Minkowski vacuum,
leads to the emission of a Minkowski quantum. Hence the rate of
production of the Minkowski quanta is simply the rate of internal
transitions of the oscillator. But, as in second order perturbation
theory, these quanta interfere
and their
energy
content is found
 at the edges of the interacting period
only. This is due to the complete neglection of the recoils of the oscillator.
(Upon taking into account the recoils by giving the oscillator a finite
mass, i.e. by quantizing the position of its center of mass,
one proves that the  Minkowski quanta
no longer interfere after a short time (a few $1/a$)\cite{Par}). We conclude
this section by giving a model independent proof that the
stationary thermal Rindler equilibrium corresponds to a
production of Minkowski quanta.

We first recall the main properties of the RSG model and then
analyse the particle content of
the emitted fluxes.

This system consists
of a massless field coupled to
a harmonic oscillator maintained in constant acceleration. Its action is
\begin{eqnarray}
S &=&\int\! dt dx\ \left[ {1 \over 2} \left[ ( \partial_t \phi)^2
- ( \partial_x \phi)^2  \right] \right.
\nonumber\\
&&\quad + \left.
 \int\! d\tau\   \left[ { 1 \over 2} \left[ ( \partial_{ \tau} q)^2 - m ^2
q^2  \right] + e( \partial_{ \tau} q)  \phi  \right]
\delta ^2 (X^{\mu } -X^{\mu
}_a (\tau ))\right]
\label{un}
\end{eqnarray}
where $X^{\mu } (\tau )$ is the accelerated trajectory
eq. (\ref{threefivexiibis}) and $e= g \sqrt{2 m}$ is a rescaled coupling
constant. Since this action is quadratic, the Heisenberg equations are
identical to the classical Euler Lagrange ones. They read:
\begin{equation}
 \partial_u  \partial_v  \phi = {e \over 4 } \theta(V) \delta
( \rho - 1/a)  \partial_{ \tau} q
\label{deux}
\end{equation}
\begin{equation}
\partial_{ \tau} ^2 q +  m ^2
q = - e \partial_{ \tau}  \phi(X^{\mu } (\tau ))
\label{trois}
\end{equation}
The left part of the field (i.e. for $V<0$)
is, by causality, identically free.
And, for $V>0$, on the left of the
accelerated oscillator trajectory, the $v$-part of the field only is
scattered. There the general solution is
\begin{equation}
\tilde \phi (u,v) = \phi (u) + \phi (v) + {e\over 2 } \tilde q(v)
\label{quatre}
\end{equation}
\begin{equation}
\tilde q(v)= q(v) + i  \int _{-\infty}^{+\infty} \! d\la\
 \psi_{ \la} e^{-i \la v}
 \left[ \phi _{ \la,R,v} + \phi _{ \la,R,u} \right]
\label{cinq}
\end{equation}
where $ \phi (u)$ and $ \phi (v)$ are the
homogeneous free solutions of  eq. (\ref{deux}); where the
/operator $ \phi _{ \la,R,v}$ is defined by
\begin{eqnarray}
 \phi _{ \la,R,v} &=&  \int\! {dv \over 2 \pi } e^{i \la v}  \phi (v)
\nonumber\\ &=& {1  \over \sqrt{4 \pi | \la |}}  \left[ \theta ( \la)
a_{\la,R} +  \theta (- \la) a_{- \la,R}^\dagger  \right]
\label{six}
\end{eqnarray}
(a similar equation defines $ \phi _{ \la,R,u}$);
where $ \psi _{ \la}$ is given by
\begin{equation}
\psi _{ \la} = { e\la \over  m ^2 -  \la ^2 -i e^2  \la /2}
\label{sept}
\end{equation}
and where $q(v)$ is a solution of
\begin{equation}
\partial_{ \tau} ^2 q +  m ^2 q + {e^2  \over 2} \partial_{
\tau}q = 0
\label{huit}
\end{equation}
The two independent solutions of  eq. (\ref{huit}) are exponentially
damped as $ \tau$
increases. Being interested by the properties at equilibrium, we drop
$q(v)$ from now on. Then, the remaining part of $ \tilde q(v)$ is a function
of the free field only. Hence, in Fourier transform, eq. (\ref{quatre})
reads
\begin{eqnarray}
\tilde  \phi _{ \la,R,u} &=&  \phi _{ \la,R,u}
\nonumber\\
\tilde  \phi _{ \la,R,v} &=&  \phi _{ \la,R,v} (1+ i{e\over 2} \psi _{ \la})+
(i{e\over 2} \psi _{ \la})  \phi _{
\la,R,u}
\label{neuf}
\end{eqnarray}
The second term in eq. (\ref{neuf}) mixes $u$ and $v$ modes. It encodes the
static Rindler polarization cloud (see \cite{Unru2} \cite{mpbrsg})
which accompanies the
oscillator and carries
neither
Minkowski nor Rindler energy. In order to simplify the following
equations, we drop it and multiply the other scattered term by two
for unitary reason -see
below. (By a simple and tedious algebra, one can
explicitly verify that this modification does not
affect the main properties
of the emitted fluxes). Then  eq. (\ref{neuf}) becomes
\begin{equation}
\tilde  \phi _{ \la,R,v} =  \phi _{ \la,R,v} (1+ ie\psi _{ \la})
\label{dix}
\end{equation}
It is useful, for future discussions, to introduce explicitly the
scattered operators  $\tilde a_{\la,R}$,
and the scattered modes $\tilde \varphi_{\la,R} (v)$
\begin{equation}
\tilde a_{\la,R} =  < \varphi_{\la,R} |  \tilde  \phi > =
a_{\la,R}(1+ ie\psi _{ \la})
\label{onze}
\end{equation}
\begin{equation}
\tilde \varphi_{\la,R} (v) =- \left[  a_{\la,R}^\dagger,  \tilde  \phi
(v)
\right]_- = (1+ ie\psi _{ \la}) \varphi_{\la,R} (v)
\label{douze}
\end{equation}
whereupon the scattered field operator $\tilde \phi (v)$ may be written as
\begin{eqnarray}
\tilde \phi (v)  &=& \int_0^{\infty}\! d\la  \left[ \tilde a_{\la,R}
\varphi_{\la,R} + h.c.  \right]
\nonumber\\  &=& \int_0^{\infty}\! d\la
\left[ a_{\la,R} \tilde \varphi_{\la,R} +  h.c. \right]
\label{treize}
\end{eqnarray}
It is now straitforward to obtain the  scattered Green function and its
Rindler energy content. If the initial (Heisenberg) state is Minkowski vacuum
the $v$-part of the scattered Green function is, for $V,V^\prime > 0$,
\begin{eqnarray}
\tilde G_+(v,v^\prime) &=& \expect{0_M}{\tilde  \phi(v)
 \tilde \phi(v^\prime)}{0_M}\nonumber\\ &=&  \int_0^{\infty}\! d\la
 |1+ ie\psi _{ \la}|^2 \left(
\beta_{\la}^2
 \varphi_{\la,R}^*(v)  \varphi_{\la,R}(v^\prime) +
\alpha_{\la}^2
 \varphi_{\la,R}(v)  \varphi_{\la,R}^*(v^\prime) \right)
\nonumber\\
&=& G_+(v,v^\prime)
\label{quatorze}
\end{eqnarray}
where $G_+(v,v^\prime)$ is the unperturbed Minkowski Green function and
where we have availed ourselves of the identity (see eq. (\ref{sept}))
\begin{equation}
|1+ ie\psi _{ \la}|^2 = 1
\label{quinze}
\end{equation}
This unitary
relation
expresses the conservation of the number of Rindler
particles.
Indeed there is no mixing of positive and negative frequencies in
eq. (\ref{onze}); in other words, the $ \beta$-term of the ''Bogoljubov"
transformation eq. (\ref{onze})
vanishes.

The identity of the Green functions in eq. (\ref{quatorze}) proves that, once
the the steady regime is established,
no flux
is, {\it
 in the
mean},
emitted.
This is
Grove theorem \cite{Grove} \cite{RSG}.

We now examine how this stationary scattering of Rindler modes is
perceived in Minkowski terms.
 The Minkowski scattered modes $ \tilde \varphi_{\la,M}$ are given by
\begin{eqnarray}
 \tilde \varphi_{\la,M} &=&  - \left[ a_{\la,R}^\dagger,  \tilde \phi (V)
\right]_-
\nonumber\\ &=&
\varphi_{\la,M} (1+ie\alpha _{\la} ^2 \psi_{\la}) -ie\alpha _{\la}
\beta_{\la} \psi_{\la} \varphi_{- \la,M} ^*
\nonumber\\ &=&  \tilde  \alpha _{\la} \varphi_{\la,M} +  \tilde
\beta_{\la} \varphi_{- \la,M} ^*
\label{seize}
\end{eqnarray}
\begin{eqnarray}
\tilde \varphi_{- \la,M} &=& \varphi_{- \la,M} (1 -ie\beta_{\la}^2
\psi_{- \la}) -ie\alpha _{\la} \beta_{\la} \psi_{- \la} \varphi_{ \la,M}
^* \nonumber\\ &=&  \tilde  \alpha _{- \la} \varphi_{- \la,M} +  \tilde
\beta_{- \la} \varphi_{\la,M} ^*
\label{dsept}
\end{eqnarray}
where $ 0 < \la < \infty $ and where we have introduced the scattered
Bogoljubov coefficients:
\begin{eqnarray}
 \tilde  \alpha _{\la} &=&  1+ie\alpha _{\la} ^2 \psi_{\la}
\nonumber\\
 \tilde
\beta_{\la} &=& -ie\alpha _{\la}
\beta_{\la} \psi^*_{\la}
\nonumber\\
\tilde  \alpha _{- \la} &=& 1 +ie\beta_{\la}^2
\psi^*_{\la}
\nonumber\\
\tilde
\beta_{- \la} &=& -ie\alpha _{\la} \beta_{\la} \psi_{\la}
\label{dhuit}
\end{eqnarray}
One verifies that the unitary relation is satisfied:
$| \tilde \alpha_{\la}|^2-| \tilde \beta_{\la}|^2=1$.
 The fact that the $ \tilde \beta $ are different from zero indicates
that each couple of jumps of the oscillator
(the absorption and subsequent emission of a Rindler quantum)
leads, in Minkowski vacuum, to the production of two Minkowski quanta.
The member $ \varphi_{- \la , M}$ is emitted when the oscillator
absorbs a rindleron and jumps into a higher level and the other one,
$ \varphi_{ \la , M}$ is emitted during the inverse process.
This is
manifest
in the mean energy flux:
\begin{eqnarray}
\bk{\tilde T_{VV}} &=&  \lim_{V^\prime
\rightarrow V}\partial_V \partial_{V^\prime} \bk{\left[ \tilde \phi(V)
\tilde \phi(V^\prime) -
\phi(V) \phi(V^\prime) \right]}
\nonumber\\ &=& 2  \int_{-\infty}^{+\infty} \! d\la\  | \tilde \beta_{\la}|^2
| \partial_V \varphi_{\la,M}|^2 +
\mbox{Re} \left[ \tilde  \alpha
_{\la} \tilde \beta_{\la}^*
\partial_V \varphi_{\la,M} \partial_V \varphi_{- \la,M} \right]
\label{dneuf}
\end{eqnarray}
whereupon the total Minkowski energy is
\begin{eqnarray}
\bk{\tilde H_M} &=&  \int_{-\infty}^{+\infty} \! dV
\bk{\tilde T_{VV}}
\nonumber\\ &=&  \int_{0}^{+\infty}  d\la\ \la (|
 \tilde \beta_{\la}|^2 +| \tilde \beta_{- \la}|^2)
  \int_{-\infty}^{+\infty} \! {dV \over 2 \pi} {1 \over a^2 |V+i
\epsilon|^2 }
\label{vingt}
\end{eqnarray}
since the integral of the second term vanishes.

Exactly as in second order perturbation theory,
there is a steady regime during which all the emitted quanta interfere
destructively
leaving no contribution to the
{\it
mean
}
flux (see eq. (\ref{quatorze})). But all non diagonal matrix
elements will be sensitive to the
created pairs. This is also the case for the
the total energy eq. (\ref{vingt}) since being diagonal in
$ \omega$
it ignores the destructive interferences (the second term of eq.
(\ref{dneuf}) whose role is to make the mean flux vanishing during the steady
regime).

In order to prove that eq. (\ref{vingt})
corresponds to a steady production of Minkowski quanta
during
the whole interacting period $\Delta \tau =T$ (infinite in eq. (\ref{vingt}))
we evaluate how many quanta are produced.
(Contrary to the energy, the total number of Minkowski quanta
is a scalar under the Lorentz group, hence not affected by the exponentially
growing Doppler shift present in the energy)
 \begin{eqnarray}
\bk{\tilde N( \Delta \tau )}  &=&  \int_{0}^{+\infty}  \! d \omega
\expect{\tilde 0_M}{ a_{ \omega} ^\dagger a_{ \omega}}{ \tilde 0_M}
\nonumber\\ &=&  \int_{0}^{+\infty}  \! d \omega
\expect{0_M}{\tilde  a_{ \omega} ^\dagger
 \tilde a_{ \omega}}{0_M}
\nonumber\\ &=&  \int_{0}^{+\infty}  \! d \omega
 \int_{-\infty}^{+\infty}  d\la\ |  \gamma_{\la,\om}( \Delta \tau )|^2
| \tilde \beta_{\la}|^2
\label{vun}
\end{eqnarray}
where $ |\tilde0_M>$ is the scattered (Schr\"odinger) state\footnote{
The simplest way to obtain this state is to find the scattering operator
$U$ such that $\tilde a_{\la,M}= U ^\dagger a_{\la,M} U$ where
$\tilde a_{\la,M}= \bk{\varphi_{\la,M} \vert \tilde \phi}$. Then $
|\tilde 0_M > =U |0_M >$.}.
The $ \tilde a_{ \omega}$ are related to the $ \tilde a_{\la,M}$
by (see eq. (\ref{threethreevii}))
\begin{equation}
 \tilde a_{ \omega} = \int_0^{\infty}\! d\omega\ \gamma_{\la,\om}
( \Delta \tau )  \tilde a_{\la,M}
\label{vdeux}
\end{equation}
where $\gamma_{\la,\om}( \Delta \tau )$ takes into account the time
dependence of the
coupling. As shown in \cite{pbt}\cite{GO}
 $\gamma_{\la,\om}( \Delta \tau )$ is
non vanishing
only for the $\om$ which enter into resonance with the oscillator
frequency $m$ during the interaction period
 $ \tau_i <  \tau <
\tau_f = \tau_i + T$. When these frequencies belong to
\begin{equation}
 \omega_i = m e^{-a \tau_i} <  \omega <  m e^{-a \tau_f}= \omega_f
\label{vtrois}
\end{equation}
$ \gamma_{\la,\om}( \Delta \tau )$ may be
replaced by $ \gamma_{\la,\om}$ (given in eq. (\ref{threethreeviii})).
Hence $ \tilde N( \Delta \tau )$ reads
\begin{eqnarray}
\bk{\tilde N( \Delta \tau )} &=&  \int_{ \omega_i}^{ \omega_f}  \!
{d \omega \over 2 \pi a \omega}  \int_{-\infty}^{+\infty}  d\la\ | \tilde
\beta_{\la}|^2
\nonumber\\ &=&  { \Delta \tau \over 2 \pi} \int_{-\infty}^{+\infty}  d\la\ |
\tilde
\beta_{\la}|^2
\label{vquatre}
\end{eqnarray}
The total energy emitted obtained from eq. (\ref{vquatre}) is
\begin{eqnarray}
\bk{\tilde H_M( \Delta \tau)}  &=&  \int_{ \omega_i}^{ \omega_f}  \! { d
\omega \over 2 \pi a} \int_{-\infty}^{+\infty}  d\la\ | \tilde
\beta_{\la}|^2
\nonumber\\
&=& \int_{ \tau_i}^{ \tau_f}  \! { d \tau \over 2 \pi}
e^{-a\tau}
m \int_{-\infty}^{+\infty} \! d\la\ | \tilde
\beta_{\la}|^2
\nonumber\\ &=& \int_{V_i}^{V_f}  \! {d V \over 2 \pi}  {1 \over a^2V^2} m
\int_{-\infty}^{+\infty} \! d\la\ |
\tilde
\beta_{\la}|^2
\label{vcinq}
\end{eqnarray}
in perfect agreement with eq. (\ref{vingt}) if the frequency width
of the oscillator in small compared to $ m$.
The rate of production (eq. (\ref{vquatre}) divided by $ \Delta \tau$) is
(small width limit) ${e^2} \alpha _{m}^2
\beta_{m}^2$ which is the rate of jumps for an inertial
oscillator in a bath at temperature $a/2 \pi$. Therefore the number of
Minkowski quanta
produced by the thermalized oscillator equals the number of internal jumps.

We now generalize these results to an arbitrary linear coupling.
We
believe
that it can be generalized, using the same type of argumentation,
 to nonlinear couplings as well. The
proof goes as follow. Any scattering of Rindler quanta by an accelerated
system which leads to a thermal equilibrium during a time much larger
than
$1/a$ can be described as in eq. (\ref{onze}) by
\begin{equation}
\tilde a_{\la,R} =  S_{\la \la^\prime} a_{\la^\prime,R}
 \label{vsix}
\end{equation}
where repeated indices are summed (or integrated) over and where the
summation over $\la^\prime$ includes both $u$ and $v$-modes (as in
eq. (\ref{neuf})).
 The matrix $S$ satisfy the
unitary relation
\begin{equation}
 S_{\la \la^{\prime\prime}}  S_{\la^{\prime\prime} \la^\prime}^\dagger
 = \delta _{\la \la^\prime}
 \label{vseven}
\end{equation}
which express the conservation of the number of Rindler quanta since
$S_{\la, \la^\prime}$ mixes positive Rindler frequencies only.
It is convenient to introduce the matrix $T$ (from now on we
do not write the indices)
\begin{equation}
S=1+iT
 \label{veight}
\end{equation}
which satisfies
\begin{equation}
2 \mbox{Im} T = TT^\dagger
\label{vnine}
\end{equation}
We introduce also the vector operator $b= \left( a_{\la,R};a_{\la,L};
a_{\la,R}^\dagger;a_{\la,L}^\dagger \right) $. Then eq. (\ref{vsix})
 can be written as
\begin{equation}
\tilde b= {\cal  S} b
\label{trente}
\end{equation}
where $ {\cal S}$ has the following block structure
\begin{equation}
 {\cal  S} =
\left( \begin{array}{cccc}
1+iT & 0 & 0 & 0 \\
0 & 1 & 0 & 0 \\
0 & 0 & 1-iT^\dagger & 0 \\
0 & 0 & 0 & 1
\end{array} \right)
\label{tone}
\end{equation}
since the $u$ and $v$-modes on the left quadrant are still free.
On the other hand, the Bogoljubov transformation eq. (\ref{threethreex})
 reads in this
notation
\begin{equation}
c ={ \cal  B} b
\label{ttwo}
\end{equation}
where $c=  \left( a_{\la,M};a_{- \la,M};a_{\la,M}^\dagger;a_{- \la,M}^\dagger
\right)$
 and where $ {\cal  B}$ is
\begin{equation}
  {\cal  B} =
\left( \begin{array}{cccc}
 \alpha &  0 & 0 & - \beta \\
 0 & \alpha & - \beta  & 0 \\
 0 & - \beta &\alpha & 0 \\
 - \beta  & 0 & 0 & \alpha \\
\end{array} \right)
\label{tthree}
\end{equation}
the diagonal matrices (in $\la$) $\alpha $ and $\beta$ being taken real.
The scattered Minkowski operators are given by
\begin{equation}
\tilde c =  {\cal  B}{ \cal  S}{ \cal  B }^{-1} c = \left(
{ \cal  S} + {\cal  B}\left[ { \cal
S},{ \cal  B }^{-1} \right]_- \right) c = {\cal  S}_M c
\label{tfour}
\end{equation}
Since  ${\cal  S}$ and  ${\cal  B}$ do not commute,  ${\cal  S}_M$ has non
diagonal elements which encode the production: \begin{equation}
  {\cal  S}_M  =
\left( \begin{array}{cccc}
 \tilde \alpha_1 &  0 & 0 & - \tilde \beta _1 \\
 0 & \tilde  \alpha_2 & \tilde  \beta^\dagger _1  & 0 \\
 0 &  \beta^\dagger_2 &\alpha _1^\dagger & 0 \\
 - \tilde  \beta_2  & 0 & 0 &\tilde  \alpha^\dagger_2 \\
\end{array} \right)
\label{tfive}
\end{equation}

where the $\tilde \alpha$ $\tilde \beta$ are given in terms of $T$ by
(see eq. (\ref{dhuit}))
\begin{eqnarray}
 \tilde  \alpha _1 &=&  1+i  \alpha T  \alpha
\nonumber\\
 \tilde
\beta _1  &=& -i  \alpha T \beta
\nonumber\\
\tilde  \alpha _2 &=& 1 +i \beta T^\dagger  \beta
\nonumber\\
\tilde
\beta_2 &=& i  \beta  T  \alpha
\label{tsix}
\end{eqnarray}
QED

\section{The Conditional Values of the Energy Momentum Tensor}

\subsection{Introduction}\label{deco}

 In Part 2 we analysed the mean energy radiated by
the atom and showed, in Section 2.4,
how it can be
decomposed into two contributions $\langle T_{VV}(U,V)
\rangle _e$ and $\langle T_{VV}(U,V) \rangle _g$ according
to the final state of the two level atom. These correspond
to the energy emitted if the atom is found in its excited
or ground state at $t=+\infty$. This decomposition is valid
for all points $U>U_a(V)$ where $U_a(V)$ is the trajectory
of the atom, i.e. in the future of the atom since for our massless
field energy flows along $V=constant$.

In this
Part, we generalize this decomposition of the energy
density into energy densities
correlated to the final state of the atom
for all $U,V$.
When $U<U_a(V) $
this describes the energy momentum of the field
configurations
which will give rise to
excitations of the
atom. In order to explicitize this proposition,
we proceed as follow. First, the generalization is introduced formally by
introducing projectors that specify the state of the atom
at $t=+\infty$. We shall then see that this construction gives rise to
nondiagonal matrix elements of $T_{VV}$ which are complex.
We need therefore to discuss their physical meaning.
To this end one
introduces an additional quantum system coupled to the operator
$T_{VV}$. The picture that emerges is then clear: to first order
in the coupling, the modification of the wave function of the additional
system is governed by these non diagonal matrix elements.
This is presented in summary fashion
in Section 3.2 and in more detail in Part 5 which is entirely
devoted to a general discussion of the procedure leading to
these conditional values of operators.
In the present Part, we discuss mainly the properties of these conditional
 energy densities.

In section 3.3, the conditional values of the
energy distribution correlated to the transitions of the accelerated atom
are described and interpreted. In Section 3.4 we
discuss a generalization of these conditional values
which does no longer refer to the transitions of the atom
and  which finds important application in the black hole
problem.

\subsection{The Conditional Energy Correlated to a Transition of the
Accelerated Atom}\label{conddec}

In Section 2.4, we had rewritten the mean energy
emitted on the left of the accelerated trajectory (i.e. $U>U_a(V)$)
as
\begin{eqnarray} \langle T_{VV}
\rangle_{\psi_-} &=&
P_e \langle T_{VV}
\rangle_e + P_g \langle T_{VV} \rangle_g
\label{ooo}
\end{eqnarray}
using the expression, eq. (\ref{state}),
 for the $\ket{\psi_-}$ at $t=\infty$ and the probability of
transition $P_e$ given in eq. (\ref{threefivexiii}).
In order to generalize this
decomposition, we first rewrite it in Heisenberg
representation by introducing the projectors $\Pi_+ =
\ket{+}\bra{+}$ and  $\Pi_- = \ket{-}\bra{-}$ onto the
excited and ground state of the atom.

In Heisenberg representation, the state of the system is
$\ket{\psi_-}=\ket{0_M} \ket{-}$ and the projector is a time dependent operator
given by
\begin{eqnarray}
\Pi_+(t) = e^{i\int^t_{-\infty} dt H_{int}} \Pi_+ e^{-i\int^t_{-\infty} dt
H_{int}}
\label{Pidet}
\end{eqnarray}
whereupon the probability to be found in the
excited
state at $t=+\infty$ is written as
\begin{eqnarray}
P_e
&=&\langle \psi_- \vert
\Pi_+(t=+\infty)   \vert \psi_-\rangle \label{ootwo}
\end{eqnarray}
Similarly, the probability to
be found in the ground state at $t=+ \infty$ is
\begin{eqnarray} P_g &=& \langle
\psi_- \vert  \Pi_-(t=+\infty)   \vert \psi_-\rangle
\label{oothree} \end{eqnarray} The conservation of
probability $P_e+P_g=1$ is realized through the
completeness of the projectors $\Pi_+(t)+ \Pi_-(t) = I$.

The conditional energies can be now obtained by decomposing
the mean using the projectors $\Pi_\pm(t)$
\begin{eqnarray}
\langle T_{VV}(U,V) \rangle_{\psi_-}
 &=& \langle \psi_- \vert \left[ \Pi_+ (+\infty) + \Pi_- (+\infty)\right]
T_{VV}(U,V)
\vert \psi_-\rangle\nonumber\\ &=& P_e { \langle \psi_-
\vert \Pi_+ (+\infty) T_{VV}(U,V) \vert \psi_-\rangle \over \langle
\psi_- \vert \Pi_+ (+\infty) \vert \psi_-\rangle } +
P_g { \langle \psi_- \vert \Pi_- (+\infty) T_{VV}(U,V) \vert
\psi_-\rangle \over \langle \psi_- \vert \Pi_- (+\infty) \vert
\psi_-\rangle }\nonumber\\ &=&  P_e \langle
T_{VV}(U,V)\rangle_e + P_g \langle  T_{VV}(U,V)\rangle_g
\label{oofour}
\end{eqnarray}
When $U>U_a(V)$ , the matrix
elements  $\langle  T_{VV}(U,V)\rangle_e$ and $\langle
T_{VV}(U,V)\rangle_g$ are the expressions obtained less
formally in eq. (\ref{tVVB}).
When
$U<U_a(V)$, these matrix elements are the desired
expressions of the energy density if the atom shall be
found at $t=+\infty$ in the excited (ground) state.

The normalization in eq. (\ref{oofour}) is chosen so
that the mean value is expressed as the probability of
making a transition times the conditional value exactly like in the usual
conditional probabilities.
We shall see in the sequel that
this decomposition into probability to end up in the excited or ground state
 times the conditional value  automatically occurs in physical
processes.

Two
important properties of the conditional values for $U<U_a(V)$ should be
noted. First
\begin{equation}
 \langle
T_{VV}(U<U_a(V),V)\rangle_e = - { P_g \over P_e} \langle
T_{VV}(U<U_a(V),V)\rangle_g
\label{-}
\end{equation}
since $\langle  T_{VV}(U,V)\rangle_{\psi_-}$
vanishes identically
for $U<U_a(V)$ because
the interaction with the accelerated atom
has not yet perturbed Minkowski vacuum.
Secondly,  $\langle  T_{VV}\rangle_e$ is complex.
This
can be seen from the explicit expression
\begin{eqnarray}
\langle  T_{VV}(U<U_a(V),V)\rangle_e &=& {1 \over P_e}
\langle \psi_-\vert e^{i\int dt H_{int}} \Pi_+  e^{-i\int
dt H_{int}} T_{VV}(U,V)\vert\psi_-\rangle\nonumber\\ &=&
{g^2 m^2 \over P_e} \langle 0_M\vert \phi_m^\dagger \phi_m
T_{VV}(U,V)\vert 0_M\rangle
{g^2 m^2 \over P_e} \cc_+^*(V) \cc_-^*(V)
\label{oofive} \end{eqnarray}
where we have used eq. (\ref{state}) and eq. (\ref{threefivexiv}).

Both the real and imaginary part of $\langle  T_{VV}\rangle_e$ have
physical meaning and intervene in physical processes. To
prove this fact one should introduce an additional quantum
system  because these matrix elements have meaning only
in quantum mechanics.

For definiteness, we take the additional system
to be a quantum oscillator sitting at $x=x_0$ and coupled to
$T_{VV}$ by the interaction hamiltonian
\begin{equation}
\int dt H_{osc.} = \int dt g^{VV}(t) p(t) T_{VV}(t,x_0)
\label{ooten}
\end{equation}
where $p(t)$ is the momentum conjugate to the position $q(t)$ of
the oscillator and $g^{VV}(t)$ is a switch function with
the correct Lorentz variance.
The initial state of the
oscillator is $\ket{osc.}$.
The state of the entire system (i.e. field + two level atom + oscillator) is
thus $\ket{\Psi}_-=\ket{\psi}_-\ \ket{osc.}$.

Then to first order in
$g^{VV}$, in the interacting picture, the mean position of the oscillator
at $t=\infty$ is given by
\begin{eqnarray}
\bk{q(t=\infty)}_{\Psi_-} &=& \bra{osc.}q(t=\infty) \ket{osc.}
\nonumber\\
&+&
\int dt  g^{VV}(t)
\bra{osc.} -i[q(t=+\infty),p(t)]_- \ket{osc.} \bk{T_{VV}(t,x_0)}_{{\psi}_-}
\nonumber\\
\label{meanchange}
\end{eqnarray}
That is, the mean change of the position is driven by the mean value
of $T_{VV}(t,x_0)$ in the state $\ket{\psi}_-$.
It corresponds to the classical response of $q(t)$ to a driving force.

But, one can also
investigate the correlations among the oscillator state and the
atom by can asking more detailed questions such that:
what is the "mean" (for the use of this
word see discussion before eq.
(\ref{tVV+}))
position of the oscillator when the two level atom is found in its excited
state?
The answer is the conditional value of $q$ obtained by
decomposing the mean according to the final state of the atom at
$t=\infty$
\begin{eqnarray}
\langle q(t=+\infty) \rangle_{{\Psi}_-} = P_e
\langle q(t=+\infty) \rangle_e + P_g
\langle q(t=+\infty) \rangle_g
\label{oo13}
\end{eqnarray}
where the conditional value $\langle q(t=+\infty) \rangle_e$ is given by
\begin{equation}
\langle q(t=+\infty) \rangle_e = { \bra{\Psi_-} \Pi_+(\infty) q(\infty)
\ket{\Psi_-} \over \bra{\Psi_-} \Pi_+(\infty) \ket{\Psi_- }}
\label{qe}
\end{equation}
To first order in
$g^{VV}$, this conditional  position of the oscillator is
\begin{eqnarray}
\langle q(t=+\infty) \rangle_e
&=& \bra{osc.}q(t=-\infty) \ket{osc.}
\nonumber\\
&+&
\int dt  g^{VV}(t)
\langle osc.
\vert -i[q(t=+\infty),p(t)]_-\vert osc.\rangle
\mbox{ Re} \left[ \langle T_{VV}(t,x_0)\rangle_e\right]
\nonumber\\
&+&
\int dt  g^{VV}(t) \langle osc.
\vert -i\left \{q(t=+\infty),p(t)\right\}_+\vert
osc.\rangle
\mbox{ Im} \left[\langle T_{VV}(t,x_0)\rangle_e\right]
\nonumber\\
\label{ootwelve}
\end{eqnarray}

Hence both the real and imaginary part of the conditional value
  of $T_{VV}$ control the
modification of the mean conditional
position. Note that
Re$\langle T_{VV}\rangle_e$ for $U>U_a(V)$ and for $U<U_a(V)$
enter exactly in the same
way in the integrals giving rise to $\langle q(t=+\infty)
\rangle_e$ as the mean value $\langle T_{VV}(t,x_0)\rangle_{\psi_-}$
drove the mean $q$ in eq. (\ref{meanchange}). The imaginary part
of $\langle T_{VV}\rangle_e$ appears in an unusual way
through an anticommutator which
depends explicitely on the state of the oscillator. In
quantum mechanics therefore, by coupling an additional system
 to the operator $T_{VV}$, one can
isolate, in a well defined manner, the energy content of the emitted particle
correlated
to a transition of the atom as well as the energy content of
the vacuum fluctuations
that shall induce the transition of the atom at  later times.

This
procedure
wherein an external quantum system is introduced to
reveal the physical significance of matrix elements like
$\langle T_{VV}\rangle_{e,g}$ will be displayed in more details in Part 5
and put in parallel with the treatment of Aharonov et al.\cite{aharo}.
The same procedure
will also be used in the black hole situation when evaluating the
conditional value of the metric correlated to
a particular final state of the radiation.

\subsection{The Properties of the Conditional Energy}\label{conded}

Having indicated by an example how both the real and
imaginary parts of $\langle T_{VV}\rangle_e$ intervene in
physical processes we now display the properties of the conditional values.
Since
$\langle T_{VV}(t,z)\rangle_g =
( \langle T_{VV}\rangle_{\psi_-} - P_e \langle T_{VV}\rangle_e )/P_g$
we shall discuss $\langle T_{VV}\rangle_e$ only.

In order to obtain exact expressions for this matrix elements,
we use again the $c_\la$ introduced in eq. (\ref{threefivexxv}).
We give now the
three
expressions for  $\langle T_{VV}\rangle_e$ : three because one finds different
expressions for
$V>0$, $U<U_a(V)$ and for
$V>0$, $U>U_a(V)$ (i.e.   before  or after
the interaction occurs), and for $V<0$ all $U$'s.
\begin{eqnarray}
\bk{T_{vv}(U<U_a, V > 0 )}_{e}
& =&
{ g^2 m^2 \over P_{e,v}}   \int\! d\la \!\int \! d\la^{\prime}\
 c_\la c_{\la^\prime}^{*}\
{1 \over (4 \pi)^2}
\ \tilde n_\la ( \tilde n_{\la^\prime} +1 )\
e^{-i(\la-\la^\prime)v} \nonumber\\
& =&
{m (N_m + 1)\over  2 \sqrt {\pi} T C_0}
 ( 1 - {i v + 2 \pi /a \over m T^2}) ( 1 + {i v \over m T^2})\
e^{ -{ (v-i \pi/a)^2 / T^2}  }
\nonumber\\ &\simeq & { m (N_m +1)\over 2 \sqrt{ \pi} T }
e^{ -{ (v-i \pi/a)^2/
T^2}  }
\label{weeki}\end{eqnarray}

\begin{eqnarray}
 \bk{T_{vv}(U>U_a, V > 0 )}_{e}
&=&{ g^2 m^2 \over P_{e,v}} \vert \int d \la c_\la
{1 \over 4 \pi}
 \tilde n_\la e^{-i\la v} \vert^2   \nonumber\\
&=& {m N_m \over  2 \sqrt {\pi} T C_0}
 \vert 1 - {i v + 2 \pi /a \over m T^2}\vert^2   e^{-{ v^2\over
T^2}} e^{ 3\pi ^2 / a^2  T^2} \nonumber\\  &\simeq& { m N_m
\over 2 \sqrt{ \pi} T }
e^{-{ v^2\over T^2}}
 \label{weekiii}\end{eqnarray}

\begin{eqnarray}
\bk{T_{v_Lv_L}(U, V < 0 )}_{e}
&=& { g^2 m^2 \over P_{e,v}} \vert
\int d \la c_\la
{1 \over 4 \pi}
\tilde n_\la e^{\pi \la /a} e^{-i\la v_L} \vert^2  \nonumber\\  &=& {m (N_m +
1)
\over  2 \sqrt {\pi} T C_0}
 \vert 1 - {i v_L +  \pi /a \over m T^2} \vert^2 e^{-{ v_L^2\over T^2}}
\nonumber\\ &\simeq& { m (N_m +1)\over 2 \sqrt{ \pi} T } e^{ - {v_L^2
\over T^2}}
\label{weekii}\end{eqnarray}
where the last two equalities in eqs. (\ref{weeki}, \ref{weekiii},
\ref{weekii})
represent the exact expressions if $c_\la$ is given by eq.
(\ref{threefivexxv}) and the
approximate expressions valid for $T>>m^{-1}$, $T>>a^{-1}$. $C_0$ is a constant
equal to \begin{eqnarray}
 C_0 &=& (N_m + 1)^{-1}\left [ (1 - { \pi \over a m T^2}) - e^{- 2 \pi m/a}
e^{ 3\pi ^2 / a^2  T^2} (1 - { 2 \pi \over a m T^2})
\right]
\simeq 1
\end{eqnarray}
These fluxes are presented in figure 4.

We now present the
complementary Rindler and
Minkowski properties of these
conditional values of $T_{vv}$.

The Rindler description is that used by a uniformly accelerated observer in
the same quadrant as the two level atom. It is best understood by making
appeal to the isomorphism of the state of the field in the right Rindler
quadrant with an inertial thermal bath.

By getting excited the two level atom has selected that the thermal bath
contains at least one particle in the mode created by $\phi_m^\dagger$.
Furthermore since energy flows along the lines $v=cst$,
$\bk{T_{vv}(U<U_a, V>0 )}_{e}$ is centered around $v=0$
with at spread $\Delta v =T$.  It carries a Rindler energy obtained by
integrating eq. (\ref{weeki})
\begin{equation}
 \int dv \bk{T_{vv}(U<U_a, V>0 )}_{e} = { \int d \la \vert c_\la
\vert^2 \tilde n_\la ( \tilde n _\la +1 ) \over \int d \la \vert c_\la
\vert^2 {1 \over \la} \tilde n_\la }
\simeq  m(N_m +1)
\label{seventytwo}
\end{equation}
The factor $N_m+1$ takes  correctly into account the Bose statistics of the
field since eq. (\ref{seventytwo}) corresponds to evalutating $\langle n^2
\rangle /\langle n \rangle$
in a thermal
distribution.

Then, by getting excited the two level atom absorbs {\it one} quantum
and
the residual energy on the future of the accelerated trajectory $U=U_a(V)$
 is (see eq. (\ref{weekiii}))
\begin{equation}
 \int dv \bk{T_{vv}(U>U_a, V>0 )}_{e} = { \int d \la
\vert c_\la
\vert^2 \tilde n_\la^2  \over \int d \la \vert c_\la
\vert^2 {1 \over \la} \tilde n_\la }
\simeq m N_m
\end{equation}

We now consider what is "seen" by a uniformly accelerated in the
left Rindler quadrant (i.e. what is the nature
of the correlations between the
transition of the atom and an additional  system
uniformly accelerated in the
left Rindler quadrant).
Before the strict correlations between the left and right quadrants, see eq.
(\ref{threethreexiii}), one expects that to the $(N_m+1)$ Rindler quanta
on the right correspond $(N_m+1)$ Rindler quanta
on the left. This can be obtained formally by considering
the Rindler energy operator $H_R$. Since Minkowski vacuum $\ket{0_M}$ is
annihilated by $H_R$ (see
eq. (\ref{threefouriv2})
the Rindler energy in the
left quadrant is equal to the energy in the right quadrant.
Indeed integrating eq. (\ref{weekii}) and using the
relation $\tilde n_\la ( \tilde n_\la + 1) =
\tilde n_\la^2 e^{2\pi \la /a}$
yields
\begin{equation}
\int dv_L \bk{T_{v_Lv_L}(U<U_a, V<0 )}_{e}
= \int dv \bk{T_{vv}(U<U_a, V>0 )}_{e}
\label{twos}
\end{equation}
The symmetry between the left and the right Rindler quadrants results in
$\bk{T_{vv}(U, V<0 )}_{e}$ being centered around $v_L=0$ with the same width
$\Delta v_L=T$.
Thus $\langle T_{VV} (U,V<0) \rangle_e$ carries also the Rindler
energy of $N_m + 1$
Rindler quanta and is almost exactly the symmetric of
$\langle T_{VV} (U<U_a(V),V>0) \rangle_e$ except for small transient
oscillations present for $V>0$
(see the explicit expressions eqs. (\ref{weeki},
\ref{weekii}) and figure 4). We also note that $\langle T_{VV} (U,V<0)
\rangle_e$ is real whereas $\langle T_{VV} (U<U_a(V),V>0) \rangle_e$ is
complex. This results from causality and can be proven in complete generality
by making appeal to a reasoning similar to that in eq. (\ref{van}). It will
have important consequences in the black hole problem, in Section \ref{back}.

The Minkowski description, i.e. that used by an inertial observer, is best
understood by rewriting the conditional value of $T_{VV}$ in terms of the
$\varphi_{\la,M}(V)$ modes, eq. (\ref{threethreevii}),
\begin{eqnarray}
 \bk{T_{VV}(U<U_a,V)}_{e}
&=& {1 \over a^2 V^2}{ g^2 m^2 \over P_{e}} \int\! d\la \!\int
\!d\la^\prime c_\la^* c_{\la^\prime}
{1 \over 4 \pi} \sqrt{  \la \la^\prime
\tilde n_{\la^\prime} ( \tilde n_\la +1)
}
 \varphi^*_{\la,M}
\varphi^*_{-\la^\prime,M}
\nonumber\\ &=&
{1 \over a^2 V^2} {m (N_m + 1)\over  2
\sqrt{\pi} T C_0}( 1 + {i \over m a T^2} \ln (-aV-i\e) -{ \pi \over m a T^2})
\nonumber\\  &\ &
 \times  ( 1 - {i \over m a T^2} \ln (-aV-i\e) -{ \pi
\over m a T^2}) e^{-\left[ \ln (-aV-i\epsilon) \right]^2 / a^2 T^2}
\label{weekiv}
\end{eqnarray}
The $i \e$  defines
$\ln (-aV-i\e)$ as $\ln \vert aV \vert $ for $V<0$ and as
$\ln \vert aV \vert
- i \pi$ for $V>0$. Upon taking the limit $\e
\rightarrow 0$ no singularity occurs.
In fact
$\bk{T_{VV}(U<U_a,V)}_{e}$ given in eq. (\ref{weekiv}) vanishes for $V=0$.
This is an accident due to the particular form of $c_\la$ chosen in
eq. (\ref{threefivexxv}) (it has
zero's for $\la=i n a$, $n= ..,-1,0,1,..$).
But, from the expression for $\cc_{\pm} (V=0)$ given
in eq. (\ref{threefivexvii}), it
results that the generic behaviour of $T_{VV}$ is to stay finite as $V
\to 0$. In more physical terms this corresponds to saying
that
the Minkowski vacuum fluctuation that induces the transition straddles
the
horizon with no clear cut separation between the pieces in the left and
right
quadrants.

Notice how the $i\epsilon$ prescription which encodes the analyticity of the
modes $\varphi_{\la,M}$ in the lower half complex plane now encodes the
vanishing of the integral
\begin{equation}
\int_{-\infty}^{+\infty}\!dV\
 \bk{T_{VV}(U<U_a,V)}_{e} =0
\label{vanenf}\end{equation}
by contour integration. The vanishing of this
integral can also be seen to result from $\ket{0_M}$ being the ground state
of $H_M$ (in similar fashion to the vanishing of eq. (\ref{pos2})).
In other words the total
Minkowski energy does
not fluctuate and is always equal to its eigenvalue zero: vacuum
fluctuations carry no energy.
Notice also how the $i\epsilon$ encodes the above mentioned
slight asymmetry between the left
 and right quadrants: $\bk{T_{VV}(U<U_a,V)}_{e}$ is real and
positive for $V<0$ whereas it is complex and oscillates for $V>0$.

In view
of the vanishing of the total Minkowski energy (eq. (\ref{vanenf})) and of
the positivity  in the region $V<0$, the
energy in the region $V>0$ must integrate to an exactly compensating real
and negative value. This is not
in contradiction with the positivity of Rindler energy in the right quadrant,
eq. (\ref{seventytwo}),
since the expressions for the Rindler and the Minkowski energy differ by the
jacobian ${ dv / dV} = {1 / aV}$. The oscillations of $T_{vv}$ for
$V>0$ that occur in eq. (\ref{weeki}) as $v \rightarrow - \infty$ (which are
negligible in
the Rindler description) are dramatically enhanced by the jacobian in such a
way that the Minkowski energy in the right quadrant becomes negative, c.f.
eq. (\ref{rayv}).

For $U>U_a$, all $V$,
after the atom has made a transition, the Minkowski energy takes the
form \begin{eqnarray}
\bk{T_{VV}(U>U_a,V)}_{ e} &=&
{1 \over a^2 V^2}{ g^2 m^2 \over
P_{e,v}} \vert \int\! d\la c_\la
\sqrt{ {
\la \tilde n_\la \over 4 \pi}} \varphi_{-\la, M}^*
\vert^2\nonumber\\
&=&{1 \over a^2 V^2} {m (N_m + 1)\over  2 \sqrt{\pi} T C_0} \vert 1 -
{i \over m a
T^2} \ln (-aV-i\e) -  {\pi\over m a T^2}  \vert^2
\nonumber\\
&&\quad\quad
  \vert e^{-{\left[ \ln (-aV-i\epsilon) \right]^2 / a^2 T^2}} \vert^2 \vert
e^{-i{m }\ln (-aV-i\epsilon)/ a } \vert^2  \label{wwsd}\end{eqnarray}
It is manifestly real and positive. This is as it should be since we are
calculating the mean value of the energy in a state that contains one
Minkowski quantum.
In particular the integral $\int\! dV \ \bk{T_{VV}(U>U_a,V)}_{ e}$ is
strictly positive (see eq. (\ref{pos1})). The time evolution has transformed
the conditional value eq. (\ref{weekiv}) which was complex and carried no
energy into a real conditional value carrying positive energy. (The change in
time of the conditional values is further discussed in Part 5).

Notice how the $i\epsilon$ prescription in eq (\ref{wwsd}) encodes the
asymmetry between the left quadrant (proportional to $N_m+1$) and the right
quadrant (proportional to $N_m$).
By absorbing the positive Rindler energy $m$, the two level atom has reduced
the negative Minkowski energy on the right thereby converting a vacuum
fluctuation into a quantum. This is summarized in figure 5.

\subsection{The Energy Correlated to a Rindler State}\label{corrR}

Up to now we have considered the final state of the atom
to  isolate certain field configurations (i.e. those
correlated to transitions of the atom). These field
configurations can also be isolated without making appeal
to the accelerated atom through the
introduction of projection operators acting directly
on field states, decomposing thereby the
mean value. To make contact with the previous section we
work in Minkowski vacuum and consider projections onto
states containing certain Rindler quanta.

We first write
unity as
\begin{eqnarray}
I = \sum_i \Pi_i
\end{eqnarray}
 where $\Pi_i$ are a complete set of projectors.
The mean value of $T_{VV}$ is then decomposed as
\begin{eqnarray}
\langle 0_M \vert T_{VV} \vert 0_M \rangle &=&
\langle 0_M \vert \sum_i \Pi_i T_{VV} \vert 0_M \rangle
\nonumber\\
&=& \sum_i P_i {
\langle 0_M \vert  \Pi_i T_{VV} \vert 0_M \rangle
\over
\langle 0_M \vert \Pi_i \vert 0_M \rangle  }
\label{r1}
\end{eqnarray}
where $P_i = \langle 0_M \vert \Pi_i \vert 0_M \rangle$
is the probability to be in the eigenspace of $\Pi_i$. As
in Section \ref{conddec} the matrix element
\begin{eqnarray}
\langle T_{VV} \rangle_{\Pi_i}=
{\langle 0_M \vert  \Pi_i T_{VV} \vert 0_M \rangle \over
\langle 0_M \vert \Pi_i \vert 0_M \rangle}
\label{TPi}
\end{eqnarray}
is the
conditional value of $T_{VV}$ if the final state is in
the eigenspace of $\Pi_i$.

In this section we shall study the properties
of some
typical conditional values in preparation for black hole
physics.
The physical relevance of these matrix elements
will be displayed in Part 4 and Part 5.

We first consider the projector
\begin{equation}
\Pi_{\la,R;\la,L} = a^\dagger_{\la,R} a^\dagger_{\la,L}
\ket{0_{RL}}\bra{0_{RL}} a_{\la,R} a_{\la,L}
\end{equation}
which projects onto the state containing one pair of
rindlerons of Rindler energy $\la$.
By availing
oneself of the identity
\begin{equation}
  a_{\la,R}^\dagger a_{\la^\prime,L}^\dagger \ket{0_{RL}} ={1 \over \al_\la
\al_{\la^\prime} } a_{\la,M}^\dagger a_{-\la^\prime,M}^\dagger
\ket{0_{RL}} + {
\be_\la \over \al_\la} \delta ( \la -
\la^\prime )\ket{0_{RL}} \end{equation}
it is
straightforward to obtain
\begin{eqnarray}
\langle \phi(V)\phi(V^\prime) \rangle_{\Pi_{\la,R;\la,L}} &=& \expect
{0_{M}}{\Pi_{\la,R;\la,L}\ \phi(V)\phi(V^\prime)}{0_M} \over  \expect
{0_{M}}{\Pi_{\la,R;\la,L}}{0_M} \nonumber\\
&=& \expect
{0_{RL}}{a_{\la,R}a_{\la,L}\ \phi(V)\phi(V^\prime)}{0_M} \over  \expect
{0_{RL}}{a_{\la,R}a_{\la,L}}{0_M} \nonumber\\ &=& {2 \over
\alpha_{\la}\beta_{\la}} \varphi_{-\la , M}^* (V)
 \varphi_{\la , M}^* (V^\prime) \ + \  {
\expect{0_{RL}}{\phi(V)\phi(V^\prime)}{0_M} \over  \scal {0_{RL}}{0_M} }
\label{threefouri} \end{eqnarray}
 It decomposes into two terms. The first depends on the quantum number $\la$
and is the contribution of the pair of rindlerons
selected by the projector $\Pi_{\la,R;\la,L}$.
It carries an energy density equal to  \begin{equation} \lim_{V^\prime
\rightarrow
V}
\partial_V \partial_{V^\prime}{2 \over \alpha_{\la}\beta_{\la}} \varphi_{-\la
, M}^* (V)
 \varphi_{\la , M}^* (V^\prime)= {\la \over 2 \pi a^2}{1 \over (V + i \e)^2}
\label{threefourii} \end{equation}

The second term is independent of $\la$ and appears because except
for the mode
$\la$, Rindler vacuum has been selected (indeed if
the projector $\Pi_{0_{RL}} =
\ket{0_{RL}}\bra{0_{RL}}$ is used only the second term
appears). It is convenient to rewrite this term as the sum of the
expectation value of $T_{VV}$ in Minkowski vacuum\footnote{
In this section and the following one we shall explicitly write the vacuum
expectation value of the energy momentum tensor $\expect{0_M}{T_{VV}}{0_M}$
even though it vanishes. It is kept only to facilitate the transcription of
these results  to the black hole problem
where the vacuum expectation of the energy is non trivial and must be
renormalized
carefully.} plus their difference
\begin{eqnarray}
\langle T_{VV} \rangle_{0_{RL}} &=&
{\expect{0_{RL}}{T_{VV}}{0_M} \over \scal {0_{RL}}{0_M} }\nonumber\\  &=&
\expect{0_M}{T_{VV}}{0_M} \ + \
\lim_{V^\prime \rightarrow V} \partial_V\partial_{V^\prime} \int_0^\infty \!
d\la\  -2 {\beta_{\la} \over \alpha_{\la} }\varphi_{-\la,M}^*(V)
\varphi_{\la,M}^*(V^\prime) \nonumber\\
&=&\expect{0_M}{T_{VV}}{0_M} \ - \ {\pi
\over 12} \left( a \over 2 \pi \right)^2 {1 \over a^2(V + i \e)^2}
\label{threefouriii} \end{eqnarray}

The Rindler
interpretation
is obtained by
considering the Rindler energy density $T_{vv}$.
Then eq. (\ref{threefourii}) gives  the energy density of the selected
 rindleron $\la/2 \pi$, the jacobian being
$\left( d V / dv \right)^2 = a^2 V^2$.
The second term, eq. (\ref{threefouriii}), is
the  Rindler vacuum energy\cite{tmunu} which is minus the thermal energy
density at a temperature $a/2 \pi$ (Minkowski vacuum
contains a thermal distribution of rindlerons).
The energy in the left quadrant is identical to
that in the right quadrant since there is a complete symmetry between the two.

The Minkowski interpretation is completely different.
Since the hamiltonian
 $H_M$ is diagonal in
$\omega $
and annihilates
Minkowski vacuum
 $H_M |0_M>=0$, both eq. (\ref{threefourii}) and  eq. (\ref{threefouriii})
contain zero Minkowski energy. Indeed, the pole prescription at the
horizon $V=0$ ensures that their integrals over the entire
domain of $V$ vanish as in eq. (\ref{vanenf}).

The projector $\Pi_{\la,R;\la,L}$
cannot be used in the black hole situation since we have no access to the
region beyond the horizon nor does it mimick the existence of a
detector confined to the right Rindler quadrant. Hence
we introduce a projector which selects the presence of a rindleron
of energy $\la$ in the right quadrant  while tracing over the state of the
field in the left quadrant:
\begin{equation}
\Pi_{\la,R} =  I_{L} \otimes
a^\dagger_{\la,R}  \ket{0_{R}}\bra{0_{R}} a_{\la,R}
 \label{threefourv} \end{equation}
where $I_{{L }}$ is the identity operator restricted to the left
quadrant and $\ket{0_{R }}$ is Rindler vacuum in the right
quadrant. The
corresponding conditional value of $T_{VV}$ is given by
\begin{equation}
\langle T_{VV} \rangle _{\Pi_{\la,R}}= {
\expect {0_M}{\Pi_{\la,R} T_{VV}}{0_M} \over  \expect {0_M}{\Pi_{\la,R}}{0_M} }
\label{threefourvi} \end{equation} It leads back
 to eq. (\ref{threefouri}) because of the EPR correlations between the two
quadrants: if there is a rindleron on the right then their
necessarily also is a rindleron on
the left (its partner) with the
opposite
 Rindler
energy. This partenaria follows from eq. (\ref{threethreexiii}) where
the operators
$a^\dagger_{\la,R}$ and $a^\dagger_{\la,L}$
appear
 in product only.
 In the
black hole problem
the equivalent
EPR correlations will mean that
 to each outgoing Hawking photon their
corresponds an ingoing partner on the other side of the horizon.

An even less restrictive projector  specifies only partially
the state of the field in the right quadrant. One chooses
that the final state contains one rindleron on the right in the mode $\la$
while tracing over all other right rindlerons and over all left rindlerons.
The resulting projector is
\begin{equation} \tilde \Pi_{\la,R}  =
I_{L} \otimes \prod_{\la^\prime \neq \la}
I_{\la^\prime,R}\otimes \ket{1_{\la,R}}\bra{1_{\la,R}}  \label{threefourvii}
\end{equation}
where $I_{\la,L}$ is the identity operator restricted to the
mode $\la$ in the left quadrant and $\ket{1_{\la, R}}$ is
the one particle state restricted to the right mode $\la$.
The
corresponding conditional value of $T_{VV}$ is \begin{eqnarray} {
\expect{0_M}{\tilde \Pi_{\la,R} T_{VV}}{0_M} \over \expect{0_M}{\tilde
\Pi_{\la,R}}{0_M}} =
{\la \over 2 \pi
a^2}{1 \over (V + i \e)^2} \ + \  \expect{0_M}{T_{VV}}{0_M}
\label{threefourviii} \end{eqnarray}
The first term is the energy of the rindleron $\la$ already obtained
in eq. (\ref{threefouri}) and eq. (\ref{threefourvi}). The second term
is simply the Minkowski vacuum expectation value since no further
specification is imposed on the final state.
This is why the probability $\expect{0_M}{\tilde
\Pi_{\la,R}}{0_M}$ to be in the eigenspace of $\tilde
\Pi_{\la,R}$ is finite. This is to be opposed to the probabilities
encountered previously (the denominators of eq. (\ref{threefouri}) and
eq. (\ref{threefourvi})) which
vanish because all the Rindler modes have been
specified
to be in their Rindler ground state.
(In physically realistic situations  only
nonvanishing probabilities
will occur. This was indeed the case for the accelerated two level atom
coupled to the field).

 Nevertheless the conditional values eq.
(\ref{threefourvi}) and eq. (\ref{threefourviii}) are related by a
unitary relation similar to eq. (\ref{r1}): by taking a set
of orthogonal projectors like $\Pi_{\la,R}$ whose combined eigenspace
is equal to the eigenspace of
$\tilde \Pi_{\la,R}$
and summing
the corresponding
conditional values multiplied by the relative probabilities that they occur,
eq. (\ref{threefourviii}) is recovered. In order to realize this unitary
relation one must select
the presence of two, three, any number of rindlerons.
The corresponding
conditional values of $T_{VV}$ are easily obtained and the contribution of each
individual selected particle is found to be independent
(if the particles are orthogonal) of the selection performed
on the other particles. In other words, for a free field the vacuum
fluctuations of orthogonal particles are independent of each other.

We finally consider the selection of a wave packet. Instead of the
projector eq. (\ref{threefourv}), we define:
\begin{equation} \Pi_{v_0,\la_0,R} = I_L \otimes a^\dagger_{v_0,\la_0,R}
\ket{0_R
} \bra{0_R} a_{v_0,\la_0,R} \end{equation} where
$a_{v_0,\la_0,R} = \int_0^{+\infty}\! d\la f(\la) a_{\la,R}$ is the
destruction operator of a wave packet of right rindlerons centered around
$v=v_0$ and $\la=\la_0$. The state $\Pi_{v_0,\la_0,R} \ket{0_M} $ is
 \begin{equation} \Pi_{v_0,\la_0,R} \ket{0_M}
= \left( \int_0^{+\infty}\! d\la f^*(\la) a^\dagger_{\la,R} \right)
\left( \int_0^{+\infty}\! d\la^\prime -
{\be_{\la^\prime} \over \al_{\la^\prime}}
f(\la^\prime) a^\dagger_{\la^\prime,L} \right)  \ket{0_{RL}}
\label{123}
\end{equation} where the EPR correlated wave packet in the left Rindler
quadrant appears explicitly.
Note the asymmetry of the wave packets: the
induced
wave packet
in the
left quadrant contains the factor $\be_{\la^\prime} / \al_{\la^\prime}$
since it originates
from
 the EPR correlations in eq. (\ref{threethreexiii}). This asymmetry plays
a fundamental role when analysing the flux emitted by the accelerated
detector and the black hole. It is responsible of the fact that
Im $ \langle T_{VV}(U<U_a) \rangle _e$ vanishes for $V<0$ only, see  eq.
(\ref{weekii}). The conditional value of $T_{VV}$ associated with this wave
packet is
\begin{eqnarray} {
\expect{0_M}{ \Pi_{v_0,\la_0,R} T_{VV} }{0_M}
\over
\expect{0_M} { \Pi_{v_0,\la_0,R} } {0_M}
}  = &
2 \left[
 \int_0^{\infty}\! d\la \int_0^{\infty}\! d\la^{\p}
{\be_{\la^\prime} \over \al_\la \al_{\la^{\prime}}^2 }
f^*(\la) f(\la^{\p}) \partial_V
\varphi_{\la,M}^* \partial_V \varphi_{-\la^{\p},M}^* \right]
 \nonumber\\ &
\times\ \left[  \int_0^{+\infty}\! d\la
{ \be_\la^2  \over \al_\la^2} \vert f(\la) \vert^2
 \right] ^{-1}
+ {
\expect{0_{RL}} {  T_{VV} }{0_M} \over \scal{0_{RL}}{0_M} }
\label{threefourx}
\end{eqnarray}

The role of $f(\la)$
in this equation is very similar to that of $c_\la$ in
Section \ref{conded} (where
$c_\la$ was the Fourier transform of the coupling to the atom, see eq.
(\ref{threefivexix})). However since $f(\la)$ in eq. (\ref{threefourx})
contains no negative frequencies, eq. (\ref{threefourx}) is singular on the
future horizon (see discussion after eq. (\ref{threefivexvi})).

\section{Black Hole Radiation}
\subsection{The Kinematics of the Collapse and the Scattered Modes}

We work in the background metric of a spherically symmetric
collapsing
star of mass $M$. Outside the star the geometry is described  by the
Schwarzschild metric
\begin{eqnarray}
ds^2 &=& (1- {2 M \over r}) dt^2 - (1- {2 M
\over r})^{-1} dr^2 - r^2 d \Omega^2 \nonumber\\
&=& (1- {2 M \over r}) du dv -r^2 d \Omega^2 \nonumber\\
v,u &=& t \pm r^*\nonumber\\
r^* &=& r + 2M \ln{ r-2M \over 2M}
\label{bh1}
\end{eqnarray}

The specific  collapse we consider is
produced by a spherically symmetric shell of pressureless
massless matter. Inside the shell space is
flat and the metric reads
\begin{eqnarray}
ds^2 &=& d \tau ^2 - dr^2 - r^2 d \Omega^2 \nonumber\\
&=& dU dv -r^2 d \Omega^2 \nonumber\\
v,U &=& \tau \pm r
\label{bh3}
\end{eqnarray}
where $v$ is the same coordinate in eq.( \ref{bh1}) and eq. (\ref{bh3})
since on ${\cal I}^-$ ($u=-\infty$) space time is flat on both sides.
The collapsing shell, taken to be thin, follows the geodesic $v=v_S$.
The connection between the two metrics is obtained by imposing the
continuity of $r$ along the shell's trajectory
\begin{equation}
dU = du (1 - {2 M \over r(u,v_S)}) = du (1 - {4 M \over v_S - U})
\end{equation}
Then by choosing $v_S=4M$ one gets
\begin{eqnarray}
du &=& - { dU \over U} \left( 4M - U
\right)
\nonumber\\  u(U) &=& U - 4M \ln ({- U \over 4M})
\label{162s}
\end{eqnarray}

In the static space time outside the star, the Klein-Gordon equation
for a mode of the form $ \varphi_{l,m}
= {1 \over \sqrt {4 \pi r^2}} Y_{lm}(\theta, \varphi)
\psi_{l}(t,r)
$
reads
\begin{equation}
\left[ \partial_t^2 -\partial_{r^*}^2 - (1- {2 M \over
r}) \left[ {l(l+1) \over r^2} + m^2 + { 2 M \over r^3} \right] \right]
\psi_l
(t,r) =0
\label{bh2}
\end{equation}
Near the horizon $r-2M << 2M$, it  becomes the wave equation
for a massless field
in $1+1$
dimensions.
By considering
only
the s-wave sector of
a massless field
and dropping the
residual
 "quantum potential" ${2M(r-2M) / r^4}$ the
conformal invariance holds everywhere, inside as well as outside
the
star.
 From now  on we shall work in this simplified
context and only discuss
briefly the differences with the more realistic four dimensional case.

The Heisenberg state is chosen to be the initial vacuum
i.e. vacuum with respect to the modes which have positive
$v$-frequency on ${\cal I}^-$. These modes are reflected at $r=0$ and read
\begin{equation}
\varphi_{\omega,0,0}(v,u) = {1 \over 4 \pi r \sqrt{\omega}}
\left( e^{-i \omega v} - e^{-i\omega U(u)} \right)
\label{modesphi}
\end{equation}
Hence, for $u > 4M$ (or even on both sides of the horizon for $-M < U < M$)
 the state of the field tends exponentially quickly (in $u$) to Unruh
vacuum\cite{Unruh}, i.e.
vacuum with respect to the modes
\begin{equation}
\exp(-i \omega v) \quad {\hbox{and}}\quad \exp({i\omega \over 4
M} e^{-u\over 4
M})
\end{equation}

The Schwarzschild $u$-modes  $\chi_{\la}(u)= {e^{-i\la u}}/( 4 \pi r
\sqrt{
\la}) $
are needed to analyse
the particle content
of the scattered modes
$\varphi_{\omega}$
on ${\cal I^+}$. In terms of $U$ they take the form
\begin{equation}
\chi_{\la}(u) =  \theta(- U){1 \over  4 \pi r
\sqrt{\la}}
 ({-U \over 4M} )^{i \la 4 M} e^{- i \la U}
\end{equation}
The exact Bogoljubov coefficients between $\varphi_\omega$ and $\chi_{\la}$
are given by
\begin{equation}
\alpha_{\omega,\la} =
\bk{\varphi_\omega,\chi_{\la}}
= {1 \over 4 \pi}
\sqrt{ \omega \over \la}
\Gamma ( 1 + i 4 M \la)
[ 4 M ( \omega - \la)]^{-i 4 M \la}
e^{\pm 2 \pi M \la}
\end{equation}
where the $\pm$ is to be understood as $+$ if $ \omega > \la$ and $-$
if $ \omega < \la$. The expression for $\beta_{\omega,\la}$ is obtained by
taking $\la$ into $-\la$. The asymptotic Bogoljubov coefficients
(relating Kruskal modes to Schwarzschild modes) are identical to the
coeficients
relating Minkowski modes to Rindler modes (see
eq. (\ref{threethreevii}) et seq.)
in the
limit $\omega \to + \infty$ which corresponds to resonance at late times
$u \to +\infty$ (see eq. (\ref{vtrois})). In this limit
the black hole emits uncorrelated (see eq. (\ref{threefivexiibis}))
quanta at the Hawking temperature $1/8 \pi M$
 since $\vert \beta_{\omega,\la}/ \alpha_{\omega,\la}\vert ^2 =
e^{-8 \pi M \la} $.

Having described the kinematics of the collapse we now turn to the
description of the energy content of the emitted quanta.
The
new difficulty lies in the
renormalization of the energy momentum tensor which must be carried out
in curved space times. We therefore turn to this point.

\subsection{Matrix Elements in Curved Space-Time}

Wald has proposed a set of eminently reasonable
conditions that a renormalized energy
momentum operator should satisfy \cite{wald}.
By an argument  similar to
Wald's (or
simply by verifying that it is in accord with his axioms), it is
 possible to deduce that  $T_{\mu \nu ({\rm ren})} (x)$ can
be
written in the following way
\begin{equation}
T_{\mu \nu ({\rm ren})} (x) =
T_{\mu \nu} (x) - t_{\mu \nu ({\rm S})} (x) I
\label{Wald4}
\end{equation}
where $T_{\mu
\nu} (x)$ is the bare energy momentum tensor.  The subtraction term  $t_{\mu
\nu ({\rm S})} (x)$ is an (infinite) conserved c-number function only of the
geometry at $x$. It can be understood\cite{mpblocal}\cite{mas}
as the (infinite) ground state
energy of the "local inertial vacuum": that state which most resembles
Minkowski vacuum at $x$. Numerous techniques have been
developed
to
calculate $t_{\mu \nu ({\rm S})}$ and we refer the reader to
\cite{birreld}
for a
review.

In a state, say the Heisenberg vacuum
$\ket{0}$, the expectation value of $T_{\mu\nu}$
 takes the form
\begin{equation}
 \expect{0}{T_{\mu \nu
({\rm ren})} (x)}{0} = \expect{0}{T_{\mu \nu } (x)}{0} -  t_{\mu \nu
({\rm S})} (x)
\label{beware}
\end{equation}
 where both terms on the r.h.s. are infinite but
their difference is finite.

If in addition one specifies the final state (i.e. on $\cal{I}^+$)
to be $\Pi \ket{0}$, where
$\Pi$ is a projector (or more generally the self
adjoint operator), the renormalized conditional value of $T_{\mu \nu}$
reads
\begin{equation}
\langle T_{\mu \nu } \rangle_\Pi = {
\expect{0}{\Pi T_{\mu \nu ({\rm ren})} } {0} \over  \expect{0}{\Pi }
{0} }
 \label{bh11}
\end{equation}
 Inserting eq. (\ref{Wald4})
 into this
expression yields
 \begin{equation}
\langle T_{\mu \nu }(x)\rangle_{\Pi} = {
\expect{0}{\Pi {T_{\mu \nu }}(x) } {0} \over  \expect{0}{\Pi } {0} } -
t_{\mu \nu ({\rm S})}(x)
\label{bh12}
\end{equation}
Then by expressing ${T_{\mu \nu }}(x)$ in terms of the operators which
annihilate the Heisenberg vacuum one obtains
\begin{equation}
\langle T_{\mu \nu }(x)\rangle_{\Pi}  = \int_0^\infty\!\! d \omega
\int_0^\infty\!\! d \omega^{\prime} {
\expect{0}{\Pi a_{\omega}^{\dagger} a_{\omega^{\prime}}^{\dagger}
 } {0} \over  \expect{0}{\Pi } {0} } \hat T_{\mu \nu }
(x)\left[ \varphi^*_\omega \varphi^*_{\omega^{\prime}}
\right] +
\expect{0}{T_{\mu \nu
({\rm ren})} (x)}{0}
\label{2terms}
\end{equation}
where $ \hat T_{\mu \nu } (x)$
is the classical differential operator
which acting on the waves $ \varphi^*_\omega$, eq. (\ref{modesphi}),
 gives their energy
density.

The renormalized conditional value contains two contributions.
The first term, the fluctuating part, depends on the particle content of the
state specified by $\Pi$. Contrariwise,  the second one is the
mean energy density of the
Heisenberg vacuum
eq. (\ref{beware}) obtained when no specification on the final state is added.

The formula eq. (\ref{bh12})
 warrants a few additional comments. First notice that their are
parts of $\bk{T_{\mu\nu}}_{\Pi}$
that are
entirely contained in the subtraction.
Most notably there is the trace anomaly and those components of the
energy
momentum tensor which are related to it by energy conservation
(in two dimensions they are $T_{uu,v}$ and $T_{vv,u}$). These parts
are
independent of $\Pi$ or, expressed differently, do not
fluctuate.

An additional (and related)  feature
concerns the absence of correlations between $T_{uu}$ and $T_{vv}$.
Not only shall this give rise to the particular structure of vacuum
fluctuations that extend back to ${\cal I}^-$, but it also implies that on the
horizon the in-going flow and the out-going flow fluctuate independently (for
instance the specification of an outgoing particle on ${\cal I}^+$ does not
affect $T_{vv}$ outside the star and in particular
 on the horizon $r=2M$). This last effect disappears partially
when considering the potential barrier that occurs in the wave equation
eq. (\ref{bh2}).

\subsection{The Conditional Value of the Energy Density}\label{bbb}

In the absence of specification of the final state, $\expect{0}{T_{\mu \nu
({\rm ren})}}{0}$ describes the mean energy content carried by Hawking
quanta. We  remind the reader that $\expect{0}{T_{\mu \nu
({\rm ren})} (x)}{0}$ is regular on the future horizon $U=0$ and that
Hawking radiation can be conceived as the matter
response  that gives regular mean energy densities on the horizon.
 We refer to ref. \cite{birreld} for further discussion of the mean flux.

In order to describe the fluctuations around the mean flux, one inquires
into the conditional energy density when the final state is a particular
out-state which arises in the rewriting of the Heisenberg state $\ket{0}$ into
states with definite energy $\la$ on $\cal{I}^+$,
see eq. (\ref{threethreexiii}).
However in the collapsing geometry,
an external observer does not have access to the region of space time
beyond  the horizon, hence the specifications of the final state
that he can perform are
restricted to an incomplete ($U<0$) region of space time and are therefore
incomplete as well.

A possible specification is,
for instance to use the projector, see eq. (\ref{threefourv}),
\begin{equation}
\Pi_{\la} =  I_{L} \otimes
a^\dagger_{\la}  \ket{B_R}\bra{B_{R}} a_{\la}
 \label{threefourva} \end{equation}
where $I_{{L }}$ is the identity operator restricted to the
inaccessible region $U>0$
and $\ket{B_R}$ is Boulware vacuum in the region $U<0$.
$a^\dagger_\la$ creates a Schwarzschild outgoing photon.
The corresponding conditional energy
reads, see eqs. (\ref{threefouri}), (\ref{threefouriii}) and (\ref{2terms}),
\begin{eqnarray}
\langle T_{\mu \nu} (x) \rangle_{\Pi_{\la}} =
{2 \over \alpha_\la \beta_\la } \hat T_{\mu \nu }(x)
\left[ \varphi^*_{\la,K} \varphi^*_{-\la,K} \right]
 &+& \left[ {
\expect{B} {T_{\mu \nu } (x)} {0} \over  \scal{B}{0}}
- \expect{0}{T_{\mu \nu} (x)}{0} \right]
\nonumber
\\
&+&\expect{0}{T_{\mu \nu
({\rm ren})} (x)}{0}
\label{bh18}
\end{eqnarray}
where the modes $\varphi_{\la,K}$ are defined from the
modes $e^{-i\om U}/\sqrt{4 \pi \om}$, eq. (\ref{modesphi}), as the Minkowski
modes
eq. (\ref{threethreevii}) are defined from the Minkowski plane waves
eq. (\ref{threethreeiii}).

The first two terms arise from the specification of the final state whereas the
third
term is the mean energy.
The first term is equal to the energy of the photon $\la$. The second
one is
the difference of energy between Boulware vacuum and the
Heisenberg vacuum.
This term appears,
as in eq. (\ref{threefouriii}),
because one has specified that, apart from $\la$, their is no other photon
emitted. This is why this term is singular on the horizon
and why
the probability to be in the eigenstate of $\Pi_{\la}$
 vanishes in the absence of backreaction
(in the semiclassical approximation, it is of  order
 $e^{-M^2}$ where $M^2$ is approximately
the total number of photons emitted).

A more reasonable specification because it has a finite probability
of occurring consists in tracing over all the photons
except the photon $\la$ which is imposed to be present (in the Rindler
problem this corresponds to the projector eq. (\ref{threefourvii})).
 Then the conditional energy is
simply, see eqs. (\ref{threefourviii}) and (\ref{2terms}),
\begin{equation}
\langle T_{\mu \nu }(x)\rangle _{\tilde \Pi_\la} =
{2 \over \alpha_\la \beta_\la } \hat T_{\mu
\nu } (x)
\left[ \varphi^*_{\la,K} \varphi^*_{-\la,K} \right]+ \expect{0}{T_{\mu
\nu
({\rm ren})} (x)}{0}
\label{bh19}
\end{equation}

Having traced over all
the other photons, the second term of eq. (\ref{bh18}) is absent in
eq. (\ref{bh19}). Nevertheless this term can also be constructed as the sum of
conditional values that
specify completely the state times the probability that they occur
 (in similar manner to the
 the unitarity relation eq. (\ref{r1})).
 In this way the difference of
energy between the Heisenberg vacuum $\ket{0}$ and
Boulware vacuum $\ket{B}$ is realized as the
sum over all possible radiated photons times the thermal probabilities that
they occur.

Finally we consider  an inertial
two level atom at large distance
from the black hole.
The specification of the state of the radiation is carried out indirectly by
requiring, as in Section 3.2, that the atom get excited.
In this case also the final radiation state is  partially
specified, since the detector is coupled to a finite set of modes.
One finds that the conditional energy contains again two terms
\begin{equation}\langle T_{\mu \nu }\rangle _{ \Pi_+} =
\bk{T_{\mu\nu}}_{e} + \expect{0}{T_{\mu
\nu
({\rm ren})} }{0}
\label{bhTmunu}
\end{equation}
where the fluctuating part, in terms of the $\phi_m$ operators (eq.
(\ref{threefiveiii}))
is, see eq. (\ref{oofive}),
\begin{equation}
\bk{T_{\mu\nu}}_{e}
= { g^2 m^2 \over P_{e}}
 \expect{0}{\phi_m^\dagger \phi_m
:T_{\mu\nu}: }{0}
\label{Tmunue}
\end{equation}
where $:T_{\mu\nu}:$ is the energy momentum operator normal ordered with
respect to the in operators defining Heisenberg vacuum $\ket{0}$.
We shall display the properties of $\bk{T_{\mu\nu}}_{e}$
in the next section.

We note already that the specification of the final radiation state
by the correlations to a transition of the two
level atom
gives rise to finite energy densities on the horizon only
if the coupling to the field decreases faster than $e^{-u/4M}$:
c.f. discussion after eq.
(\ref{threefourx})
and (\ref{threefivexvi}). Therefore, the
specification of a mode (eq. (\ref{bh19})) or a wave packet made out of
positive $\la$ frequencies only gives rise inevitably to
singular energy densities on the horizon.

It is
also
interesting to speculate
about the
nature of the in-going vacuum fluctuations when they cross the future horizon
$U=0$. They could be analysed by
selecting the presence of ingoing quanta near the horizon.
A ''natural" set of
 modes to select near the horizon are Kruskal $v$-modes. One is
therefore led to consider the Kruskal vacuum fluctuations in Schwarzschild
vacuum, which is similar to considering Minkowski fluctuations in Rindler
vacuum. If space time were the full Schwarzschild manifold, these would
present a singularity on the past horizon that could be smoothed out using
wave packets. Since space time is not the full
Schwarzschild manifold (there is no past horizon)
the star's surface
will play the role of past
horizon and one expects large energy densities
in the outermost layers of the star.

\subsection{From Vacuum Fluctuations to Black Hole Radiation}\label{vacfl}

We now turn to
the
the fluctuating part of the conditional energy correlated to the
transition of the two level atom, i.e. the term $\bk{T_{\mu\nu}}_{e}$ of eq.
 (\ref{bhTmunu}).

For s-waves, when one neglects
the residual potential of the
d'Alembertian  eq. (\ref{bh2}), $\langle
T_{\mu\nu}\rangle_e$ is completely independent of
the geometry.
Furthermore, the conformal invariance of the field makes
the mapping of the results obtained in the Rindler problem to the
present problem
straightforward.
Let us choose therefor the time dependent coupling $f(t)$ of our detector
given by the gaussian switch off of Section 3.2, see eq. (\ref{threefivexix})
and eq. (\ref{threefivexxv}). More precisely,
the coupling is
such that the atom, of resonance frequency $\la = m$ will be excited
around the retarded time
$u=u_0$.
Hence its
Fourier
components
are, see  eq. (\ref{threefivexxv})
\begin{equation}
 c_\la =
\int \!dt\ f(t) e^{-imt}e^{i\la t} =
D { \la \over m} e^{i \la u_0}e^{-(\la - m )^2 T^2 /2}
 (1 -
e^{-2\pi \la /a})
\end{equation}
The spread in time is $\Delta t = \Delta u = T$
 and $u_0$ is taken well inside
the region $u>0$, see eq. (\ref{162s}),
 where the isomorphism of the scattered waves and the
Kruskal modes is achieved.

If the two level atom is found excited after
the switch off, due to the vanishing of the modes at $r=0$, see eq.
(\ref{modesphi}), and the light-like character of the propagation, the
correlated radiation state eschews from
a spherically symmetric vacuum fluctuation on ${\cal
I}^-$.
This fluctuation
is located in a region
\begin{equation}
\vert v  -
v_\infty\vert=\vert \Delta U\vert=\vert {\Delta u e^{- u_0 / 4 M}}\vert
\simeq T e^{- u_0 / 4 M}
\label{blaack}
\end{equation}
where $v = v_\infty$ ($=0$ in our
collapse) is the light ray that
shall become the
future horizon $U=0$.
Indeed this localization is furnished by the
$v$ dependence of the
conditional energy density
 on ${\cal I}^-$ which reads (see eqs. (\ref{Tmunue}) and (\ref{weekiv}))
\begin{eqnarray}
\bk{T_{vv}({\cal I^-},v)}_{e}
&=& { g^2 m^2 \over P_{e}} \int\! d\la \!\int
\!d\la^\prime c_\la^* c_{\la^\prime}
{1 \over 4 \pi \sqrt{\la\la^\prime} }\sqrt{
\tilde n_{\la^\prime} ( \tilde n_\la +1)
}\hat T_{\mu
\nu }
\left[
 \varphi^*_{\la,K}
\varphi^*_{-\la^\prime,K}\right]
\nonumber\\
&\simeq & {1 \over 4 \pi r^2}{16 M^2 \over v^2} {m  \over  2 \sqrt{\pi} T }
(N_{m} +1)\ \exp \left[
-\left[
{ 4 M  \over T}  (  \ln ({-v-i\e
\over 4M}) + u_0 )
\right]^2\right]\quad
\label{epsilfr}
\end{eqnarray}
Where the width of the gaussian factor gives eq. (\ref{blaack}).
As in eq. (\ref{weekiv}), the $\e$ specification of the log ensures
that the total energy carried by this fluctuation vanishes, see eq.
(\ref{vanenf}).
Thus
we see that the analysis of the fluctuations by an inertial observer near
${\cal I}^-$ is isomorphic with what was called the Minkowski interpretation
in Part 4.
 As in the accelerated case, the energy density is
enhanced
by the jacobian $du/dU= e^{u / 4 M}$ centered around $u=u_0$ which appears here
as
$1/v^2$ when the reflection at $r=0$ is taken into account. Hence after a
$u$-time of the order of $4M \ln M$, the energy density in $T_{vv}$
(rescaled by $ 4 \pi r^2$)
become``transplanckian" and located within a
distance $\Delta v$ much smaller than the Planck length
(If one does not rescale $T_{\mu\nu}$
the transplanckian energies only exist in a region of finite $r$
which nevertheless increases exponentially with $u_0$).
The dramatic consequences that these transplanckian energies might introduce
are discussed in ref. \cite{EMP}. In that article it is argued that the
nonlinearity of general relativity cannot accommodate these densities and
that  a taming mechanism must exist if Hawking radiation does exist.

After issuing from ${\cal I}^-$, the vacuum fluctuation
 contracts until it reaches $r=0$ and then reexpands along $U=const$
lines. Upon crossing the
surface of the star in a region $\Delta U \simeq T e^{-u_0/4M}$
centered on the horizon,
it separates
into a piece (the partner)
that
falls into the singularity,
carrying a negative Schwarzschild
energy equal to $- m(1+N_m)$ (see eqs. \ref{weekii} and \ref{twos}),
and a piece carrying positive
energy equal to $ m(1+N_m)$ that keeps expanding and escapes to ${\cal
I}^+$ to constitute the quantum that induces the
transition of the atom (see eq. \ref{weeki} and figure 4).
The analysis performed by an inertial asymptotic observer near the
detector, on ${\cal
I}^+$,
is isomorphic with
the Rindler interpretation of Section 3.3 since the gravitational red shift
replace exactly the role of the Doppler accelerated one in
the accelerated place.
For instance, one finds readily that the total energy carried by the
fluctuating $\bk{T_{uu}({\cal I^+},u)}_{e}$ is indeed $m(1+N_m)$ as in
eq. (\ref{seventytwo}).

Similarly, if the two level atom is found
in its ground state
 after the switch off,
its wave function
is correlated to the absence of the Hawking photon specified by $c_\la$.
In that case, one would find near ${\cal I}^-$ a vacuum fluctuation whose
energy content is exactly the opposite of the previously considered
case (times $P_g/P_e$).
Near ${\cal I}^+$ it
would therefore
contain a negative energy flux of total energy $-
 m (N_m + 1) P_e/ (1- P_e)$  encoding the fact that their are quanta
absent from the thermal flux emitted by the black hole.

If more realistically, we take
a two-level atom coupled locally to the field
(i.e. coupled to all the modes $l>0$),
 it
will
select
particles coming out of the black hole in its direction. Then the picture
that emerges is essentially the same as for
an s-wave
except that on ${\cal I}^-$ the vacuum fluctuation is localized
on the antipodal point of the detector.
The  created quantum and its partner,
 are on the same side and
not antipodal
(with respect to each other)
because they have opposite energy.

We now turn to the description in the intermediate regions in order to
interpolate between the descriptions between
${\cal I}^-$ and ${\cal I}^+$.
One possible interpolation
consists in using a set of static observers at constant $r$.
Then the ''Rindler" description would be used everywhere outside the
star. However a difficulty arises in this scheme if one really considers a
set of material ''fiducial" \cite{sus1}
detectors
at constant $r$. For upon interacting
with the field and thermalizing at the local temperature
$\sqrt{r\over(r-2M)}{1 \over 8 \pi M}$ the
detectors will emit large amounts of
ultraviolet Kruskal
"real"
quanta (see Section 3.3 wherein it is shown how the accelerated atom
transforms vacuum fluctuations into "real" quanta). The backreaction of
these
on mass shell quanta
cannot be neglected and, as already stated, cannot be evaluated
owing to the transplanckian energy
they carry.

An alternative interpolation
 consists in giving the value of $T_{\mu\nu}$ in the local
inertial coordinate system (Riemann normal coordinates). This stems from the
idea that
local
physics
should be describe locally in such a coordinate system. This approach has
 been used
in defining the subtraction necessary to
renormalize the energy momentum tensor
\cite{birreld}\cite{mpblocal}\cite{mas}. In the two dimensional model the
local inertial coordinates are easy to construct. Since $\tilde u = r(u,v)$
is an affine parameter along the geodesics
$v={\rm constant}$, a natural way to represent the outgoing flux outside the
star is as
\begin{equation}
T_{\tilde u \tilde u} (\tilde u) = \left( {d u(r,v) \over dr} \right )^2
T_{uu}(u(r,v))\label{localc}
\end{equation}
This is represented in  a
Penrose diagram in figure 6 and Eddington-Finkelstein coordinates in figure 7.
The inertial coordinate $\tilde u$ will come up
very naturally in the next section.

\subsection{The Gravitational Back Reaction}\label{back}

Up to now in this Part we have presented the properties of the
conditional values of $ T_{\mu\nu}$. We now investigate how these matrix
elements intervene in physical processes. An example of the role of these
matrix elements was given in section \ref{conddec}.
Here we shall specifically discuss
gravitational back reaction effects to black hole radiation.

A simple way to understand the role of the
conditional values of $ T_{\mu\nu}$ is to imagine a change in the background
geometry $g_{\mu\nu} \to g_{\mu\nu} + \delta g_{\mu\nu}$. This change
modifies both the mean values of the flux as well as more detailed properties
such as the probability to find a specific photon on ${\cal{I}}^+$.
We focus on this later change.

The new probability can be computed, in the interacting picture,
around the background $g_{\mu\nu}$ with the Hamiltonian given by
\begin{equation}
\int d^4x H_{int} = -{1\over 2} \int d^4x \sqrt{-g} \delta g^{\mu\nu}
T_{\mu\nu}
\label{backreact}
\end{equation}
The new probability is then, see eq. (\ref{r1}),
\begin{eqnarray}
P_{g+\delta g}
= \langle 0 \vert
e^{i\int d^4x H_{int}}
\Pi
e^{-i\int d^4x H_{int}}
 \vert 0 \rangle
\label{rr1}
\end{eqnarray}
where $\Pi$ specifies the state of the radiation field on ${\cal I}^+$.
To first order in $\delta g_{\mu\nu}$
 the relative change in
probability is
\begin{eqnarray}
{
P_{g+\delta g}  -P_{g}  \over P_{g} }
&=& {
\langle  0 \vert \Pi
\left( -i\int d^4x H_{int}\right)
 \vert 0 \rangle
\over \langle 0 \vert  \Pi \vert 0 \rangle } \ +\ \mbox{ c.c.}\nonumber\\
&=&
\int d^4x \sqrt{-g} \delta g^{\mu\nu}
\mbox{ Im} \left[ \langle T_{\mu\nu} \rangle_{\Pi} \right]
\label{rr2}
\end{eqnarray}
It is thus the imaginary part of $\langle T_{\mu\nu} \rangle_{\Pi}$
only which controls the change in probability induced by $\delta g_{\mu\nu}$.
Furthermore, since the background part of conditional energies (the second
term of eq. (\ref{2terms})) is by construction real, only the fluctuating
part, which depends explicitly of the selected quantum,
 contributes to $P_{g+\delta g}  -P_{g}$.

To illustrate how the various properties of
the fluctuating part of $\langle T_{\mu\nu} \rangle_{\Pi}$ intervene
in such an expression, let us take
the simple example wherein $\delta g_{\mu\nu}$ is due to the infall of an
additional
light like shell of mass $\delta m$ at time $v=v^\prime$ with $v^\prime
\geq v_S$. Then for $v_S<v<v^\prime$ the metric, eq. (\ref{bh1}) and eq.
(\ref{162s})), is unchanged
\begin{equation}
ds^2 = \left( 1 -{2M\over r}\right) dv^2 -2dvdr -r^2 d^2\Omega
\label{rr3}\end{equation}
whereas for $v>v^\prime$ it is
\begin{equation}
ds^2 = \left( 1 -{2M+ 2\delta m \over r}\right) dv^2 -2dvdr -r^2 d^2\Omega
\label{rr4}\end{equation}
where we have used for obvious convenience the Eddington Finkelstein
 coordinates $v$ and $r$.

The change in the action $S=\int d^4x \sqrt{-g} {1\over 2}
g^{\mu\nu} \partial_\mu \phi \partial_\nu \phi$ is for s-waves
\begin{eqnarray}
\delta S = - \int d^4 x H_{int} &=& \int_{v^\prime}^{+\infty}\!dv
\int_0^\infty\!dr \ 4 \pi r^2 (\delta m/r) \partial_r\phi \partial_r\phi
\nonumber\\
&=& \delta m \int_{v^\prime}^{+\infty}\!dv
\int_0^\infty\!dr \ 4 \pi r  T_{\tilde u \tilde u}(r,v)
\label{rr5}\end{eqnarray}
where $ T_{\tilde u \tilde u}$ is given by eq. (\ref{localc}).
As emphasized
at the end of section \ref{vacfl} it is the energy
momentum in Riemann normal coordinates which appears automatically in such
problems since the response to a local change in the geometry is local as
well.

In order to compute the change in probability due to $H_{int}$, we first
recall that the imaginary part of $\langle  T_{\tilde u \tilde u}
\rangle_{\Pi}$ vanishes on
the other side of the horizon, for
$r<2M$ (see discussion after eq. (\ref{twos})). From eq. (\ref{rr5}) we
understand
that this is dictated by causality: a change in $g_{\mu\nu}$ in the region
$r<2M$ cannot affect the probability to find a specific Hawking photon on this
side.

In the case when $v^\prime = v_S$, one simply has a shell of  mass $M+\delta
m$. In this case we know exactly the change in probability since the
probability of finding a photon of frequency $\la$ is \begin{equation}
P_{M+\delta m}
= |\be_{\om,\la}|^2/|\alpha_{\om,\la}|^4
=e^{-8 \pi \la (M+\delta m)} (1 - e^{-8 \pi \la (M+\delta m)})
\end{equation}
hence
\begin{equation}
{\delta P\over P} = -8\pi \la \delta m {(1 - 2e^{-8 \pi \la M})
\over (1 - e^{-8 \pi \la M})}
\end{equation}
Thus in this case the transplanckian character of the energy density is
washed out by the integration in eq. (\ref{rr5}). This can be verified
explicitly by evaluating the integral near $r=2M$ and  $v=v^\prime$.
There, one can
replace $dr$ by $-dU/2$ whereupon by making appeal to the vanishing of the
integrals $\int\! dU\! \langle T_{UU}\rangle_\Pi$ and $\int\! du
\mbox{Im}\langle T_{uu}\rangle_\Pi$ one finds
that the integral at fixed $v$ $\int\!
dr 4 \pi r \mbox{Im}
\langle T_{\tilde u\tilde u} \rangle_\Pi$ does not scale like $e^{u/4M}$.

Furthermore at large $r$ the integral, eq. (\ref{rr2}),
vanishes once more since $\int\! du
\mbox{Im}\langle T_{uu}\rangle_\Pi$ vanishes.
Hence if the additional mass
crosses the photon trajectory when it is on mass shell at $r>>4M$ there is no
modification of the probability of creating the photon.

We have thus obtained a local description of the quantum matter response to a
modification of the classical background geometry. To address the quantum
gravitational back reaction to the creation of a Hawking photon one should
treat
$\delta g_{\mu\nu}$ as a quantum operator\cite{THooft}\cite{bmpps}.
For spherically symmetric radiation and
spherical symmetric gravitational fields the relation between $\delta
g_{\mu\nu}$ and $T_{\mu\nu}$ is a constraint, i.e.
$\delta g_{\mu\nu}$ is completely fixed by $T_{\mu\nu}$. One can then envisage
the backreaction as an iterative scheme which ultimately should be treated self
consistently.

The first step in this procedure is very simple. It consists in
taking $\delta g_{\mu\nu}$ to
be the "position coordinate" of the additional system (the
oscillator) introduced in Section 2.5.
Then, as for the oscillator, see eq. (\ref{ootwelve}),
the mean value of $\delta g_{\mu\nu}$ is obtained by integrating
Einstein's equations
with the mean energy momentum tensor as a source.
This corresponds to the linear approximation to the semiclassical solution
\cite{Massar2}.
But one can also evaluate the "mean" conditional
value of $\delta g_{\mu\nu}$. This conditional
change in the metric
 is obtained by integrating Einstein's equations
with $\bk{T_{\mu\nu}}_{\Pi}$ as source.
\footnote{See however the different ways in which Re$\langle T_{uu}\rangle_\Pi$
and Im$\langle T_{uu}\rangle_\Pi$ enter in eq. (\ref{ootwelve}). This might
lead to interesting effects.}
 Since the total energy carried by the conditional
value of $T_{\mu\nu}$
vanishes from ${\cal I}^-$ till the emergence of the fluctuation from the star
after reflection on $r=0$, $\bk{\delta g_{\mu\nu}}_{\Pi}$, the conditional
value of $\delta g_{\mu\nu}$, will vanish
outside the interval $\Delta v$ eq. \ref{blaack} centered around $v_S$.
Within that interval the precise
shape of  $\bk{\delta g_{\mu\nu}}_{\Pi}$
 will depend on the particular choice of
selected wave packet by the projector ${\Pi}$.
On the contrary, outside the star, for $r>4M$ and
$u> u_0$ (i.e. in the middle
of the two members of the pair), $\bk{\delta g_{\mu\nu}}_{\Pi}$ will encode the
mass loss
$\omega$ and in fact describes a new classical (real valued)
Schwarzschild space where the mass is
$M-\omega$.

The next step consists in taking
 into account the effect of $\delta g_{\mu\nu}$
on the production of the Hawking photon itself. This gravitational self
interaction can be encoded in
 a interaction
hamiltonian of the form $H_{int} = T_{\mu \nu} D^{\mu \nu
\alpha\beta}T_{\alpha\beta}$ where $D$ is the linearized gravitation
propagator.
To calculate $\delta P/P$ due to this self interaction one must confront the
infinities which arise in matter loops. This will not be done here.

However there are some simple question wherein $H_{int}$ does come in which do
not involve loops. One such question is the effect of the creation of a first
Hawking photon on subsequent ones. Suppose one calculates
$P_{\Pi_{\la_1,u_1;\la_2,u_2}}$ where $\Pi_{\la_1,u_1;\la_2,u_2}$ specifies the
presence on ${\cal I}^+$ of a photon of frequency $\la_1$ located near $u_1$
and another photon of frequency $\la_2$ located near $u_2$. Then to first order
in $H_{int}$, the relative change in the probability to find the two photons is
\begin{equation}
{ \delta P_{\Pi_{\la_1,u_1;\la_2,u_2}} \over
P_{\Pi_{\la_1,u_1}}P_{\Pi_{\la_2,u_2}} }
=
2 \mbox{ Im}\left[
{\bra {0}\Pi_{\la_1,u_1;\la_2,u_2} H_{int} \ket{0} \over
\bra {0}\Pi_{\la_1,u_1;\la_2,u_2}  \ket{0}}
\right]
\end{equation}
where, in the absence of gravitational coupling, the probability of
finding two photons factorizes into the individual
probabilities.
In this expression once more there are infinite loops. However upon taking
derivatives the quantity
$(d/d\la_1)(d/ d\la_2) (\delta P / P)$ is finite. The  effects
of particle $\la_1,u_1$ on particle $\la_2,u_2$ are
isolated from other
effects. This finite quantity can then
be expressed in terms of the
products of the one-particle conditional values
\begin{eqnarray}
\langle T_{\mu\nu}(x)\rangle_{\Pi_{\la_1,u_1}} D^{\mu \nu \alpha \beta}(x,y)
\langle T_{\alpha\beta}(y)\rangle_{\Pi_{\la_2,u_2}}\nonumber\\
\langle \partial_\mu \phi(x)\partial_\alpha
\phi(y)\rangle_{\Pi_{\la_1,u_1}} D^{\mu \nu \alpha \beta}(x,y)
\langle \partial_\nu \phi(x)\partial_\beta \phi(y)\rangle_{\Pi_{\la_2,u_2}}
\end{eqnarray}
In conclusion we have shown that the conditional values of $T_{\mu\nu}$ enter
into tree graphs. In order to understand the role of the transplanckian
frequencies in Hawking radiation and how they are tamed by quantum gravity one
should confront loops and the infinities they involve. We hope to report on
this in subsequent work.

\section {Post Selection, Weak Measurement}

\subsection{Introduction}

In section \ref{conddec} we showed succinctly  how the conditional values of
$T_{VV}$  control the first order perturbation onto an additional
system. The aim of this section is to present a self contained
discussion devoted this result.
We shall work in complete
generality and make no explicit reference to the accelerated
system nor to black hole radiation.

This analysis of the conditional values (i.e. non diagonal
matrix element of an
operator)  was first carried out by Aharonov et al. \cite{aharo}
in the context of
measurement theory. In
essence they studied the first order backreaction
 onto an
additional system
which they took to be a measuring device. But the formalism is
more general. In the case of pair production the additional system could
be the external electric or gravitational field which is now
described quantum mechanically. Moreover this formalism can be used to
study the self interaction of
the pairs without introducing the additional system. This is because,
when the first order (or weak)  approximation is
valid, the backreaction takes a simple and universal form governed by  a
c-number, the conditional value of the operator which controls the
interaction (called by Aharonov et al. the ``weak value'').

In section 5.2 we implement the specification of the
final state(s) in a  rather formal way by acting  with projection
operators  which select
the desired final state(s). Aharonov et al.  call this
specification of the final state, a ``post-selection''.

In section 5.3 we show
how  post-selection may be realized operationally following the rules of
quantum mechanics by coupling the system to be studied ($S$)
an  additional system in a metastable
state   (the "post--selector" $PS$) which will make a transition only if the
system is in the required  final state(s).
The conditional value of an operator obtained in this manner changes as time
goes by from an off diagonal matrix element
to an expectation value, thereby making
contact with more familiar physics. This extended formalism finds important
application when considering the physics of the accelerated detector since
the accelerated detector itself plays the role of post selector.

\subsection{Conditional (or Weak) Values}

The approach developed by
Aharonov et al.\cite{aharo} for studying pre- and post-selected
ensembles
consists in performing at an intermediate time a "weak measurement"
on $S$.

The system to be studied ($S$) is in the state $\ki$ at
time $t_i$ (or
 more generally is described by a density matrix $\rho_i$).
The unperturbed time
evolution of this pre-selected state
 can be described by the following density matrix
\begin{equation} \rho_S(t) = U_S(t,t_i)\ki \bi U_S(t_i,t) \label{weaki}
\end{equation}  where $U_S = \exp (- i H_S t)$ is the time evolution operator
for the system $S$. The post-selection at time $t_f$ consists in specifying
that the system belongs to a certain subspace, ${\cal H}_S^{0}$, of
${\cal H}_S$.
Then the probability to find the system in
this subspace at time $t_f$ is \begin{equation}
P_{\Pi_S^{0}} = Tr_S \Bigl [ \Pi_S^{0} \rho(t_f) \Bigr] = Tr_S \Bigl[
\Pi_S^{0} U_S(t_f,t_i)\ki \bi U_S(t_i,t_f) \Bigr] \label{weakii}
\end{equation}   where
$\Pi_S^{0}$ is the projection operator onto ${\cal H}_S^0$ and $Tr_S$ is
the trace over the states of system $S$.
In the special cases wherein the specification
of the final state is to be in a pure state $\ket{\psi_f}$ (i.e.
$\Pi_S^0=\ket{\psi_f}\bra{\psi_f}$)
then the probability is simply given by the overlap
\begin{equation}
P_{f}= \vert \bra{\psi_f} U_S(t_f,t_i)\ket{\psi_i}\vert^2
\end{equation}

Following Aharonov et al. we introduce an additional system, called the "weak
detector" ($WD$), coupled to $S$.
The
interaction hamiltonian between $S$ and $WD$ is taken to be of the form
$H_{S-WD} (t) = \e f(t) A_S B_{WD}$ where $\e$ is a coupling constant,
$f(t)$ is a c-number function, $A_S$ and $B_{WD}$ are hermitian operators
acting
on $S$ and $WD$ respectively.

Then to first order in $\e$ (the coupling is weak), the evolution of the
coupled
system $S$ and $WD$ is given by
 \begin{eqnarray}
\rho(t_f)& =&\ket{\Psi (t_f)}\bra{\Psi(t_f)}\nonumber\\
\mbox{where}\quad \quad \quad\quad&&\nonumber\\
\ket{\Psi(t_f)}& =& \Bigl [
U_S(t_f,t_i)U_{WD}(t_f,t_i) - i\e \int_{t_i}^{t_f}\! dt \
 U_S(t_f,t)U_{WD}(t_f,t)  f(t) A_S B_{WD}\times \Bigr. \nonumber\\
&& \quad \quad \Bigl.
  U_S(t,t_i)U_{WD}(t,t_i) \Bigr] \ki \ket{WD}
\label{weakiii}
 \end{eqnarray}
where $U_S$ and $U_{WD}$ are
the free evolution operators for $S$ and  $WD$ and  $\ket{WD}$ is the initial
state of  $WD$. Upon post-selecting at $t=t_f$ that $S$ belongs to the
subspace ${\cal H}_S^{0}$ and tracing over the  states of the
system $S$, the reduced density matrix describing the $WD$ is obtained
\begin{eqnarray}
 \rho_{WD}(t_f) &=& Tr_S
 \Bigl [ \Pi_S^0 \rho(t_f)
\Bigr]\end{eqnarray}
To first order in $\e$, it takes a very simple
form \begin{eqnarray}
 \rho_{WD}(t_f)&\simeq &
P_{\Pi_S^{0}}\ket{\Psi_{WD}(t_f)}
\bra{\Psi_{WD}(t_f)}\nonumber\\
\mbox{where}\quad \  \ \quad&&\nonumber\\
\ket{\Psi_{WD}(t_f)}&=&
\left[ U_{WD}(t_f,t_i) -
 i\e
\int_{t_i}^{t_f}\! dt \ U_{WD}(t_f,t)  f(t) \bk{A_{S}(t)}_{\Pi_S^{0}}
B_{WD}U_{WD}(t,t_i)
 \right] \ket{WD}\nonumber\\
&&\label{weakiv}
 \end{eqnarray}
where $P_{\Pi_S^{0}}$ is the probability to be in subspace ${\cal H}_S^{0}$
and
  \begin{equation}
\bk{A_{S}(t)}_{\Pi_S^{0}} = {Tr_S
\Bigl [ \Pi_S^{0} U_S(t_f,t) A_S U_S(t,t_i) \ki\bi U_S(t_i,t_f) \Bigr]
\over Tr_S \Bigl [ \Pi_S^{0} U_S(t_f,t_i)\ki\bi U_S(t_i,t_f) \Bigr]}
\label{weakv}
\end{equation}
 is a c-number called  the weak or conditional value of $A$. If
one specifies completely the final state, $\Pi_S^{0} =
\ket{\psi_f}\bra{\psi_f}$  then
the result of Aharonov et al.
obtains:
\begin{equation}
\bk{A_{S}(t)}_{\psi_f} = {\bra{\psi_f}
U_S(t_f,t) A_S U_S(t,t_i) \ki  \over \bra{\psi_f} U_S(t_f,t_i) \ki}
\label{weakvi}
 \end{equation}
The principal feature of the above formalism is its independence on the
internal structure of the $WD$. The first order backreaction of $S$
onto $WD$ is
universal: it is always controlled by the c-number $\bk{A_{S}(t)}_{\Pi}$,
the ''weak
value of $A$". Therefore
if $S$ is coupled to itself by an
interaction hamiltonian, the backreaction
will be controlled by the
weak value of $H_{\rm int}$
in first order perturbation theory.
For instance the modification of the probability that the
final state belongs to ${\cal H}_S^{0}$ is
given by the imaginary part
of $\bk{H_{\rm int}}_{\Pi_S^{0}}$.
Indeed
\begin{eqnarray}
P_{\Pi_S^0}^\prime &=& Tr_S \left[
\Pi_S^0 ( 1 - i \int\! dt\  H_{\rm int} )
\rho_i ( 1 + i \int\! dt\  H_{\rm int} )\right]\nonumber\\
&=&P_{\Pi_S^0} (1 - 2 { \mbox{ Im} \bk{H_{\rm int}}_{\Pi_S^{0}}})
\end{eqnarray}

The weak value of $A$ is complex. By  performing a
series of
measurements on $WD$
and by varying the coupling function $f(t)$, the real and imaginary part of
$\bk{A_{S}(t)}_{\Pi}$
 could in principle be determined. Here the word ''measurement"
must be understood in its usual quantum sense: the average over repeated
realizations of the same situation. This means that the weak value of $A_S$
should also be understood as an average. The fluctuations around
$\bk{A_{S}(t)}_{\Pi}$
are encoded in the second order terms of eq.~(\ref{weakiii}) which have been
neglected in eq.~(\ref{weakiii}).

To illustrate the role of the real and imaginary parts of $\bk{A_{S}}_{\Pi}$,
we
recall the example of Aharonov et al. consisting of a  weak
detector which has one degree of freedom $q$,
 with a gaussian initial state
$\scal{q}
{WD}=e^{-q^2/
2\Delta^2}$,$-\infty < q < + \infty$ (see also the example of Section 3.2).
 The unperturbed hamiltonian of $WD$ is taken to vanish
(hence $U_{WD}(t_1,t_2) = 1$)
 and the interaction hamiltonian is  $H_{S-WD} (t) =\epsilon
\delta(t-\tilde t) p A_S$ where $p$ is the momentum conjugate to $q$. Then
after the post-selection the density matrix
of the $WD$ is given by $P_{\Pi} \ket{WD(t_f)}\bra{WD(t_f)}$,
see eq. (\ref{weakiv}),
  where, to
first order in $\epsilon$, $\ket{WD(t_f)}$ is
\begin{eqnarray}
\ket{WD(t_f)} &=&\left(1-i\epsilon p\bk{A_{S}(\tilde t)}_{\Pi} \right)
\ket{WD}
 \end{eqnarray}
whereupon the conditional value of $q$ and $p$ are
\begin{eqnarray}
\bk{q}_{\Pi}&=& \epsilon \mbox{Re} \bk{A_{S}(\tilde t)}_{\Pi}\nonumber\\
\bk{p}_{\Pi}&=& \epsilon \mbox{Im} \bk{A_{S}(\tilde t)}_{\Pi}/\Delta^2
 \label{weakviii}
 \end{eqnarray}
 Thus the real part of $\bk{A_{S}(\tilde t)}_{\Pi} $ induces a
translation of the center of the gaussian, the imaginary part
a change in the
momentum. Their effect on the $WD$ is therefore measurable.
The validity of the first order approximation requires
$\e |\bk{A_{S}(\tilde t)}_{\Pi} | / \Delta <<1$.

It is instructive to see how unitarity is realized in the above formalism.
Take
 $\Pi^j_S$
to be  a
complete orthogonal set of projectors acting on the
Hilbert space of $S$. Denote by $P_j$ the probability that
the final state of the system belong to the subspace spanned by $\Pi^j_S$ and
by $\bk{A_{S}(t)}_{\Pi^j_S} $ the corresponding weak
value of $A$. Then the mean value of
$A_S$ is  \begin{equation} \bi A_S(t) \ki =
\sum_j P_j \bk{A_{S}(t)}_{\Pi^j_S}
  \label{weakix}
\end{equation}
Thus the mean
backreaction if no post-selection is performed is the average over the
post-selected backreactions (in the linear response approximation).
Notice that the imaginary parts of the weak
values necessarily cancel since the l.h.s. of eq.~(\ref{weakix}) is
real. Equation (\ref{weakix}) is the short cut used in the main text to
obtain with minimum effort the weak values.

\subsection{Physical Implementation of Post Selection}

Up to now the postselection has been implemented by
inserting by hand
the projector $\Pi^0_S$.
Such a projection may be
realized operationally by introducing
an additional
quantum
system, a ''post-selector" ($PS$),
coupled in such a way
 that it will make a transition if and only if the system $S$
is in the required final state. Then by considering
only that subspace of the
final states in which $PS$ has made the
transition, a post-selected state is specified. This
quantum description of the
post-selection
is similar in spirit to the measurement theory developed in ref.
\cite{Vonn}: by introducing explicitly the measuring device in the
hamiltonian the collapse of the wave function ceases to be a necessary
concomitant of measurement theory.

We shall consider the very simple model of a $PS$ having two states,
initially in the ground state, and coupled to the system by an interaction
of the form \begin{equation}
H_{S-PS} = \la g(t) ( a^\dagger Q_S + a Q_S^\dagger)
\label{weakx}
\end{equation}
where $\la$ is a
coupling constant, $g(t)$ a time dependent function, $a^\dagger$  the operator
that induce transitions from the ground state to the exited state of the
$PS$, $Q_S$ an operator acting on the system $S$. The
postselection is performed at $t=t_f$ and consists in finding the $PS$ in the
exited state.

For simplicity we shall work to second order in $\la$ (although in principle
the interaction of $PS$ with $S$ need not be weak).
In the interaction
representation, the wave function of the combined system $S+WD+PS$ is
to order $\epsilon$ and to order $\lambda^2$
\begin{eqnarray}
\lefteqn{{\cal T} e^{-i \int\! dt \ H_{S-WD}(t) + H_{S-PS}(t)}
\ket{\psi_i}\ket{WD}\ket{0_{PS}}=}\nonumber\\
& & \left[
1 -i \int dt \left( H_{S-WD}(t) + H_{S-PS}(t) \right)\right.
\nonumber\\
& & \ -{1 \over 2}
\int dt \int dt^\prime
{\cal T}\left[ H_{S-PS}(t)  H_{S-PS}(t^\prime) \right]
-
\int dt \int dt^\prime
{\cal T}\left[ H_{S-WD}(t)  H_{S-PS}(t^\prime) \right]
\nonumber\\
& & \left.\
+{i \over 2}
\int dt \int dt^\prime\int dt^{\prime\prime}
{\cal T}\left[ H_{S-PS}(t)  H_{S-PS}(t^\prime)
 H_{S-WD}(t^{\prime\prime})\right]
\right]
\ket{\psi_i}\ket{WD}\ket{0_{PS}}\quad
\label{weakxi}
\end{eqnarray}
where $\ket{0_{PS}}$ is the ground state of $PS$ and ${\cal T}$ is
the time ordering operator.
The probability of finding the $PS$ in the excited state at $t=t_f$ is,
at order $\la^2$,
\begin{equation}
P_{e}
= \la^2  \bra{\psi_i} \int dt g(t) Q^\dagger_S
\int dt^\p g(t^\p) Q_S \ket{\psi_i}
\end{equation}
Upon imposing that the
$PS$ be in its excited state at $t=t_f$ the
resulting
wave function  is, to order $\epsilon$ and $\la^2$,
\begin{eqnarray}
&&\quad \left [
-i \int dt \la g(t) Q_S(t)  \right.
\nonumber\\
&& \left. -
\int dt \int dt^\prime
{\cal T}
\left[\epsilon f(t) A_S(t) B_{WD}(t) \la g(t^\prime) Q_S(t^\prime)
\right]
\right]
\ket{\psi_i}\ket{WD}a^\dagger\ket{0_{PS}}
\label{weakxii}\end{eqnarray}
Making a density matrix out of the state (\ref{weakxii}),
tracing over the states of $S$ and $PS$
yields the reduced density matrix $\rho_{WD} =
P_{e}\ket{\Psi_{WD}}\bra{\Psi_{WD}}$ of $WD$, where, to order $\epsilon$
$\ket{\Psi_{WD}}$ is
\begin{equation}
\ket{\Psi_{WD}}=\left[
1 - i \epsilon \int_{t_i}^{t_f} d\tilde t f(\tilde t) B_{WD}(\tilde t)
\bk{A_{S}(\tilde t)}_{e} \right] \ket{WD}
\label{weakxiii}
\end{equation}
and
\begin{equation}
\bk{A_{S}(\tilde t)}_{e} =
{ \bi \int dt g(t) Q_S^\dagger (t)  \int dt^\p
g(t^\p) {\cal T } \left[ A_S(\tilde t) Q_S(t^\p) \right]
\ki \over \bi \int dt g(t) Q_S^\dagger (t) \int
dt^\p g(t^\p) Q_S(t^\p) \ki} \label{weakxiv} \end{equation}
Note how the weak value of $A_S$ results from the quantum mechanical
interference of the two terms in eq.~(\ref{weakxii}).

There are
several important cases when the time ordering in eq.~(\ref{weakxiv})
simplifies.
If $g(t)$ is non vanishing only after $t=\tilde t$, ie. WD interacts with S
before S interacts with PS, then $\bk{A_{S}(\tilde t)}_{e}$  takes a
typical (for a weak value) asymmetric form
\begin{equation}
 \bk{A_{S}(\tilde t)}_{e} =  {
\bi \int dt g(t) Q_S^\dagger (t)
 \int dt^\p g(t^\p) Q_S(t^\p)  A_S(\tilde t) \ki \over \bi \int dt g(t)
Q_S^\dagger
(t) \int dt^\p g(t^\p) Q_S(t^\p) \ki} \label{weakxv} \end{equation}
If in addition $g(t)=\delta(t-t_f)$
and $Q_S=\Pi_S^0$,
eq.~(\ref{weakv}) is recovered using $ (\Pi_S^0)^2 = \Pi_S^0$ and
$t_f >\tilde t$. This
is expected since in this case the post--selector has simply gotten
correlated to the system in the subspace ${\cal H}^0_S$

If on the other hand $g(t)$ is non vanishing only before $t=\tilde t$, ie. S
interacts with PS before WD interacts with S, then the time
ordering operator becomes trivial once more and eq.~(\ref{weakxiv}) takes the
form \begin{equation}
\bk{A_{S}(\tilde t)}_{e} =  { \bi \int dt g(t)
Q_S^\dagger (t) A_S(\tilde t)
 \int dt^\p g(t^\p) Q_S(t^\p)   \ki \over \bi \int dt g(t) Q_S^\dagger (t)
\int dt^\p g(t^\p) Q_S(t^\p) \ki} \label{weakxvi}
\end{equation}
This is by construction the
 expectation value of $A_S$ if the $PS$ has made a transition.
It is
necessarily real contrary to  eq.~(\ref{weakxv}).

Finally, the weak value of $A_S$ if the $PS$ has not made a
transition can also be computed.
Once more the two cases discussed in eqs (\ref{weakxv}) and (\ref{weakxvi})
are particularly simple: if $g(t)$ is non vanishing only after $t=\tilde t$ one
finds \begin{eqnarray}
\bk{A_{S}(\tilde t)}_{d} &=&{1 \over 1 - P_{e}  }\Biggl(
\bra{\psi_i}A_S\ket{\psi_i} \nonumber\\
& &-\la^2 \mbox{ Re}  \left[
\bi \int dt g(t) Q_S^\dagger (t)
 \int dt^\p g(t^\p) Q_S(t^\p)  A_S(\tilde t) \ki\right]\Biggr)
\label{weakxvii}
\end{eqnarray}
On the other hand if $g(t)$ is non vanishing only before $t=\tilde t$ one finds
\begin{eqnarray}
\bk{A_{S}(\tilde t)}_{d}  &=&{1 \over 1 - P_{e}  }\Biggl(
\bra{\psi_i}A_S(\tilde t)\ket{\psi_i} \nonumber\\
& & -{1 \over 2} \la^2 \mbox{ Re} \left[
\bi A_S(\tilde t) \int dt
 \int dt^\p {\cal T} g(t) Q_S^\dagger (t) g(t^\p) Q_S(t^\p)   \ki\right]
\Biggr)
\label{weakxviii} \end{eqnarray}
These are related to the mean value of $A_S$
and to eq.~(\ref{weakxiv}) through the unitary relation eq.~(\ref{weakix}):
if $g(t)$ is non vanishing only after $t=\tilde t$
\begin{eqnarray}
 P_{e}\bk{A_{S}(\tilde t)}_{e}
+ (1 - P_{e} )
\bk{A_{S}(\tilde t)}_{d} = \bi A_S(\tilde t) \ki \label{weakwviiii}
\end{eqnarray}
and if $g(t)$ is non vanishing only before $t=\tilde t$
\begin{eqnarray}
\lefteqn{ P_{e}\bk{A_{S}(\tilde t)}_{e}
+ (1 - P_{e} )\bk{A_{S}(\tilde t)}_{d} \ =
\bi e^{i\int \! dt \ H_{S-PS}}
A_S (\tilde t)
e^{-i\int \! dt \ H_{S-PS}}
\ki
\ =}
\nonumber\\
& &\bi A_S(\tilde t) \ki
  +\la^2
\bi \int dt g(t) Q_S^\dagger (t) A_S(\tilde t)
\int dt^\p g(t^\p) Q_S^\dagger (t^\p)\ki \ -\nonumber\\
& &  -\la^2 \mbox{ Re } \bi A_S(\tilde t) \int dt
\int dt^\p {\cal T} g(t) Q_S^\dagger (t) g(t^\p) Q_S^\dagger (t^\p)\ki
\label{weakwxx}
\end{eqnarray}
where the r.h.s is the average value of $A_S$ before
(eq.~(\ref{weakwviiii})) and after (eq.~(\ref{weakwxx})) the detector
has interacted with $S$.
\\
\\
{\bf Acknowledgements.}
\noindent The authors would like to thank R. Brout,
F. Englert, S. Popescu and Ph. Spindel for very helpful
discussions.

\section{Appendix: The $\cal{D}$ Term}\label{app}

We recall the this term arises from the following
decomposition of the second $g^2$ Born term in
$\ket{\psi_-(t=+\infty)}=e^{-i \int dt dx H_{int}}\ket{0_M} \ket{-}$:
\begin{eqnarray}
& & - g^2 m^2
\int_{-\infty}^{+\infty}\!
d\tau
\int_{-\infty}^{\tau}\!
d\tau^\p f(\tau) e^{-im\tau}  \phi(\tau)
  f^*(\tau^\p) e^{+im\tau^\p}  \phi(\tau^\p)\ket{0_M}
\ket{-}\nonumber \\
&=& - {g^2 m^2 \over 2}
\int_{-\infty}^{+\infty}\!
d\tau
\int_{-\infty}^{+\infty}\!
d\tau^\p f(\tau)f^*(\tau^\p)
e^{-im(\tau-\tau^\p)} \phi(\tau) \phi(\tau^\p)
\left[ 1 + \epsilon(\tau - \tau^\p)\right]
\ket{0_M}
\ket{-}\nonumber \\
&=& -{g^2 m^2\over 2}
\phi_m^\dagger \phi_m \ket{0_M}\ket{-}
-  g^2 m^2
{\cal D}  \ket{0_M}\ket{-}
\label{stateA}
\end{eqnarray}
where
\begin{equation}
{\cal D} = {1 \over 2}
\int_{-\infty}^{+\infty}\!  d\tau_2
\int_{-\infty}^{+\infty}\!
  d \tau_1 f(\tau_2) f^*(\tau_1)
\epsilon(\tau_2 - \tau_1) e^{-i m( \tau_2- \tau_1)}
\phi(\tau_2) \phi(\tau_1)
\label{calDA}
\end{equation}
and where $\epsilon(\tau_2 - \tau_1) = \theta(\tau_2 - \tau_1) -
\theta(\tau_1 - \tau_2 )$.

To explicitize the role of the ${\cal D}$ term, it is appropriate
to compute the energy density carried by it when the initial state is
$\ket{0_M}\ket{-}$.
One finds, to  order $g^2$,
\begin{eqnarray}
\bk{T_{VV}(V)}_{{\cal D}}&=&  -g^2 m^2
\mbox{ Re}\left[ \bra{0_M} T_{VV}(V) {\cal D} \ket{0_M}\right]
\nonumber \\
&=&  -g^2 m^2
 \bra{0_M} \left[ T_{VV}(V),{\cal D}\right]_- \ket{0_M}
\nonumber \\
&=& - {g^2 m^2 \over 2}
\int_{-\infty}^{+\infty} \! d\tau_2
\int_{-\infty}^{+\infty} \! d \tau_1
f(\tau_2)  f^*(\tau_1)
\epsilon(\tau_2 - \tau_1)
 e^{-i m( \tau_2- \tau_1)}
\nonumber \\
& &\quad \quad \quad \quad \quad \quad \quad \quad \quad \quad
\bra{0_M} \left[T_{VV}(V), \phi(\tau_2)\phi(\tau_1) \right]_- \ket{0_M}
\label{Comm}
\end{eqnarray}
where we have used the antihermitian property of ${\cal D}$:
${\cal D}^\dagger =
-{\cal D}$.
$\bk{T_{VV}(V)}_{{\cal D}}$ enjoys the following properties.

1. Being a commutator, $\bk{T_{VV}(V)}_{{\cal D}}$ is causal (see
eq. (\ref{van})), and vanishes
in the left quadrant $V<0$ contrary to $\bk{T_{VV}(V)}_{e}$ and
$\bk{T_{VV}(V)}_{g}$.

2. $\bk{T_{VV}(V)}_{{\cal D}}$ carries no Minkowski energy since
 the hamiltonian $H_M$, eq. (\ref{threethreev}), annihilates Minkowski vacuum.

3. $\bk{T_{VV}(V)}_{{\cal D}}$ carries no Rindler energy since
$H_R$ (the boost generator given in eq. (\ref{threefouriv})) annihilates
Minkowski vacuum.
Therefore by virtue of 1.,
the Rindler energy in the right quadrant ($V>0$) vanishes
\begin{equation}
\int_{-\infty}^{+\infty} \! dv\ \bk{T_{vv}(v)}_{{\cal D}} =0
\label{Com2}
\end{equation}
Thus $\bk{T_{vv}(v)}_{{\cal D}}$ is, at most, an energy density repartition.

4. When $f(\tau)=1$ for all $\tau$, $\bk{T_{vv}(v)}_{{\cal D}}$ vanishes
identically.
To prove
this one evaluates
the commutator in eq. (\ref{Comm}) and one finds
\begin{eqnarray}
\bk{T_{vv}(v)}_{{\cal D}}= 2g^2 m^2
\int_{-\infty}^{+\infty} \! d\tau_2\
\epsilon(\tau_2 - v)
\mbox{Re} \left[ f(\tau_2)   f^*(v)
e^{-i m( \tau_2 -v)}
\bra{0_M} \phi(\tau_2) i\partial_v \phi(v) \ket{0_M}
\right]
\nonumber\\
\label{Com3}
\end{eqnarray}
where we have used the commutation relation
\begin{equation}
\left[T_{vv}(v), \phi(\tau_2)\phi(\tau_1) \right]_- =
-2i \delta(\tau_1 -v)
\phi(\tau_2)
\partial_v \phi(v)
-2i \delta(\tau_2 -v)
\partial_v \phi(v)
\phi(\tau_1)
\label{Com4}
\end{equation}
and the antisymmetric character of $\epsilon(\tau_2 - \tau_1)$.
Since the expectation value $\bra{0_M} \phi(\tau_2)\phi(v) \ket{0_M}$ is
evaluated along the
accelerated trajectory eq. (\ref{threefivexiibis}),
it is a function of $\tau_2-v$ only. Therefore
the integrand of eq. (\ref{Com3}) is an odd function of $\tau_2-v$
and the integral vanishes.
Hence $\bk{T_{vv}(v)}_{{\cal D}}$ is
an energy repartition which is concerned only with
the transients induced by the switch on and off effects.

5. When $f(\tau)$ is a slowly varying function with respect to
both $1/m$ and $1/a$ (c.f. the discussion associated with eq. (\ref{golder}),
$\bk{T_{vv}(v)}_{{\cal D}}$ is smaller than the contribution of
Re$[\bk{T_{vv}(v)\phi_m\phi_m^\dagger}]$ by a factor $1/aT$ except
near the edges of the interaction period where $f(\tau)$ almost
vanishes.
This can be seen by developing $f(\tau_2)$ given in eq. \ref{Com3} in a series
around $\tau_2 =v$ and evaluating the magnitude of the first non vanishing
term, i.e. one
treats the variations of the
switch off function $f(\tau)$ as an adiabatic effect.
One finds that indeed the $\cal{D}$ is smaller than
Re$[\bk{T_{vv}(v)\phi_m\phi_m^\dagger}]$ except when $\tau > aT^2$.

\vfill \newpage

Figure Captions

Figure 1.

\noindent The Minkowski coordinates $t,z$ and $U,V$. The
left (L) and right (R) Rindler quadrants. The Rindler coordinates $\tau ,
\rho$ in R and the trajectory of a uniformly accelerated atom.

Figure 2.

\noindent The absolute value of the switch function $f(\tau)$ given in eq.
(\ref{fapprox}) for $m=2a$ and $T=3 a^{-1}$. $\tau$ is given in units of
$a^{-1}$.

Figure 3.

\noindent The mean Rindler energy density ($\langle
T_{vv}(v)\rangle_{therm.}$) emitted to order $g^2$ at thermal equilibrium
 is represented
for  $m=2a$ and $T=3a^{-1}$. $v$ is given in units of $a^{-1}$ and $T_{vv}$ in
arbitrary units since the flux is proportional to the coupling $g$.
One sees the vanishing of the flux in the steady regime and the positivity
of the transients.
In the Minkowski description they are enenhanced by the jacobian $dV/dv$ to
make the total
Minkowski energy emitted positive.

Figure 4.

\noindent The conditional value $\bk{T_{vv}}_e$ if the two level
atom is initially in its ground state and ends up in its excited state. The
parameters are the same as in Figure 2 and 3: $m=2a$ and $T = 3 a^{-1}$. The
$v$ axis is given in units of $a^{-1}$ and $T_{vv}$ in units of $a^2$.
For  $U<U_a, V>0$, $\langle T_{vv} \rangle _e$ is complex and oscillates. The
real part has a central positive bump which encodes that their is a
rindleron carrying positive energy which will induce the transition of the
atom. For $V<0$, $\langle T_{vv} \rangle _e$
is real and positive. It describes the partner of the rindleron which will
be absorbed by the atom. The oscillations of $\langle T_{vv}
(U<U_a, V>0) \rangle _e$ are such that the total Minkowski energy of the
vacuum fluctuation vanishes. For $U>U_a, V>0$, $\langle T_{vv} \rangle _e$
is positive and of order $N_m$. In order to represent it we have had to
change the vertical scale.

Figure 5.

\noindent A schematic picture of the energy fluxes $\langle T_{vv} \rangle
_e$. We have represented in dark grey the regions where
$\langle T_{vv} \rangle_e$ is $O(N_m+1)$ and in light grey the regions
where it is $O(N_m)$.

Figure 6.

\noindent The local description of a vacuum fluctuation
giving rise to a Hawking photon emitted around $u=u_0$ is
represented in a Penrose diagram. The shaded areas
correspond to the regions where
$T_{\tilde u \tilde u}(\tilde u)$ is non vanishing.
$v=v_S$ is the trajectory of the collapsing spherically
symmetric shell of massless matter.

Figure 7.

\noindent The same as in figure 6 drawn in
Eddington-Finkelstein coordinates.


\begin{thebibliography}{99}


\bibitem{hawk2}S.W. Hawking, Nature {\bf 248}, 30 (1974)\\
Commun. math. Phys. {\bf 43}, 199 (1975).

\bibitem{Unruh} W.G. Unruh, Phys. Rev. {\bf D14} (1976) 287.

\bibitem{birreld}N.D. Birrel and P.C.W. Davies, {\it Quantum Fields in Curved
Space},
Cambridge University Press (1982).


\bibitem{UnWa}W. G. Unruh  and  R. M. Wald, Phys. Rev. D {\bf 29} (1984) 1047


\bibitem{Grove} P. G. Grove, Class. Quantum Grav. {\bf 3} (1986) 801

\bibitem{RSG} D. Raine, D. Sciama and P. Grove, Proc. R. Soc. {\bf A435}
(1991)


\bibitem{Unru2}W. G. Unruh, Phys. Rev. D {\bf 46} (1992) 3271



\bibitem{Hint} F. Hinterleitner, Ann. Phys. (N.Y.) {\bf 226} (1993) 165


\bibitem{mpbrsg} S. Massar, R. Parentani and R. Brout, Class. Quantum
Grav.
{\bf 10} (1993) 385

\bibitem{AM}J. Audretsch and R. M\"uller, Phys. Rev. D {\bf 49} (1994)
 4056; Phys. Rev D {\bf 49} (1994) 6566;
Phys. Rev A {\bf 50} (1994) 1755



\bibitem{bmpps} R. Brout,
 S. Massar,
  S. Popescu,
   R. Parentani and Ph. Spindel, ``{\it Quantum Source of the Back
Reaction on a
 Classical Field}" Preprint ULB-TH 93/16, UMH-MG 93/03, (1993).


\bibitem{aharo}Y. Aharonov, D. Albert, A. Casher and L. Vaidman, Phys.
Lett.
A {\bf 124} 199 (1987),\\
Y. Aharonov and L. Vaidman, Phys. Rev. A {\bf 41} 11 (1990).

\bibitem{Vonn}J. von Newmann, {\em Mathematical Foundations of Quantum
Mechanics}, Princeton University Press, Princeton (1955)


\bibitem{Bardeen} J.M. Bardeen, Phys. Rev. Letters {\bf 46} (1981)
382.

\bibitem{PT} R. Parentani and T. Piran,  Phys. Rev. Lett. {\bf 73} (1994)
2805


\bibitem{Massar2}  S. Massar, {\em The semi classical back reaction to black
hole evaporation} preprint ULB-TH 94/19, gr-qc/9411039

\bibitem{THooft} G. 't Hooft, Nucl. Phys. B {\bf 256} (1985) 727


\bibitem{Jacobson} T. Jacobson, Phys. Rev.
 {\bf D44} (1991) 1731, {\bf D48} (1993) 728.


\bibitem{EMP}F. Englert, S. Massar and R. Parentani,
 Class. Quantum. Grav. {\bf 11} (1994) 2919

\bibitem{sus1} L. Susskind, Phys. Rev. D {\bf 49} (1994) 6606


\bibitem{sus2} L. Susskind,
``{\it Some speculations about Black Hole Entropy in String Theory}"
Preprint RU-93-44, hep-th/9309145 (1993).

\bibitem{STUg} L. Susskind, L. Thorlacius  and J. Uglum, Phys. Rev. D {\bf 48}
(1993) 3743

\bibitem{tHooft}C. R. Stephens, G. 't Hooft and B. F. Whiting,
Class. Quant. Grav. {\bf 11} (1994) 621

\bibitem{THooft2}G. 't Hooft,
{\em Horizon Operator Approach to Black Hole Quantization}
preprint THU-94/02 (1994) gr-qc/9402037

\bibitem{Engl}F. Englert, ``{\it Operator weak values and black hole
complementarity}'', preprint ULB-TH 03/95, to be published in the annals of
the Oskar Klein Centenary Symposium

\bibitem{Verlinde} K. Schoutens, H. Verlinde,
 E. Verlinde, ``{\it Black Hole Evaporation and Quantum Gravity}"
Preprint CERN-TH.7142/94, PUPT-1441, (1994), hep-th/ 9401081.

\bibitem{Wilceck} P. Kraus and F. Wilczek,
Nucl. Phys. B433 (1995) 403

\bibitem{Boul2}D. G. Boulware, Annals of Physics {\bf 124} (1980) 169

\bibitem{tmunu} R. Parentani,
  Class. Quantum Grav. {\bf 10} (1993) 1409.

\bibitem{pbt}R. Parentani and R. Brout, Int. J. Mod. Phys. D{\bf 1}, 169
(1992).

\bibitem{GO} R. Brout, S. Massar, R. Parentani and Ph. Spindel, {\em A
Primer for Black Hole Quantum Physics} (1995) ULB-TH 95/02, UMH-MG 95/01 and
LPTENS 95/03 {\em submitted to Phys. Rep.}

\bibitem{Par}R. Parentani, {\em The Recoils of the Accelerated Atom
and the Decoherence of its Fluxes}, preprint (1995) LPTHENS 95/02

\bibitem{Wald} R. Wald, Commun. Math. Phys. {\bf 45} (1975) 9.

\bibitem{Grove2} P. G. Grove
in {\it The Origin of Structure in the Universe} edited by E.
Gunzig and P. Nardone, Kluwer Academic Publishers (Netherlands) 1993

\bibitem{wald} R. Wald, Commun. Math. Phys., {\bf 54},1 (1977),
Phys. Rev. D, {\bf 17}, 1477 (1978),

\bibitem{mpblocal} S. Massar, R. Parentani and R. Brout, Class. Quantum
Grav. {\bf 10} (1993) 2431

\bibitem{mas} S. Massar, Int. J. Mod. Phys. D {\bf 3} (1994) 237

\end{thebibliography}
\end{document}